%% file: main.tex
\documentclass[twoside, 12pt, colorlinks = true, linkcolor = black, urlcolor  = blue, citecolor = blue, anchorcolor = blue]{article}

\usepackage{epsfig}

\def\ss{\mbox{\boldmath $\sigma$}}
\newcommand{\be}{\begin{equation}}
\newcommand{\ee}{\end{equation}}
\newcommand{\bea}{\begin{eqnarray}}
\newcommand{\eea}{\end{eqnarray}}

\usepackage[utf8]{inputenc}
\usepackage{amsmath,amssymb,amsfonts,bm}
\usepackage{stmaryrd,latexsym}
\usepackage{color}
\usepackage{graphicx}
\usepackage{float} 
\usepackage{epstopdf}
\usepackage[sort&compress,numbers]{natbib}
\usepackage{doi}%<----------
\usepackage[]{hyperref}
\usepackage{threeparttable}
\usepackage[normalem]{ulem} % [normalem] prevents the package from changing the default behavior of `\emph` to underline.
\usepackage{relsize}

\def\babar{\mbox{\slshape B\kern-0.1em{\smaller A}\kern-0.1em
    B\kern-0.1em{\smaller A\kern-0.2em R}}}
\newcommand{\panda}{$\overline{\text{P}}$ANDA}

\newcommand{\rmA}{{\rm A}}
\newcommand{\rmB}{{\rm B}}

\newcommand{\re}{\textrm{Re}}
\newcommand{\im}{\textrm{Im}}
\newcommand{\sth}{s_{\textrm{th}}}

\newcommand{\kstarkk}{K^*\bar{K}K}
\newcommand{\dzstar}{D^{*0}}
\newcommand{\dzstarbar}{\bar D^{*0}}
\newcommand{\dzbar}{\bar D^0}
\newcommand{\dzsdzsbar}{D^{*0}\bar D^{*0}}
\newcommand{\x}{X(3872)}

\newcommand{\sigh}{\Sigma_c^{(*)}\bar{D}^{(*)}}
\newcommand{\sigd}{\Sigma_c\bar{D}}
\newcommand{\sigdstar}{\Sigma_c\bar{D}^*}
\newcommand{\sigstard}{\Sigma_c^*\bar{D}}
\newcommand{\eg}{{\it e.g.}}
\newcommand{\ie}{{\it i.e.}}
\newcommand{\order}[1]{\mathcal{O}\left(#1\right)}

\begin{document}

\title{Threshold cusps and triangle singularities in hadronic reactions}

\author{Feng-Kun\ Guo,$^{1,3,}$\footnote{\texttt{fkguo@itp.ac.cn}} \quad
Xiao-Hai\ Liu,$^{2,}$\footnote{\texttt{xiaohai.liu@tju.edu.cn}} \quad
Shuntaro\ Sakai$^{1,}$\footnote{\texttt{shsakai@itp.ac.cn}}\\[2mm]
$^1${\it CAS Key Laboratory of Theoretical Physics, Institute of Theoretical Physics,}\\ {\it Chinese Academy of Sciences, Beijing 100190, China}\\
$^2${\it Center for Joint Quantum Studies and Department of Physics,}\\ {\it School of Science, Tianjin University, Tianjin 300350, China}\\
$^3${\it School of Physical Sciences, University of Chinese Academy of Sciences,}\\ {\it Beijing 100049, China}
}

\maketitle

\begin{abstract} 

The spectrum of hadrons is the manifestation of color confinement of quantum chromodynamics. Hadronic resonances correspond to poles of the $S$-matrix. 
Since 2003, lots of new hadron resonant structures were discovered in the mass regions from light mesons to hadrons containing a pair of a heavy quark and an antiquark. Many of them are candidates of exotic hadrons, and they are usually observed as peaks in invariant mass distributions. However, the $S$-matrix also has kinematical singularities due to the on-shellness of intermediate particles for a process, such as two-body thresholds and triangle singularities (TSs), and they can produce peaks as well. On the one hand, such singularities may be misidentified as resonances; on the other hand, they can be used as tools for precision measurements. In this paper, we review the threshold cusps and various triangle singularities in hadronic reactions, paying attention to their manifestations in phenomena related to exotic hadron candidates. 

\end{abstract}

% \begin{keyword}

% Hadron spectroscopy \sep triangle singularity \sep threshold cusp \sep exotic hadrons \sep multiquark states

% \end{keyword}

\newpage

\tableofcontents

\input{section1}

\input{section2}

\input{section3}
\input{section4}

\input{section5}
\input{section6}
\input{section7}

\input{appendix}

\section*{Acknowledgements}

We would like to thank all our collaborators to share their insights with us. We are grateful to Ulf-G. Mei{\ss}ner for a careful reading of the manuscript.
This work is supported in part by the National Natural Science Foundation of China (NSFC) and  the Deutsche Forschungsgemeinschaft (DFG) through the funds provided to the Sino-German Collaborative Research Center  CRC110 ``Symmetries and the Emergence of Structure in QCD"  (NSFC Grant No. 11621131001), by the NSFC under Grants No. 11835015, No.~11947302, No. 11975165 and No. 11961141012, by the Chinese Academy of Sciences (CAS) under Grants No. QYZDB-SSW-SYS013 and No. XDPB09, and by the CAS Center for Excellence in Particle Physics (CCEPP).
S.S. is also supported by the 2019 International Postdoctoral Exchange Program and by the CAS President's International Fellowship Initiative (PIFI) under Grant No. 2019PM0108.

\bigskip

\bibliographystyle{apsrev4-1}
\bibliography{ts}

\end{document}

%% file: section1.tex
\newpage
\section{Introduction }
\label{sec:1}

The dynamics of the strong interaction is described by quantum chromodynamics (QCD). It is a non-Abelian gauge theory with quarks and gluons, which carry the color quantum number, as the basic degrees of freedom. Although the Lagrangian of QCD looks remarkably simple and at high energies the theory can be solved by a perturbative expansion in a series of the strong coupling constant, it becomes notoriously difficult in the low-energy regime which is completely dominated by nonperturbative dynamics. The characteristic energy scale for this regime is $\Lambda_\text{QCD}$, which is of the order of a few hundreds of MeV.
The most salient nonperturbative feature of QCD at low energies is the color confinement phenomenon: the asymptotic strongly interacting particles that can be detected in experiments are not the colorful quarks and gluons, but the colorless hadrons composed of them. Therefore, understanding the spectroscopy of mesons and baryons is crucial to gain deeper insights into the strong interaction in the nonperturbative regime.

For a long time, the quark model was successful in classifying all the observed hadrons into different flavor SU(3) multiplets~\cite{GellMann:1964nj,Zweig:1981pd}, with the baryons and mesons formed from three quarks ($qqq$) and quark-antiquark ($q\bar q$) pairs, respectively. After the birth of QCD, QCD-based dynamics of quarks was introduced into the quark model through the exchange of gluons and/or pions, and such quark models are normally constituent (or valence) quark models, see, \eg, Refs.~\cite{Chodos:1974je,DeRujula:1975qlm,Eichten:1978tg,Eichten:1979ms,Manohar:1983md,Godfrey:1985xj,Capstick:1986bm}.
Assignments of many observed mesons and baryons into $q\bar q$ and $qqq$ multiplets can be found in the review on quark model in the Review of Particle Physics (RPP) by the Particle Data Group (PDG)~\cite{Tanabashi:2018oca}. 
However, in addition to the conventional quark-antiquark mesons and three-quark baryons, in the quark model notation there can be other configurations of color singlets, such as glueballs made purely of gluons, hybrid states consisting of both quarks and gluon excitations, and multiquarks. Hadrons with such configurations are called exotic hadrons. In fact, the possibility of having more quarks in a hadron, such as $qq\bar q\bar q$ and $qqqq\bar q$, were already mentioned in the seminal papers of quark model~\cite{GellMann:1964nj,Zweig:1981pd}. 
The dynamical quark models were generally successful when the results were confronted with the observed spectrum, with a few exceptions such as the lightest scalar meson nonet, including $f_0(500)$, $K_0^*(700)$, $f_0(980)$ and $a_0(980)$. These were then suggested to be tetraquark states in the 1970s~\cite{Jaffe:1975fd,Jaffe:1976ig} and later on as hadronic molecules~\cite{Weinstein:1982gc,Oller:1997ti,Kaiser:1998fi,Locher:1997gr,Nieves:1999bx,Pelaez:2015qba,Baru:2003qq}.
The spectroscopy of exotic hadrons and searching for them in various high energy experiments have been one of the central issues in the study of low-energy strong interactions since then~\cite{Jaffe:1975fd,Jaffe:1976ig,Jaffe:1976ih}. 
For any type of the above mentioned exotic hadron configurations, one would expect that there should be the whole family of ground and excited states just like the normal mesons and baryons. 
However,  unambiguous confirmation of exotic hadrons was lacking.  This is partly due to the limits of the experiments in the last century, and also reflects how little the excited hadron spectrum has been really understood from QCD.

Tremendous experimental progress has been made in the last two decades because of the operation of the modern generation of experiments such as the $B$ factories \babar{} and Belle, the high-luminosity electron-positron collision experiments such as BESIII, and the experiments at hadron colliders including CDF, D0, ATLAS, CMS and LHCb. 
The year of 2003 witnessed the observations of the $\Theta(1540)$~\cite{Nakano:2003qx}, the $D_{s0}^*(2317)$~\cite{Aubert:2003fg}, the $D_{s1}(2460)$~\cite{Besson:2003cp}, and the $\x$\footnote{The $\x$ is named $\chi_{c1}(3872)$ according to its quantum numbers $I^G(J^{PC})=0^+(1^{++})$ by the PDG~\cite{Tanabashi:2018oca}. Similarly, the vector charmonium-like states $Y(4260)$ and $Y(4660 )$ mentioned below are called $\psi(4260)$ and $\psi(4660)$, respectively. This naming scheme does not mean that the PDG assumes them to be normal $c\bar c$ charmonium states. Here we follow the $XYZ$ naming scheme that is still used in most of the relevant publications.}~\cite{Choi:2003ue}. 
These discoveries triggered lots of theoretical and experimental studies. Although the pentaquark candidate $\Theta(1540)$ died away due to experiments with higher statistics~\cite{Hicks:2012zz},\footnote{A conventional explanation of the observed peaks was given in Refs.~\cite{MartinezTorres:2010zzb,Torres:2010jh}.} the others got confirmed in following experiments. 
More new resonance-like structures were observed in the subsequent years. Notable examples include the charmonium-like states $Y(4260)$~\cite{Aubert:2005rm} and $Y(4660)$~\cite{Wang:2007ea}, the charged structures in the charmonium mass region $Z_c(4430)$~\cite{Choi:2007wga}, $Z_c(3900)$~\cite{Ablikim:2013mio,Liu:2013dau} and $Z_c(4020)$~\cite{Ablikim:2013wzq}, the charged bottomonium-like structures $Z_b(10610)$ and $Z_b(10650)$~\cite{Belle:2011aa}, and the pentaquark candidates with hidden charm $P_c(4312)$, $P_c(4440)$ and $P_c(4457)$~\cite{Aaij:2015tga,Aaij:2019vzc}. Most of these new structures were observed in the heavy-flavor sector. In particular, the heavy quarkonium-like ones are often called $XYZ$ states in the literature due to the undetermined internal structure. 

On the one hand, these discoveries enlarged the known QCD spectrum to a large extent; on the other hand, they became a nice showcase of the intricate nonperturbative nature of QCD at low energies\footnote{Hadron spectroscopy is classified as a low-energy QCD problem even for systems containing heavy quarks. Here the ``low energy" should be understood as the energy from which  the heavy quark mass has been subtracted from the system, and it is of the order $\Lambda_\text{QCD}$, a few hundreds of MeV.}: most of them fall off the expectations from quark model, which despite being just a model had provided useful guidance in classifying a large amount of hadrons into various multiplets. 
Therefore, they are regarded as prominent candidates of exotic hadrons. However, how the spectrum of exotic hadrons should be organized and even what types of exotic hadrons can be well defined are still unclear. Partly because of this, the observation of each of these new structures leads to different models such as compact tetraquarks (or pentaquarks), hadronic molecules, hybrid states, hadro-charmonia, and kinematic effects, etc.  
Nevertheless, a deeper understanding of how the hadron spectrum, in particular that of the excited hadrons above (or at least close to) strong decay thresholds, is organized can shed light on the color confinement problem of QCD. For that, we first need to uncover the pattern of the observed structures. It is possible that some of these structures do not really correspond to the existence of a new hadron resonance, but are mainly due to effects of special kinematics such as threshold cusps and/or triangle singularities (to be generally called kinematical effects). Such kinematical effects are the foci of this review article. 
For more comprehensive reviews of the new hadronic structures and the corresponding model explanations, we refer to Refs.~\cite{Jaffe:2004ph,Swanson:2006st,Klempt:2007cp,Klempt:2009pi,Brambilla:2010cs,Chen:2016qju,Chen:2016spr,Esposito:2016noz,Hosaka:2016pey,Richard:2016eis,Lebed:2016hpi,Dong:2017gaw,Guo:2017jvc,Ali:2017jda,Olsen:2017bmm,Kou:2018nap,Kalashnikova:2018vkv,Cerri:2018ypt,Liu:2019zoy,Brambilla:2019esw}.  

In many cases, hadron resonances are observed as narrow or broad peaks in the invariant mass distributions of certain hadronic final states. Their masses and widths are normally obtained by fitting to the invariant mass distributions, or Dalitz plot distributions if the statistics is high enough, using the isobar model with the Breit--Wigner parameterization for resonances~\cite{Tanabashi:2018oca} (or the Flatt{\'e} parameterization for near-threshold states~\cite{Flatte:1976xu}) for resonances. 
However, there are traps in this way of identifying resonances. 

Firstly, in the ideal case when a narrow resonance is well isolated from the others and there is no background, it does show up as a peak with the peaking position and peak width roughly correspond to its mass and width, respectively. However, the realistic situation is often much more complicated: there can be coupled channels with thresholds near the resonance; there can be other resonances not far away; and there are always contributions from the non-resonant background.   
Depending on the interference with all these other contributions, a resonance may even show up as a dip, see, \eg, Ref.~\cite{Taylor:1972}. The case of the $f_0(980)$ provides a nice example. While it shows up as a sharp peak in the $\pi\pi$ invariant mass distributions for the processes $J/\psi\to \phi\pi^+\pi^-$~\cite{Ablikim:2004wn} and $D^+\to\pi^+\pi^-\pi^+$~\cite{Aitala:2000xu}, it appears as a dip for the process $J/\psi\to\omega\pi^+\pi^-$~\cite{Ablikim:2004qna}\footnote{In this case, there is in fact a small peak corresponding to the $f_0(980)$, which is absent in the $\pi\pi$ scattering cross section, in the $\pi\pi$ invariant mass distribution. It is not evident and is followed by the dip at around 1~GeV. For further details, see the discussions in Refs.~\cite{Roca:2004uc,Lahde:2006wr,Liu:2009ub}.} and the $\pi\pi$ elastic scattering cross section for isospin $I=0$ and $J=0$. 

Secondly, not all peaks in invariant mass distributions (or bands in Dalitz plots) are  due to the existence of a resonance with a mass around the peak energy. 
Peaks in invariant mass distributions arise often because the transition amplitude or the $S$-matrix has nearby singularities in the complex energy plane.
Resonances correspond to one kind of singularities, \ie, poles of the $S$-matrix. Their origin is dynamical. The amplitude develops a pole because the interaction between quarks and gluons or among hadrons has the right strength. Hence, the pole position relies on details of the interaction. Poles are necessarily a nonperturbative phenomenon. 
In addition to the dynamical poles, the $S$-matrix also has kinematical  singularities in the sense that their locations are determined completely by kinematical variables, such as masses and energies of the involved particles, instead of the interaction strength. 
They occur because intermediate particles between the initial and final states can become real propagating particles.  Such singularities are called Landau singularities, and their locations are determined by the Landau equations~\cite{Landau:1959fi}.\footnote{The equations were also derived in Bjorken's PhD thesis~\cite{Bjorken:1959fd} and by Mathews~\cite{Mathews:1959zz} and Nakanishi~\cite{Nakanishi:1959}.}
The simplest case is the square-root branch points at normal two-body thresholds, which always produce cusps at thresholds of the $S$-wave channels in the energy distributions, known as the Wigner cusp~\cite{Wigner:1948zz}.
A more complicated type is the so-called triangle singularity (TS) due to three on-shell intermediate particles in a loop diagram. The TS is a logarithmic singularity. Under certain conditions which go under the name of the Coleman--Norton theorem~\cite{Coleman:1965xm}, the singularity can be located on the physical boundary, \ie, in the physical region, when the decay widths of intermediate particles are neglected. 
Then, being logarithmic, the singularity may produce observable effects (a peak or a dip depending on the interference with the other contributions) if the involved interactions are strong enough. Sometimes, such effects  mimic the behavior of a resonance, and lay traps on the way of establishing an unambiguous hadron spectroscopy.  
It is thus important to distinguish  kinematic singularities from genuine resonances.

In fact, whether TSs can produce observable effects has been discussed since the 1960s~\cite{Peierls:1961zz,Goebel:1964zz,Aitchison:1964zz,Schmid:1967ojm,Goebel:1982yb,Anisovich:1983rj,He:1984ev}. 
However, due to limited processes that were accessible in experiments at that time (such as the nucleon-nucleon and pion-nucleon reactions), no convincing evidence was found for peaks produced by TSs.
Nowadays, with the much larger scope of high-energy experiments, especially for processes involving heavy quarks, quite a few experimentally observed peaks were suggested to be due to TSs. In particular, some of them are prominent candidates of exotic hadrons. Furthermore, there are also predictions of in which processes and at what energy region TSs are expected to play an important role. 
Resonant structures that were suggested to be due to TSs (or the singularities were expected to play a sizable role), the related processes and the involved intermediate particles in the triangle loops are listed in Table~\ref{tab:list-meson} for mesons, and in Table~\ref{tab:list-baryon} for baryons.

%-------------------------------
\begin{table}[tbhp]
\begin{center}
\begin{threeparttable}
\caption{List of processes in which TSs were suggested to play a significant role. In the first column, the resonances or resonance-like structures (energies) that receive a large TS contribution are listed. For the predicted ones awaiting experimental confirmation, the approximate relevant energies are given, marked with underlined dots. 
The intermediate particles in the triangle diagrams are listed in the third column, and the charge conjugated channels are implicit for the loops of mesons whenever necessary. We use I(F) to represent that the TS effects are responsible to produce a resonance-like structure in the initial (final) states which couple to the first (last) two of the three intermediate particles. In this table, we use the following short names for resonances: $a_0$ for $a_0(980)$; $f_0$  for $f_0(980)$; $K_2^*$ for $K_2^*(1430)$; $D_{s0}^*$ for $D_{s0}^*(2317)$; $D_{s1}$ for $D_{s1}(2460)$; $D_1$ for $D_1(2420)$; $D_2$ for $D_2(2460)$; $X$ for $\x$. }
\label{tab:list-meson}
\begin{tabular}{lllll}
\hline\hline
Structures & Processes & Loops & I/F & Refs. \\\hline
$\rho(1480)$~\cite{Bityukov:1986yd,Bityukov:1987bh} & $\pi^- p \to  \phi\pi^0 n $ & $\kstarkk$ & I & \cite{Achasov:1987ku,Achasov:1989ma} \\
$\eta(1405/1475)$~\cite{Augustin:1989zf,Bai:1990hs,Bai:1998eg,Bai:2000ss,BESIII:2012aa} & $\eta(1405/1475)\to\pi f_0 
$ 
& $\kstarkk$ & I & \cite{Wu:2011yx,Aceti:2012dj,Wu:2012pg,Achasov:2015uua,Du:2019idk}\tnote{a,b}  \\ 
$f_1(1420)$~\cite{Barberis:1998by} & $f_1(1285)\to \pi a_0/\pi f_0
$ & $\kstarkk$ & I & \cite{Wu:2012pg,Aceti:2015zva,Debastiani:2016xgg,Achasov:2016wll}\tnote{b} \\ 
$a_1(1420)$~\cite{Adolph:2015pws,Akhunzyanov:2018lqa} & $a_1(1260)\to f_0\pi\to3\pi$ & $\kstarkk$& I & \cite{Ketzer:2015tqa,Aceti:2016yeb,Akhunzyanov:2018lqa} 
 \\ 
$1.4$~GeV~\cite{Ablikim:2018pik}
& $J/\psi\to \phi\pi^0\eta/\phi\pi^0\pi^0$ & $\kstarkk$ & I & \cite{Jing:2019cbw}\tnote{b} \\
 \dotuline{$1.42$ GeV} & $B^-\to \dzstar\pi^-f_0 (a_0),\tau\to \nu_\tau \pi^- f_0 (a_0)$  & $\kstarkk$& I & \cite{Pavao:2017kcr,Dai:2018rra} \\
  &  $D_s^+\to\pi^+\pi^0f_0 (a_0),\bar B_s^0\to J/\psi\pi^0 f_0 (a_0)$  &$\kstarkk$ &I  & \cite{Sakai:2017iqs,Liang:2017ijf} \\
 $f_2(1810)$~\cite{Tanabashi:2018oca} & $f_2(1640)\to \pi\pi\rho$ & $K^*\bar K^* K$ & I & \cite{Xie:2016lvs} \\ 
 \dotuline{$1.65$ GeV} & $\tau\to \nu_\tau \pi^- f_1(1285)$ & $K^*\bar K^* K$ & I & \cite{Oset:2018zgc} \\
 \dotuline{1515 MeV} & $J/\psi\to K^+K^-f_0\left(a_0\right)$ & $\phi\bar K K$& I & \cite{Liang:2019jtr}  \\
 \hline
 \dotuline{2.85 GeV, 3.0 GeV} & $B^-\to K^-\pi^-D_{s0}^*/K^-\pi^-D_{s1}$ & $K^{*0} D^{(*)0}K^+$ & I & \cite{Liu:2015taa,Sakai:2017hpg} \\
 \dotuline{5.78~GeV} & $B_c^+\to \pi^0\pi^+B_s^0$ & $\bar K^{*0} B^+ \bar K$ & F & \cite{Liu:2017vsf}
 \\\hline
\dotuline{$[4.01,4.02]$ GeV} & $[\dzstarbar\dzstar]\to \gamma X$ & $\dzstar\dzstarbar D^0$ & I & \cite{Guo:2019qcn} \\
\dotuline{$4015$ MeV} & $e^+e^-\to \gamma X$  &$\dzstar\dzstarbar D^0$& I & \cite{Braaten:2019gfj,Braaten:2019gwc} \\
 \dotuline{$4015$ MeV}& $B\to K X\pi$, $pp/p\bar p\to X\pi+$anything & $\dzstar\dzstarbar D^0$  & I & \cite{Braaten:2019yua,Braaten:2019sxh} \\
$\Upsilon(11020)$~\cite{Santel:2015qga,Abdesselam:2015zza} & $e^+e^-\to Z_b\pi$ & $B_1(5721)\bar B B^*$ & I &\cite{Wang:2013hga,Bondar:2016pox} \\
 \dotuline{$3.73$~GeV}& $X\to \pi^0\pi^+\pi^-$ & $\dzstar\dzbar D^0$ & F & \cite{Achasov:2019wvw}  \\
 \dotuline{$[4.22,4.24]$ GeV} & $e^+e^-\to \gamma J/\psi\phi$/$\pi^0 J/\psi\eta$  &$D_{s0(s1)}^{*} \bar D_s^{(*)} D_s^{(*)}$& F & \cite{Liu:2015cah} \\  \dotuline{$[4.08,4.09]$ GeV} & $e^+e^-\to \pi^0 J/\psi\eta$  &$D_{s0(s1)}^{*} \bar D_s^{(*)} D_s^{(*)}$& F & \cite{Liu:2015cah} \\
 $Z_c(3900)$~\cite{Ablikim:2013mio,Liu:2013dau} & $e^+e^-\to J/\psi\pi^+\pi^-$ & $D_1\bar D D^*$ & F & \cite{Wang:2013cya,Wang:2013hga,Liu:2013vfa,Liu:2014spa,Pilloni:2016obd,Gong:2016jzb}\tnote{c} \\
  & & $D_0^*(2400)\bar D^* D$ & F & \cite{Szczepaniak:2015eza,Szczepaniak:2015hya} \\
 $Z_c(4020,4030)$~\cite{Ablikim:2013wzq,Ablikim:2017oaf} & $e^+e^-\to \pi^+\pi^-h_c(\psi')$ & $D_{1(2)}\bar D^{(*)}D^{(*)}$ & F & \cite{Liu:2014spa} \\
 $X(4700)$~\cite{Aaij:2016iza,Aaij:2016nsc} & $B^+\to K^+  J/\psi\phi $ & $K_1(1650) \psi'\phi$ & F & \cite{Liu:2016onn} \\
 $Z_c(4430)$~\cite{Choi:2007wga,Aaij:2014jqa} & $\bar{B}^0\to  K^-\pi^+J/\psi$ & $\bar K^{*0} \psi(4260)\pi^+$ & F& \cite{Nakamura:2019btl}\\
 $Z_c(4200)$~\cite{Chilikin:2014bkk,Aaij:2019ipm} & $\bar{B}^0\to K^-\pi^+\psi(2S)$ & $\bar{K}^*_2\psi(3770)\pi^+$ &F & \cite{Nakamura:2019btl}\\
  & $\Lambda_b^0\to p\,\pi^-J/\psi$ &    
 $N^*\psi(3770)\pi^-$ &F & \cite{Nakamura:2019btl}\\
 $X(4050)^\pm$~\cite{Mizuk:2008me} & $\bar{B}^0\to K^-\pi^+\chi_{c1}$ & $\bar K^{*0} X\pi^+$ & F & \cite{Nakamura:2019emd}\\
 $X(4250)^\pm$~\cite{Mizuk:2008me} & $\bar{B}^0\to K^-\pi^+\chi_{c1}$ & $\bar{K}^*_2\psi(3770)\pi^+$ & F & \cite{Nakamura:2019emd}\\
$Z_b(10610)$~\cite{Belle:2011aa} & $e^+e^-\to \Upsilon(1S)\pi^+\pi^-$ &$B_J^*\bar B^* B$ & F & \cite{Szczepaniak:2015eza} \\
\hline\hline
\end{tabular}
\vspace{-2mm}
\begin{tablenotes}
\begin{minipage}[t]{16.5 cm}
\noindent\item[a] The width effect from the intermediate $K^*$ is considered in Refs.~\cite{Achasov:2015uua,Du:2019idk}.
\noindent\item[b] TS enhanced isospin breaking effects are discussed in these references.
\noindent\item[c] The $D_1\bar D D^*$ triangle diagrams have also been considered in the analysis of Ref.~\cite{Albaladejo:2015lob}.
\end{minipage}
\end{tablenotes}
\end{threeparttable}
\end{center}
\end{table}  
%-------------------------------

%-------------------------------
\begin{table}[tbh]
\begin{center}
\begin{threeparttable}
\caption{Similar to Table~\ref{tab:list-meson}, but for processes of baryons or producing baryon-like structures.  }
\label{tab:list-baryon}
\begin{tabular}{lllll}
\hline\hline
Structures & Processes & Loops & I/F & Refs. \\\hline
$2.1$~GeV~\cite{Moriya:2013hwg} & $\gamma p^+\to N^*(2030)\to K^+\Lambda(1405)$ & $K^*\Sigma\pi$ & I & \cite{Wang:2016dtb} \\
\dotuline{$2.1$~GeV} & $\pi^- p^+\to K^0\Lambda(1405),\; pp\to p K^+\Lambda(1405)$ & $K^*\Sigma\pi$ & I & \cite{Bayar:2017svj} \\
\dotuline{1.88~GeV} & $\Lambda_c^+\to\pi^+\pi^0\pi\Sigma$ & $\bar K^* N\bar K $ & I & \cite{Dai:2018hqb,Xie:2018gbi}\tnote{a} \\
$N(1700)$~\cite{Tanabashi:2018oca} & $N(1700)\to \pi \Delta$ & $\rho N \pi$ & I &  \cite{Roca:2017bvy} \\
$N(1875)$~\cite{Tanabashi:2018oca} & $N(1875)\to \pi N(1535)$ & $\Sigma^*K\Lambda$ & I &  \cite{Samart:2017scf} \\
$\Delta(1700)$~\cite{Ajaka:2008zz,Kashevarov:2009ww,Gutz:2014wit} & $\gamma p \to \Delta(1700)\to \pi N(1535) \to p \pi^0\eta$ & $\Delta\eta p$ & I & \cite{Debastiani:2017dlz} \\
$2.2$~GeV~\cite{Pal:2017ypp} & $\Lambda_c^+\to \pi^0\phi p$ & $\Sigma^* K^* \Lambda$ & F & \cite{Xie:2017mbe} \\
$1.66$~GeV~\cite{Yang:2015ytm,Shen:2018talk} & $\Lambda_c^+\to \pi^+ K^-p$ & $a_0\Lambda\eta,\Sigma^*\eta\Lambda$ & F & \cite{Liu:2019dqc} \\
$P_c(4450)$~\cite{Aaij:2015tga} & $\Lambda_b^0\to K^- J/\psi p$ & $\Lambda(1890)\chi_{c1}p$ & F & \cite{Guo:2015umn,Liu:2015fea,Guo:2016bkl,Bayar:2016ftu}\tnote{b} \\
& & $N(1900)\chi_{c1}p$ & F & \cite{Guo:2016bkl} \\
 peaks relevant for $P_c$& $\Lambda_b^0\to K^- J/\psi p$& $\bar D_{sJ}\Lambda_c^{(*)}\bar D_{}^{(*)}$ & F &\cite{Liu:2015fea,Aaij:2019vzc} \\
\hline\hline
\end{tabular}
\begin{tablenotes}
\begin{minipage}[t]{16.5 cm}
\noindent\item[a] TS enhanced isospin breaking effects are discussed in these references.
\noindent\item[b] Various possible combinations of a $\Lambda^*$ hyperon and a charmonium are considered in Refs.~\cite{Guo:2016bkl,Bayar:2016ftu}.
\end{minipage}
\end{tablenotes}
\end{threeparttable}
\end{center}
\end{table}     
%-------------------------------

In this article, we will review the manifestation of threshold cusps and TSs in hadronic reactions. Special attention will be paid to the conditions when such kinematic singularities are potentially important. The Landau equation will be briefly reviewed in Section~\ref{sec:2}. 
The two-body threshold cusp and how its effects are connected to the interaction strength of the final state interaction are discussed in Section~\ref{sec:3}, where the question of whether some of the near-threshold peaks are just due to threshold cusps is also  addressed. Section~\ref{sec:4} is devoted to discuss TSs in depth, covering the Coleman-Norton theorem and the Schmid theorem. The Argand plot for an amplitude with a TS as well as amplitude analysis including TSs will also be discussed.
In Section~\ref{sec:5}, we will review TSs that were suggested to produce peaking structures in initial states. Those leading to peaks in the invariant mass distributions of final states will be discussed in Section~\ref{sec:6}. We end up with a brief summary and outlook in Section~\ref{sec:7}.

Let us mention that here we focus more on applications of the kinematical singularities in hadronic reactions. For the mathematical foundation and more detailed discussions on the TS and other Landau singularities, we refer to the monographs~\cite{Eden:1966dnq,Chang:1983,Gribov:2009,Anisovich:2013gha} and the recent lecture notes by one of the main players in the early days~\cite{Aitchison:2015jxa}. The hadron structures for which threshold cusps and TSs were suggested to play an important role are often also good candidates of hadronic molecules. For a review devoted to various aspects of hadronic molecules, we refer to Ref.~\cite{Guo:2017jvc}. 

\bigskip

%% file: section2.tex
\section{Landau equations}
\label{sec:2}

\begin{figure}[tb]
\begin{center}
    \includegraphics[width=0.25\textwidth]{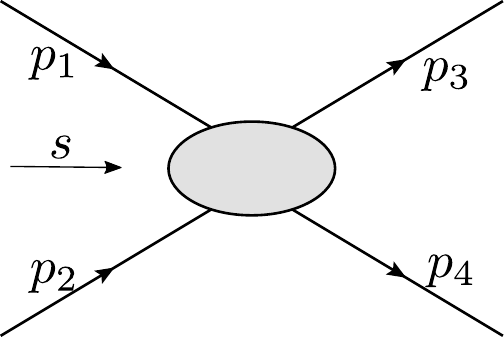}
\begin{minipage}[t]{16.5 cm}
\caption{Two-body scattering.
\label{fig:scattering}}
\end{minipage}
\end{center}
\end{figure}

Let us consider a physical process, for which the initial and final states can be connected through all possible intermediate states allowed by symmetry. When the energy is large enough the intermediate particles can be real and travel a long distance before they rescatter. Correspondingly, the amplitude gets an imaginary part because of the on-shell intermediate particles. Let us consider a two-body scattering shown in Fig.~\ref{fig:scattering}. 
Defining the Mandelstam variables in the usual way, 
\begin{equation}
  s = (p_1 + p_2)^2, \quad t = (p_1 - p_3)^3, \quad u = (p_1 - p_4)^2,
\end{equation}
the two-body scattering amplitude is a function of $s$ and $t$ with $u$ constrained by the identity $s+t+u= \sum_i p_i^2 = \sum_i m_i^2$. Defining the scattering angle $\theta$ as the angle between $\vec{p}_1$ and $\vec p_3$ in the $1+2$  (or $3+4$)  center-of-mass (c.m.) frame, $t$ can be expressed in terms of $s$ and $\cos\theta$ as
\begin{equation}
t(s,\cos{\theta}) = m_1^2+m_3^2 - 2 E_1 E_3 
+2 p_{\rm 1cm} p_{\rm 3cm} \cos\theta,
\end{equation}
where  
\begin{align}
  E_1 &=\frac{1}{2\sqrt{s}}(s+m_1^2-m_2^2), \quad
  E_3=\frac{1}{2\sqrt{s}}(s+m_3^2-m_4^2), \nonumber\\
  p_{\rm 1cm} & = \frac{1}{2\sqrt{s}}\sqrt{\lambda(s,m_1^2,m_2^2)}, \quad
  p_{\rm 3cm}  = \frac{1}{2\sqrt{s}}\sqrt{\lambda(s,m_3^2,m_4^2)}
\end{align}
are the energies and the magnitudes of the three-momenta of particles 1 and 3, respectively,  in the $1+2$  (or $3+4$) c.m. frame, with  $\lambda(x,y,z) = x^2 + y^2 + z^2 - 2xy - 2yz -2 xz$ the K\"all\`en function. 

The unitarity of the $S$-matrix, $S S^\dag =1$, leads to the unitary relation for the $T$-matrix defined as $S = 1 + i T$,
\begin{equation}
 T - T^\dag  = i T\, T^\dag .
\end{equation}
For the partial-wave scattering amplitude, 
\begin{equation}
  T_L(s) = \frac{1}{2} \int_{ - 1}^{ + 1} d\cos\theta \, P_L(\cos\theta) T(s,t,u),
\end{equation}
with the subindex $f(i)$ denoting the final (initial) state, considering only two-body intermediate states, the unitary relation becomes (for a clear discussion of the unitary relation, see, \eg, the classic book by Martin and Spearman~\cite{MartinSpearman})
\begin{equation}
  \im\, T_{L,fi}(s) = \sum_{a} T_{L,fa}(s)\, \rho_a(s)\, T_{L,ai}^*(s), \label{eq:uni}
\end{equation}
where  $s$ is the total energy squared in the c.m. frame. Here, the relation holds for coupled channels, and $\rho_a(s)$ is the two-body phase space factor for the intermediate channel $a$,
\begin{align}
  \label{eq:rho}
  \rho_a(s) &= \frac{k_a}{8\pi\sqrt{s} } \theta(\sqrt{s} - m_{a1} - m_{a2} ), \\
  k_a &=  \frac1{2\sqrt{s} } \sqrt{ \lambda(s, m_{a1}^2, m_{a2}^2) } = \frac{1}{2 \sqrt{s} } \sqrt{[s -(m_{a1} + m_{a2})^2] [s -(m_{a1} - m_{a2})^2]} \, , 
  \label{eq:qcm}
\end{align}
with $m_{a1}$ and $m_{a2}$ the masses of the internal particles in channel $a$, $k_a$ the c.m. momentum, and $\theta(x)$ the Heaviside step function. One sees that when the total energy in the c.m. frame is equal to the threshold of the intermediate states, $\sqrt{s} = m_{a1} + m_{a2} $, the phase space factor has a square-root branch point. As a result of Eq.~\eqref{eq:uni}, the scattering amplitude must have a branch point at threshold as well. 
It is because of the on-shellness of the intermediate particles, and is the simplest kinematical singularity. More complicated singularities such as the anomalous threshold~\cite{Nambu:1957vx,Karplus:1958zz} due to three internal particles can be analyzed by using Feynman diagrams, and the validity goes beyond perturbation theory~\cite{Landau:1959fi,Eden:1966dnq}.  The general conditions for kinematical singularities due to on-shell intermediate particles are described by the Landau equations~\cite{Landau:1959fi,Bjorken:1959fd} which we derive briefly as follows (see also, \eg, Refs.~\cite{Bjorken:1965zz,Eden:1966dnq,Chang:1983,Gribov:2009}).

\subsection{Derivation}
\label{sec:landau_derivation}

\begin{figure}[tb]
\begin{center}
    \includegraphics[width=0.3\textwidth]{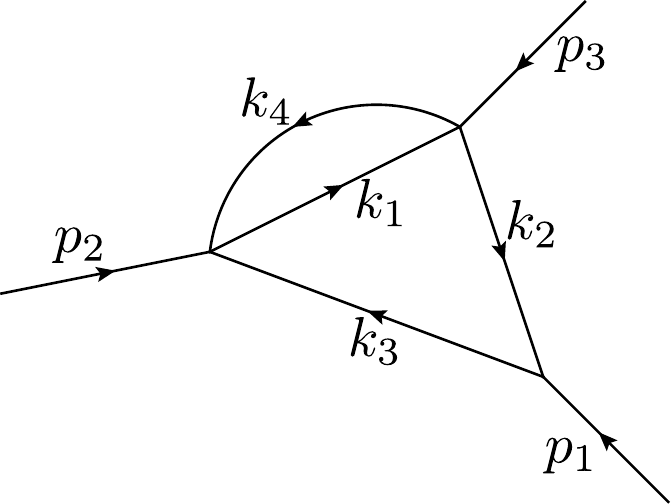}
\begin{minipage}[t]{16.5 cm}
\caption{Illustration of a three-point two-loop diagram with four internal particles ($l=2$, $n=4$).
\label{fig:feyn_2loop}}
\end{minipage}
\end{center}
\end{figure}

The kinematical singularities arise only due to the singularities of propagators of the intermediate particles. Thus, without loss of generality, we can consider the following $l$-loop integral with $n$ propagators,
\begin{equation}
  I(p_1,\ldots,p_m) = \int \frac{d^4q_1\ldots d^4q_l}{[(2\pi)^4i]^l} \frac{1 }{ (k_1^2 - m_1^2 + i\epsilon)\ldots (k_n^2 - m_n^2 + i\epsilon) },
\end{equation}
where $p_1,\ldots,p_m$ are the momenta of external particles, $q_1,\ldots, q_l$ are the loop momenta to be integrated over, and $k_1,\ldots,k_n$ are the intermediate momenta which are linear in the external and loop momenta. Introducing Feynman parameters, it can be rewritten as 
\begin{align}
  I(p_1,\ldots,p_m) = \int_0^1 \prod_{i = 1}^n d\alpha_i \,\delta\left(\sum_{i} \alpha_i - 1\right) \int \prod_{j = 1}^l  \frac{d^4q_j}{(2\pi)^4i} \frac{1}{ \left[ J(\alpha,q,p) + i\epsilon \right]^n },
\end{align}
where 
\begin{align}  
 \, J(\alpha,q,p) = \sum_{i = 1}^{n} \alpha_i (k_i^2 - m_i^2) . \label{eq:J}
\end{align}
The internal momenta $k_i$'s are linear in the loop momenta $q$'s and external momenta $p$'s. For  instance, let us consider the example shown in Fig.~\ref{fig:feyn_2loop}.
%---------------------------------------
%---------------------------------------
Taking the loop momenta to be 
\begin{equation}
  q_1 = k_1,\quad q_2=k_2,\quad k_3 = p_1+q_2, \quad k_4=p_3+q_1-q_2, 
  \label{eq:kexample}
\end{equation}
one gets
\begin{align}
  J(\alpha,q,p) =\;& (\alpha_1 + \alpha_4) q_1^2 + (\alpha_2 +\alpha_3 +\alpha_4 ) q_2^2 - 2\alpha_4 q_1\cdot q_2 + 2 \alpha_4 p_3\cdot q_1 + 2 (\alpha_3p_1 -\alpha_4p_3 )\cdot q_2 \nonumber\\
  & + \alpha_3 p_1^2 + \alpha_4 p_3^2 - \sum_i \alpha_i m_i^2 .
\end{align}
In general, $J$ is a quadratic function of the loop momenta, and one can write 
\begin{align}
  J(\alpha,q,p) &= \sum_{i,j} q_i a_{ij} q_j + 2 \sum_i b_i(\alpha,p) q_i - C(\alpha,p) \nonumber\\
  &= \sum_{i,j} (q_i + l_i) a_{ij} (q_j + l_j) - C(\alpha,p) - \sum_{i,j} l_i a_{ij} l_j \,,
  \label{eq:Jquad}
\end{align}
with $b_i(\alpha,p)$, $C(\alpha,p)$ and $l_i = \sum_j (a^{-1})_{ij}b_j$ functions of the Feynman parameters $\alpha$ and the external momenta $p$. The matrix $a$ (function of only the Feynman parameters) is real and symmetric, and can be diagonalized by an orthogonal matrix. Thus, the first term in the last equation can be rewritten in a form of $\sum_i \tilde{q}_i^2 \tilde{a}_i$, where $\tilde q_i$'s form the new loop momentum basis and $\tilde{a}_i$'s are the eigen-values of the matrix $a$. This allows a variable change of the loop integrals from $q$ to $\tilde q$.  Making the Wick rotation, $q^0 = i q_{E}^4$, and performing the integration over all of the loop momenta one by one using 
\begin{equation}
  \int \frac{d^4q}{(2\pi)^4 i} \frac1{(c\,q^2-\Delta)^n} = (-1)^n\int\frac{d^4q_E}{(2\pi)^4} \frac1{(c\,q^2_E+\Delta)^n} = \frac{(-1)^n}{(4\pi)^2}\frac{\Gamma(n-2)}{\Gamma(n)}  \frac{1}{c^2\, \Delta^{n-2}},
\end{equation}
where $c$ is independent of the loop momentum $q$,
one gets
\begin{align}
  I(p_1,\ldots,p_m) = N \int_0^1 \prod_{i=1}^n d\alpha_i \frac{\delta\left(\sum_i\alpha_i-1\right)}{(\det a)^2} \frac1{  \Delta^{n-2l} },
  \label{eq:I_feyn}
\end{align}
where $N= ( - 1)^n(4\pi)^{-2l} \left[\Gamma(n - 2)/\Gamma(n)\right]^l  $ is just a number independent of $\alpha$ and momenta,  and 
\begin{equation}
    \Delta(\alpha,p) = -J(\alpha,q,p)\big|_{q_i = -l_i} = C(\alpha,p) + \sum_{i,j} l_i a_{ij} l_j .
    \label{eq:Delta}
\end{equation}

Singularities of the integral $I$ must come from the singularities of the integrand, \ie, the zeros of $\Delta$ in Eq.~\eqref{eq:I_feyn} assuming $\det a\neq0$. However, a singularity of the integrand does not necessarily become a singularity of the integral. This is because if we treat the integral variables as complex, then the integral contour can be deformed to avoid the singularity of the integrand. As a result, the integral is analytic. 
In general, there are two cases when the integration contour cannot be deformed to avoid the singularity, and the integral becomes singular (see, \eg, Refs.~\cite{Eden:1966dnq,Chang:1983}):
\begin{itemize}
    \item The singularity of the integrand is at the endpoint of the integration region, so that it is impossible to deform the contour to avoid it. In this case, the integral develops a singularity called the endpoint singularity.
    \item When there are two or more singularities of the integrand at the same point (separated by an infinitesimal distance), and if they are located on opposite sides of the integration contour, the contour is then pinched by them and cannot be deformed further. In this case, the integral also develops a singularity, which is called the pinch singularity.
\end{itemize}

\begin{figure}[tb]
\begin{center}
    \includegraphics[width=0.48\textwidth]{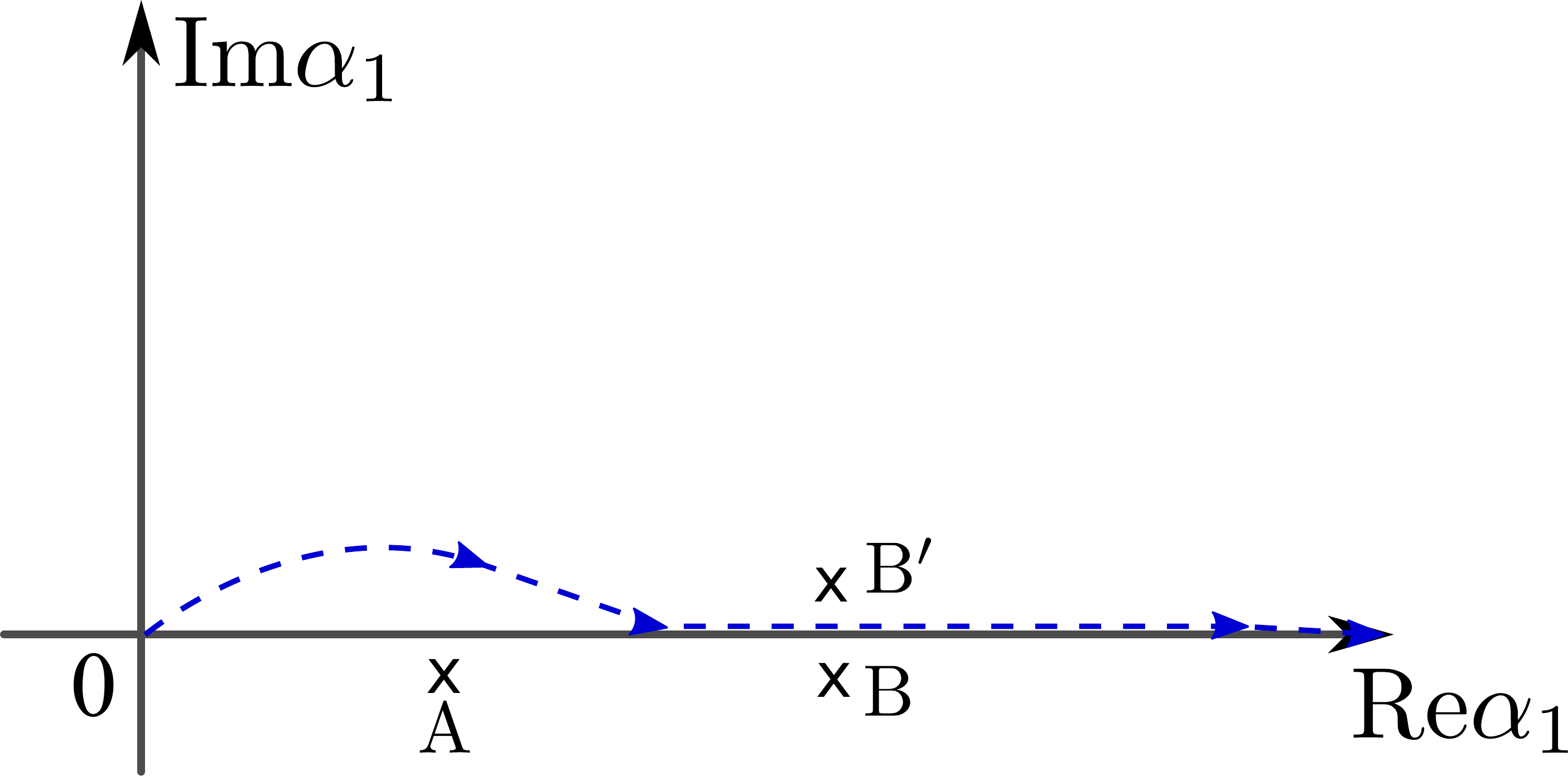}
\begin{minipage}[t]{16.5 cm}
\caption{Illustration of how the integration contour in the complex $\alpha_1$ plane can be distorted to avoid the singularity of $J$ at A but pinched by two singularities at B and B$'$ separated from each other by an infinitesimal distance.
\label{fig:pinch}}
\end{minipage}
\end{center}
\end{figure}
For instance, we consider the integral over $\alpha_1$ in Eq.~\eqref{eq:I_feyn}. If the external momenta and $\alpha_2,\ldots,\alpha_l$ are such (denoting them collectively as $(p_A,\alpha_A)$) that $\Delta$ has a zero at point $\alpha_1=$A as shown in Fig.~\ref{fig:pinch}, the contour can be deformed away from it, and the integral over $\alpha_1$ is analytic at the point $(p_A,\alpha_A)$ in the hyperspace of the external momenta and $\alpha_2,\ldots,\alpha_l$. While if the zero of $\Delta$ is at the endpoint $\alpha_1 = 0$, or there are two zeros at the same point (B and B$'$ in the figure separated by an infinitesimal distance), \ie, $\Delta\big|_B=0$ and $\partial \Delta/\partial\alpha_1\big|_B =0$, the integral will be singular.

Therefore, assuming that $\det a$ does not vanish, the conditions for the general integral $I$ to have singularities are given by 
\begin{align}
  \left\{
    \begin{aligned}
  \Delta &= J(\alpha,q,p)\big|_{q_i=-l_i} = 0 , \\
  \alpha_i &= 0,\quad {\rm or}~~~ \frac{\partial \Delta}{\partial \alpha_i} = \frac{\partial J}{\partial \alpha_i}\bigg|_{q_i = -l_i} = 0 .
    \end{aligned}
  \right.
  \label{eq:landau0}
\end{align}
The former equation in the second line is due to endpoint singularities, and the latter one is due to pinch singularities.
From Eq.~\eqref{eq:Jquad} and Eq.~\eqref{eq:J}, the condition $q_i = -l_i$ for the first line is equivalent to 
\begin{equation}
  0 = \frac{\partial J}{\partial q_i} = \sum_{j = 1}^n \frac{\partial J}{\partial k_j} \frac{\partial k_j}{\partial q_i} = 2 \sum_{j{\rm\;in\;the\;}q_i {\rm\;loop} } \pm\alpha_j k_j ,
\end{equation}
where the last sum runs over the momentum inside the loop integrating over $q_i$, and the sign depends on whether the direction of $k_i$ is the same ($+$) or opposite ($-$) to the flow of the loop momentum. For instance, for the diagram in Fig.~\ref{fig:feyn_2loop}, from Eq.~\eqref{eq:kexample}, the above equation leads to $\alpha_1 k_1 + \alpha_4 k_4 = 0$ and $\alpha_2 k_2 + \alpha_3 k_3 - \alpha_4 k_4 = 0$.
Fro Eq.~\eqref{eq:J}, one has $\partial J/\partial \alpha_i = k_i^2-m_i^2$. Therefore, Eqs.~\eqref{eq:landau0} can be rewritten as
\begin{align}
  \left\{
    \begin{aligned}
   & \sum_ i \pm \alpha_i k_i^\mu = 0 \quad\text{for each loop}, \\
   & \alpha_i = 0\quad {\rm or} ~~ k_i^2 - m_i^2 = 0 \quad \text{for each}~i,
    \end{aligned}
  \right.
  \label{eq:landau}
\end{align}
which are known as the Landau equations for singularities of the $S$-matrix. 
One sees that for each of the intermediate particles, either it is on its mass shell ($k_i^2=m_i^2$), or it does not have any contribution to the singularity ($\alpha_i=0$). 
The latter case may be considered as the corresponding propagator is eliminated, and the loop diagram reduces to a graph with one propagator less. For a given loop diagram, the singularities with all intermediate particles being on shell are the leading Landau singularities, and those with some of propagators dropped give the subleading Landau singularities. 
It is worthwhile to emphasize that although the singularities are derived using Feynman diagrams, its validity goes beyond perturbation theory~\cite{Landau:1959fi,Eden:1966dnq}.

Notice that there are $4l+n$ equations in Eqs.~\eqref{eq:landau}: $4l$ for the first line and $n$ for the second line. Together with the constraint $\sum_i\alpha_i=1$, there are $4l+n+1$ constraints for $4l+n$ variables $q_i^\mu$ and $\alpha_i$. Therefore, there may only be a solution for very specific values of the external momenta for given masses of internal particles (or specific values of masses of internal particles for given external momenta). 
This becomes a crucial point when discussing phenomenological implications of Landau singularities, because it implies that the singularity location is highly sensitive to the values of the involved kinematical variables. For a detailed discussion about the pinch singularities in the momentum space for the triangle diagram, we refer to Sec.~\ref{sec:coleman-norton}.

\subsection{Application to the two-body threshold singularity}
\label{sec:landau_threshold}

%---------------------------------------
\begin{figure}[tb]
\begin{center}
    \includegraphics[height=2.5cm]{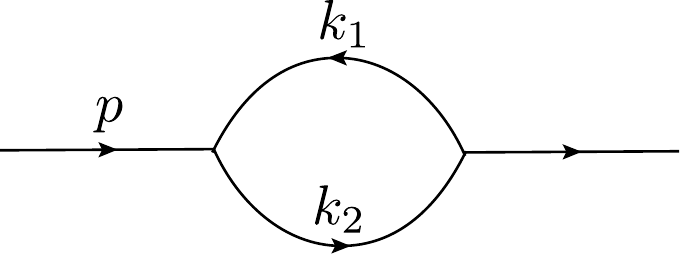}
\begin{minipage}[t]{16.5 cm}
\caption{ 
  The simplest loop diagram with two internal particles.
\label{fig:feyn_1loop}}
\end{minipage}
\end{center}
\end{figure}  
%---------------------------------------
Let us apply the Landau equations to the simple two-point loop with two internal particles shown in Fig.~\ref{fig:feyn_1loop}.
The Landau equations in this case read
\begin{align}
  \left\{ 
    \begin{aligned}
     & \alpha_1 k_1^\mu + \alpha_2 k_2^\mu = 0 ,\\
     & k_1^2 - m_1^2 = k_2^2 - m_2^2 = 0.
    \end{aligned}
  \right.
\end{align}
Taking into account the constraint $\alpha_1 + \alpha_2 =1$, and contracting the first line by $k_{1\mu}$ and $k_{2\mu}$ separately, one gets 
\begin{align}
  \left\{ 
    \begin{aligned}
     & 2 \alpha_1 m_1^2 + (1-\alpha_1) \left(m_1^2 + m_2^2 - p^2\right) = 0 ,\\
     & 2 (1-\alpha_1) m_2^2 + \alpha_1 \left(m_1^2 + m_2^2 - p^2\right) = 0 .
    \end{aligned}
  \right.
\end{align}
Solving the equations for the Feynman parameter $\alpha_1$ and the external momentum $p^2$, one readily finds two solutions:
\begin{align}
  \alpha_1 &= \frac{m_2}{m_1+m_2}, \quad p^2 = (m_1+m_2)^2 ;\\
  \alpha_2 &= \frac{m_2}{m_2-m_1}, \quad p^2 = (m_1-m_2)^2 .
\end{align}

The solution given in the first line means that when the c.m. energy of the two internal particles, \ie, $E=\sqrt{p^2}$, is equal to the threshold $m_1+m_2$, there is a singularity. This is the two-body threshold singularity mentioned at the beginning of this section. It is a square root branch point. Since square root is a double-valued function, one can construct two Riemann sheets for it with the cut stretched between the threshold and infinity along the positive energy axis. The Riemann sheet connected to the upper edge of the positive energy axis is called the first or physical Riemann sheet, and the other one is called the unphysical or second (if there is only one channel)
Riemann sheet. One notices that in this case, the value of $\alpha_1$ is in the physical range $[0,1]$, \ie, both $\alpha_1$ and $\alpha_2$ are positive.

For the solution in the second line, $\alpha_1$ is either larger than 1 ($\alpha_2<0$) or smaller than 0, and thus is not in the physical range. This means that the corresponding singularity is not on the physical Riemann sheet of $p^2$. The singularity in this case, $(m_1-m_2)^2$, is called the pseudo-threshold.

\subsection{Application to the triangle singularity}
\label{sec:landau_triangle}

%---------------------------------------
\begin{figure}[tb]
\begin{center}
    \includegraphics[width=0.66\textwidth]{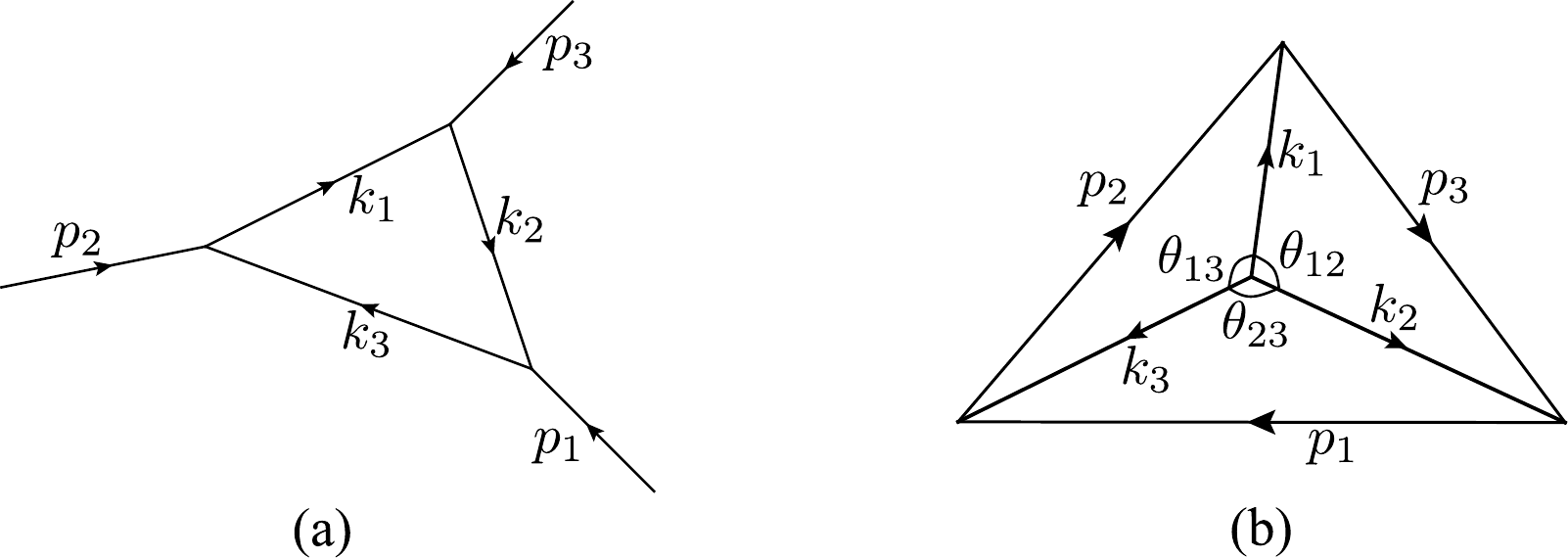}
\begin{minipage}[t]{16.5 cm}
\caption{ A triangle diagram (a) and its dual diagram (b). In the dual diagram, all lines are in the same plain. The lengths of the edges of the triangle are given by $\sqrt{p_i^2}$, and the lengths of the internal lines are given by $\sqrt{k_i^2} = m_i$.
\label{fig:feyn_triangle}}
\end{minipage}
\end{center}
\end{figure}  
% ---------------------------------------

Let us now consider the next simplest case, the triangle diagram shown in Fig.~\ref{fig:feyn_triangle}~(a). We start from  
\begin{align}
  \left\{ 
    \begin{aligned}
     & \alpha_1 k_1^\mu + \alpha_2 k_2^\mu +\alpha_3 k_3^\mu  = 0 ,\\
     & k_1^2 - m_1^2 = k_2^2 - m_2^2 = k_3^2 - m_3^2 =0.
    \end{aligned}
  \right. 
  \label{eq:landau_tri}
\end{align}
Contracting the first line with $k_{1\mu}$, $k_{2\mu}$ and $k_{3\mu}$, respectively, and using the second line one finds
\begin{align}
  \left\{ 
    \begin{aligned}
     & \beta_1  + \beta_2 y_{12} + \beta_3 y_{13} = 0 ,\\
     & \beta_1 y_{12} + \beta_2 + \beta_3 y_{23} = 0 ,\\
     & \beta_1 y_{13} + \beta_2 y_{23} + \beta_3 = 0 ,
    \end{aligned}
  \right.
\end{align}
with $\beta_i = \alpha_i m_i$, and 
\begin{equation}
  y_{ij} = \frac{k_i\cdot k_j}{m_i m_j} = \frac{m_i^2 + m_j^2-p_k^2}{2m_i m_j} \quad (i\neq j\neq k),
  \label{eq:y}
\end{equation}
where $p_i$ and $k_i$ are defined according to Fig.~\ref{fig:feyn_1loop}~(b). For the above equations to have a solution, we must have
\begin{equation}
  \left|  
  \begin{matrix}
    1 & y_{12}^{} & y_{13}^{} \\
    y_{12}^{} & 1 & y_{23}^{} \\
    y_{13}^{} & y_{23} ^{}& 1
  \end{matrix}
  \right| = 1 + 2\, y_{12}^{}\, y_{23}^{}\, y_{13}^{} - y_{12}^2 - y_{23}^2 - y_{13}^2 = 0 ,
  \label{eq:landau_y}
\end{equation}
which gives the leading singularities of the triangle diagram. For a given pair of $p_i^2$ and $p_j^2$, the third one can be solved. Let us suppose that $p_1^2$ and $p_2^2$ are given, then $y_{23}$ and $y_{13}$ follow from Eq.~\eqref{eq:y}, and there are always a pair of solutions if we treat the momenta and thus $y_{ij}$ as complex variables. They are given by
\begin{equation}
  y_{12}^{} = y_{13}^{} y_{23}^{} \pm \sqrt{ (1-y_{13}^{2}) (1-y_{23}^{2}) } .
  \label{eq:ysol}
\end{equation}

However, in general the solutions of Eq.~\eqref{eq:landau_y} do not present singularities of the amplitude in the physical region. 
In the derivation of the Landau equations, the second line of Eq.~\eqref{eq:landau0} only means that the integrand has two coalescing singularities, \ie, zeros of $\Delta(\alpha,p)$ in Eq.~\eqref{eq:Delta}, at the same point, but one does not know whether they pinch the integration contour. For instance, they may be on the same side of the integration contour, or do not lie in the region for $\alpha_i\geq 0$. In fact, one can show that as along as the coalescing zeros of $\Delta(\alpha,p)$ are at $\alpha_i\geq 0$, they must pinch the contour~\cite{Coleman:1965xm}. That is because in the vicinity of the two-fold zeros $\bar \alpha_i$, $\Delta(\alpha,p)$ can be written as $\Delta(\alpha,p) = \frac12\sum_{i,j} (\alpha_i-\bar \alpha_i) (\alpha_j - \bar \alpha_j) \Delta_{ij}$ with $\Delta_{ij} = \partial^2\Delta(\bar \alpha,p) /(\partial\alpha_i \partial\alpha_j)$ a real symmetric matrix. Changing integration variables from $\alpha_i$ to $\alpha_i'$ by a real orthogonal transformation so that $\Delta_{ij}$ is diagonalized to $\delta_{ij}\Delta_{ij}'$, one may write $\Delta(\alpha,p) = \frac12\sum_{i,j} (\alpha_i'-\bar \alpha_i')^2\Delta_{ii}'$. The zeros of $\Delta(\alpha',p)-i\epsilon$ must be located on opposite sides of the real $\alpha_i'$ axes, and thus pinch the integral contour as long as $\alpha_i\geq0$.

Equations~\eqref{eq:landau_tri} can also be solved geometrically in the special case: $|m_i - m_j|< \sqrt{p_k^2}< m_i +m_j$ for $i\neq j \neq k$, \ie, $|y_{ij}|<1$, in which case all of the involved particles are stable.\footnote{The solution goes beyond these conditions, see the discussion in Chapter 2 of Ref.~\cite{Eden:1966dnq}.} 
The energy-momentum conservation $p_1^\mu+p_2^\mu+p_3^\mu=0$ means that the three external momentum (taking as in the Euclidean space) can form a closed triangle with the length of each edge given by $\sqrt{p^2}$. The energy-momentum conservation for each vertex in Fig.~\ref{fig:feyn_triangle}~(a) means that the corresponding momenta (for instance, $k_1^\mu, k_2^\mu$ and $p_3^\mu$) can form a triangle as well. 
Equation $\alpha_1 k_1^\mu + \alpha_2 k_2^\mu +\alpha_3 k_3^\mu  = 0 $ means that the three internal momenta also lay in the same plane, which is the same as that of the external ones. The situation is shown in Fig.~\ref{fig:feyn_triangle}~(b), which is called the dual diagram of the Feynman diagram in (a)~\cite{Karplus:1958zz}. 
The on-shell conditions $k_i^2=m_i^2$ mean that the lengths for the internal lines are given by the masses.
Consequently, for the angles $\theta_{ij}$ between $k_i$ and $k_j$ shown in the graph, one has
\begin{equation}
  \theta_{12} + \theta_{13} + \theta_{23} = 2\pi,\quad \text{and}~~ \cos\theta_{ij} = y_{ij} .
\end{equation}
Using identities for trigonometric functions, it is easy to verify that the above equation is equivalent to Eq.~\eqref{eq:landau_y}. For instance, Eq.~\eqref{eq:ysol} may be written as the following trigonometric identity:
\begin{equation}
  \cos(\theta_{13} + \theta_{23}) = \cos\theta_{13} \cos\theta_{23} \pm \sqrt{ (1-\cos^2\theta_{13})(1-\cos^2\theta_{23}) } .
  \label{eq:cos}
\end{equation}
Thus, the dual diagram completely solves the Landau equations~\cite{Landau:1959fi,Taylor:1960zz}.

The condition of $\alpha_i\geq0$ corresponds to $\theta_{ij}\leq\pi$ in the dual diagram, \ie, the point that $k_1,k_2$ and $k_3$ meet together is inside the triangle~\cite{Taylor:1960zz}. This can be easily understood as if the point is outside the triangle, the other ends of the $k_i$ vectors (the triangle vertices) would be on the same side of the meeting point, which requires one of $\alpha_i$'s to be negative in order to fulfill $\sum_i \alpha_i k_i=0$. It is easy to see from Eq.~\eqref{eq:cos} that among the two solutions, only the one with a minus sign between the two terms satisfies $\theta_{ij}\leq\pi$ (so that $\sin\theta_{ij} = \sqrt{1-\cos^2\theta_{ij}}$), 
\begin{equation}
  y_{12}^{} = y_{13}^{} y_{23}^{} - \sqrt{ 1-y_{13}^{2}} \sqrt{1-y_{23}^{2} }.
  \label{eq:landauphysical}
\end{equation}
The other solution in Eq.~\eqref{eq:ysol} is not in the physical region.
Applying the above equation to the deuteron form factor by taking all the internal particles to be nucleons with a mass $m$ and $p_1^2=p_2^2=M^2$ with $M$ being the deuteron mass, one gets the anomalous threshold in this case at $p_3^2 = 4 M^2-M^4/m^2$.

The physical picture of TSs in the decay region where one of the internal particles is unstable  will be presented in Section~\ref{sec:coleman-norton}, where the Coleman--Norton theorem~\cite{Coleman:1965xm} is discussed, which shows that the Landau equations together with positive $\alpha_i$ and real internal momenta are necessary and sufficient conditions for a singularity to be on the physical boundary.

\subsection{Character of Landau singularities}
\label{sec:character}

The Landau equations allow us to determine the location of a singularity. It is also important to know how singular the amplitude is.
We have seen at the beginning of this section that the two-body threshold cusp is a square root branch point. 
The general conclusion about the character of the leading landau singularity for a given diagram with $l$ loops and $n$ propagators is given by~\cite{Landau:1959fi,Gribov:2009}
\begin{equation}
  \mathcal{A} \sim (s_0-s)^{(4l-n-1)/2} \log(s_0-s)
\end{equation} 
if $4l-n-1$ is even and nonnegative, and 
\begin{equation}
  \mathcal{A} \sim (s_0-s)^{(4l-n-1)/2} 
\end{equation}      
for the other cases, 
where $s_0$ denotes the location of the leading Landau singularity in terms of some variable $s=p_i^2$ with $p_i^\mu$ an external four-momentum.

%---------------------------------------
\begin{figure}[tb]
\begin{center}
% \begin{minipage}[t]{8 cm}
    \includegraphics[width=0.5\textwidth]{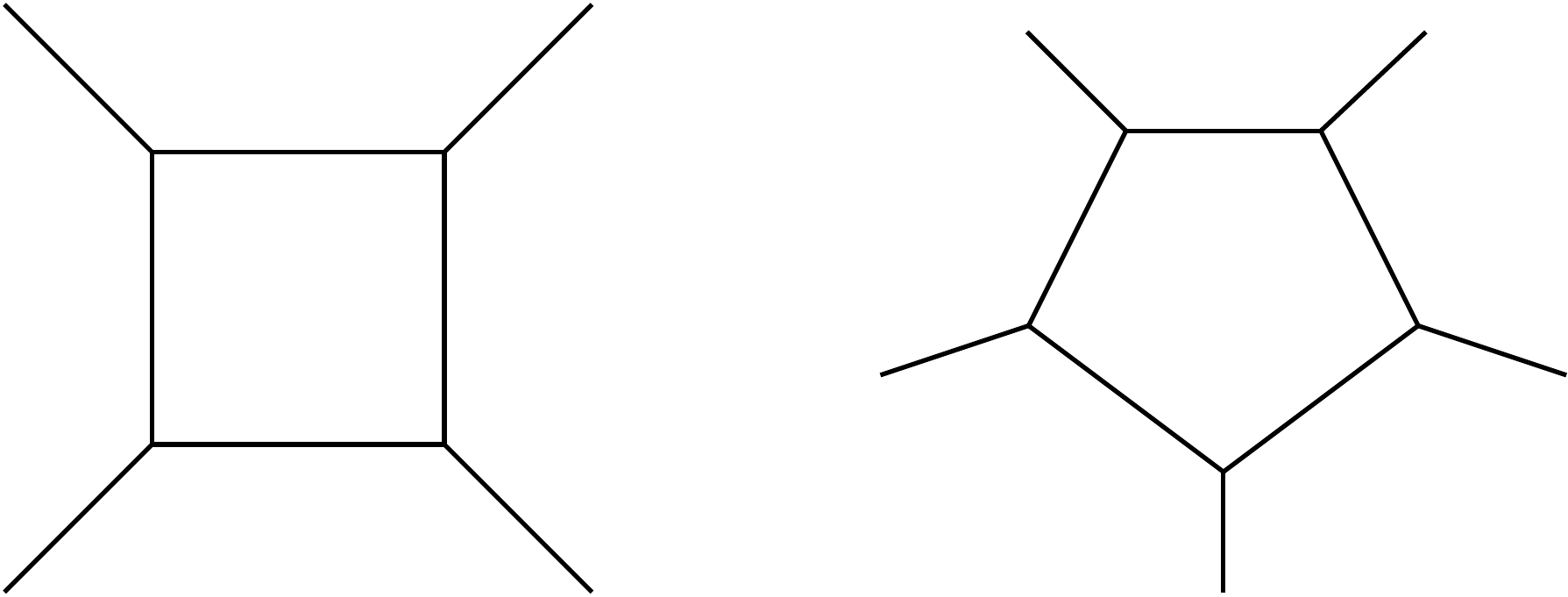}
% \end{minipage}
\begin{minipage}[t]{16.5 cm}
\caption{ A box diagram and a pentagon diagram. 
\label{fig:feyn_boxpenta}}
\end{minipage}
\end{center}
\end{figure}  
% ---------------------------------------

Thus, for a two-point one-loop diagram, $l=1$ and $n=2$, the singularity behaves as a square root branch point, $\sqrt{s_0-s}$; for a one-loop triangle diagram, 
$l=1$ and $n=3$, the leading singularity behaves as a logarithmic branch point, $\log(s_0-s)$. The singularity is more singular if we add more propagators. For the box and pentagon diagrams in Fig.~\ref{fig:feyn_boxpenta}, from the equations above it is easy to find that the corresponding leading singularities behave as a singular branch point, $(s_0-s)^{-1/2}$, and a pole, $(s_0-s)^{-1}$, respectively~\cite{Gribov:2009}. However, it would not be more singular even if more propagators are added.
Gribov argues that the strongest singularity an amplitude can have is most likely a pole~\cite{Gribov:2009}.

\bigskip

%% file: section3.tex
\section{Two-body threshold cusps}
\label{sec:3}

In this section, we discuss the two-body threshold singularity. As we have mentioned above, it is a square root branch point. Consequently, it shows up as a cusp exactly at the corresponding threshold in the energy distribution (for $S$-wave rescattering). 
Since the position of the threshold branch point is fixed, its shape should reflect the interaction in the near-threshold region.

Let us consider the elastic two-body scattering amplitude
\begin{equation}
	T_L(s) = \frac{e^{i\delta_L(s)}\sin\delta_L(s)}{\rho(s)} = \frac{8\pi\sqrt{s}}{k \cot\delta_L(s) - i\,k} ,\label{eq:TLtb}
\end{equation}
where $\delta_L$ is the phase shift in the  $L^\text{th}$ partial wave, and $\rho(s)$ and $k$ are the two-body phase space factor and the magnitude of the c.m. momentum, respectively, as given in Eq.~\eqref{eq:rho}.
We have the effective range expansion when the momentum is small, 
\begin{equation}
	k\cot\delta_L(s) = k^{-2L} \left( -\frac1{a_L} + \frac12 r_{e,L} k^2 + \order{k^4} \right),
\end{equation}
with $a_L$ and $r_{e,L}$ the scattering length and effective range, respectively. In the immediate vicinity of the threshold region, we may consider only the scattering length term, and have 
\begin{equation}
		T_L(E) \simeq 8\pi (m_1+m_2)\frac{ (2\mu E)^L}{-1/a_L - i(\sqrt{2\mu E})^{2L+1}},
	\label{eq:Tthreshold}
\end{equation}
where the three-momentum has been approximated by the nonrelativistic expression $k\simeq\sqrt{2\mu E}$ with $E=\sqrt{s} - m_1 -m_2$ and $\mu = m_1 m_2/(m_1 + m_2)$ the reduced mass. One finds that the first derivative of the scattering amplitude with respect to $E$ is discontinuous at threshold for the $S$ wave, which means that there is a cusp  at threshold. For higher partial waves, the first derivative is continuous, and thus there is no cusp at threshold for them. 

\begin{figure}[tb] 
\begin{center}
\includegraphics[height=6cm]{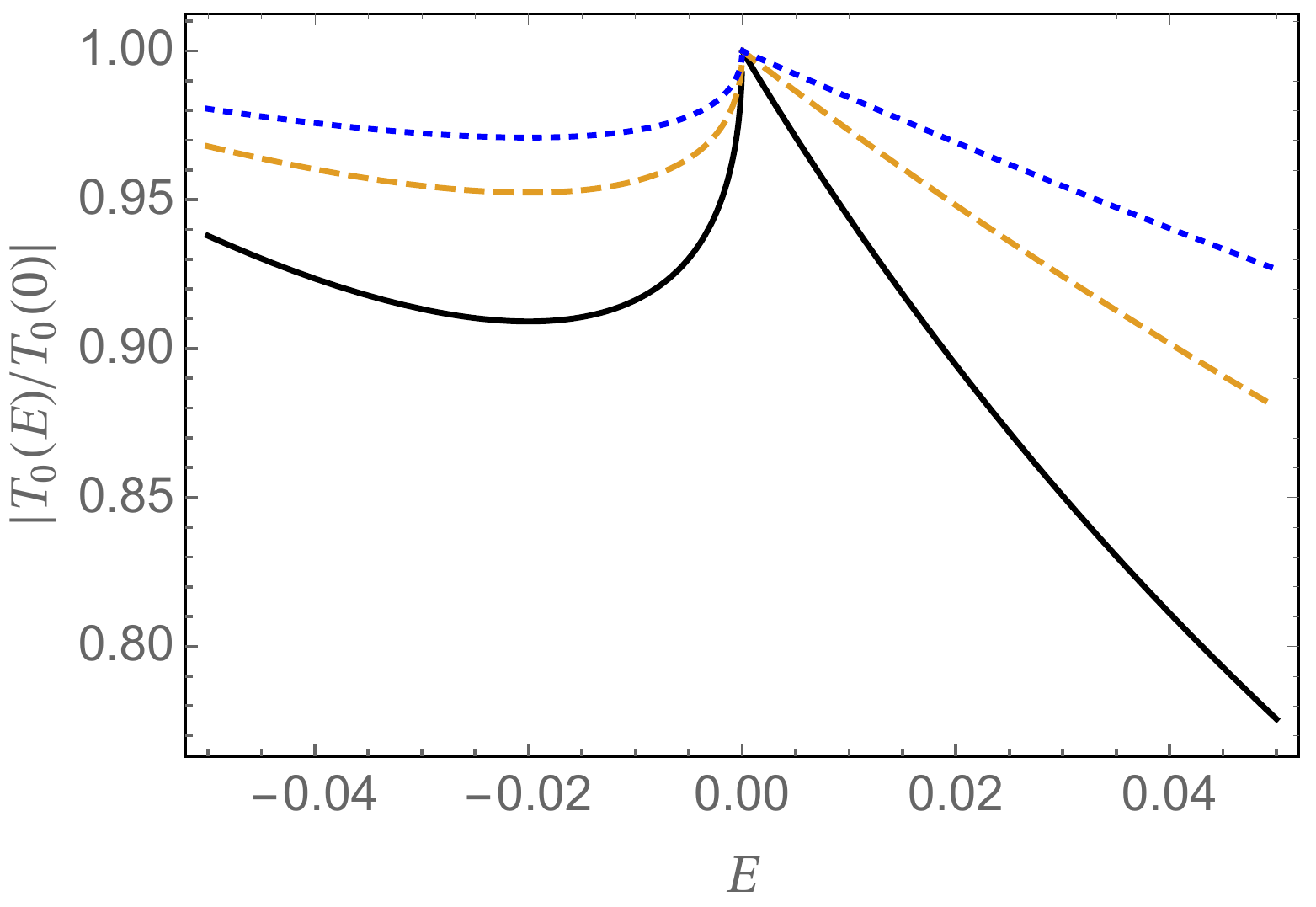}  
\begin{minipage}[t]{16.5 cm}
\caption{Dependence of the threshold cusp on the $S$-wave interaction strength. Here, we use the first two terms in the effective range expansion, and take $\mu = 1$ and $r_{e,0} = 5$ (in arbitrary units) for illustration. For the three curves from top to bottom, $a_0=-0.3$, $-0.5$ and $-1.0$, respectively.
\label{fig:cusp_illustration}}
\end{minipage}
\end{center}
\end{figure}

We show in Fig.~\ref{fig:cusp_illustration} the cusp behavior for the absolute value of the $S$-wave elastic scattering amplitude.
It is easy to see from Eq.~\eqref{eq:Tthreshold} how the strength of the cusp is dictated by the interaction strength at the threshold, \ie, the scattering length: for small $a_0$, the denominator of Eq.~\eqref{eq:Tthreshold} is dominated by the scattering length term which is a constant, and thus the amplitude should behave rather smoothly; for large $a_0$, the denominator is dominated by the $-i\sqrt{2\mu E}$ term, and then the cusp becomes evident. 
The above amplitude under the scattering length approximate always has a pole at $k = i/a_0$. For small $a_0$, the pole does not correspond to a realistic pole of the full $S$ matrix since it is far away from the threshold and beyond the region where such an approximation can be applied. For large $a_0$, the pole is near threshold, and needs to be considered seriously.

%--------------------------------
%--------------------------------

%--------------------------------
\begin{figure}[tb] 
\begin{center}
\includegraphics[height=3.5cm]{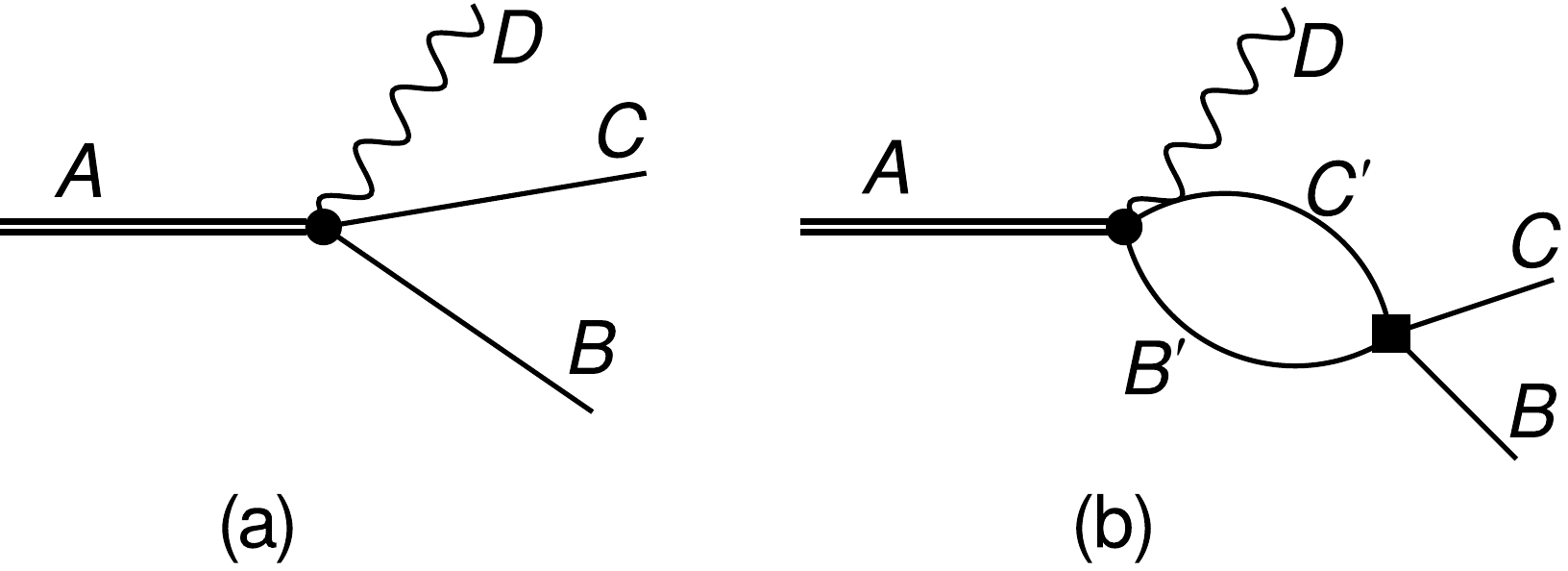}  
\begin{minipage}[t]{16.5 cm}
\caption{A decay process $A\to B\,C\,D$. (a): direct transition at tree level; (b): via final state interaction $B'C'\to BC$. 
\label{fig:fsi}}
\end{minipage}
\end{center}
\end{figure}  
%--------------------------------

Of course, for the threshold cusp to be observed, the intermediate two particles must rescatter into final states with a lower threshold, as shown in Fig~\ref{fig:fsi}. In this case, the cusp strength is controlled by the rescattering $B'C'\to BC$ at the threshold of the intermediate particles $B'C'$. Before we move to discuss concrete examples, let us emphasize that threshold cusp (and the related final state interaction) and resonance are not mutually exclusive to explain a near-threshold structure. On the contrary, a sharp cusp might hint at a near-threshold pole and thus the existence of a resonance.\footnote{
In practice, the threshold cusp can appear with different shapes depending on the relative phase to background (see, \eg, Sec.~XVIII of Ref.~\cite{lifshitz1965quantum}).}

\subsection{Determination of  scattering lengths from threshold cusps }
\label{sec:scatteringlength}

Experimentally, the $\pi\pi$ scattering lengths can be measured in several ways.
The angular distributions of the $K_{e4}$ decay is sensitive to the $\pi\pi$ phase
shifts which are related with the scattering lengths. The first
experiment along these lines was carried out by the Geneva-Saclay
Collaboration in the seventies of the last century~\cite{Rosselet:1976pu}.  A similar method was
employed  by the E865 and NA48/2 Collaborations~\cite{Pislak:2001bf,Batley:2007zz,Ananthanarayan:2000ht,Pislak:2003sv}.
The pionium lifetime can also be related to the $\pi\pi$
scattering lengths,\footnote{A pionium is a bound state of $\pi^+$ and $\pi^-$ formed mainly due to the Coulomb force. Its size characterized by the Bohr radius, $2/(\alpha M_\pi)$, is about 400~fm. Thus, its properties are affected by the strong interaction at the longest distance, and  precise measurements of them (and those of other hadronic atoms) provide knowledge of the relevant scattering lengths. The physics of hadronic atoms are nicely reviewed in Refs.~\cite{Deloff:2003ns,Gasser:2007zt,Gasser:2009wf}. } and the experimental result~\cite{Adeva:2005pg} is well consistent with the prediction from chiral perturbation theory~\cite{Colangelo:2001df}. For a brief review
of the $\pi\pi$ scattering and a list of experimental measurements, see
Ref.~\cite{Gasser:2009zz}.

%--------------------------------
\begin{figure}[tb] 
	\begin{center}
		\includegraphics[height=9.cm]{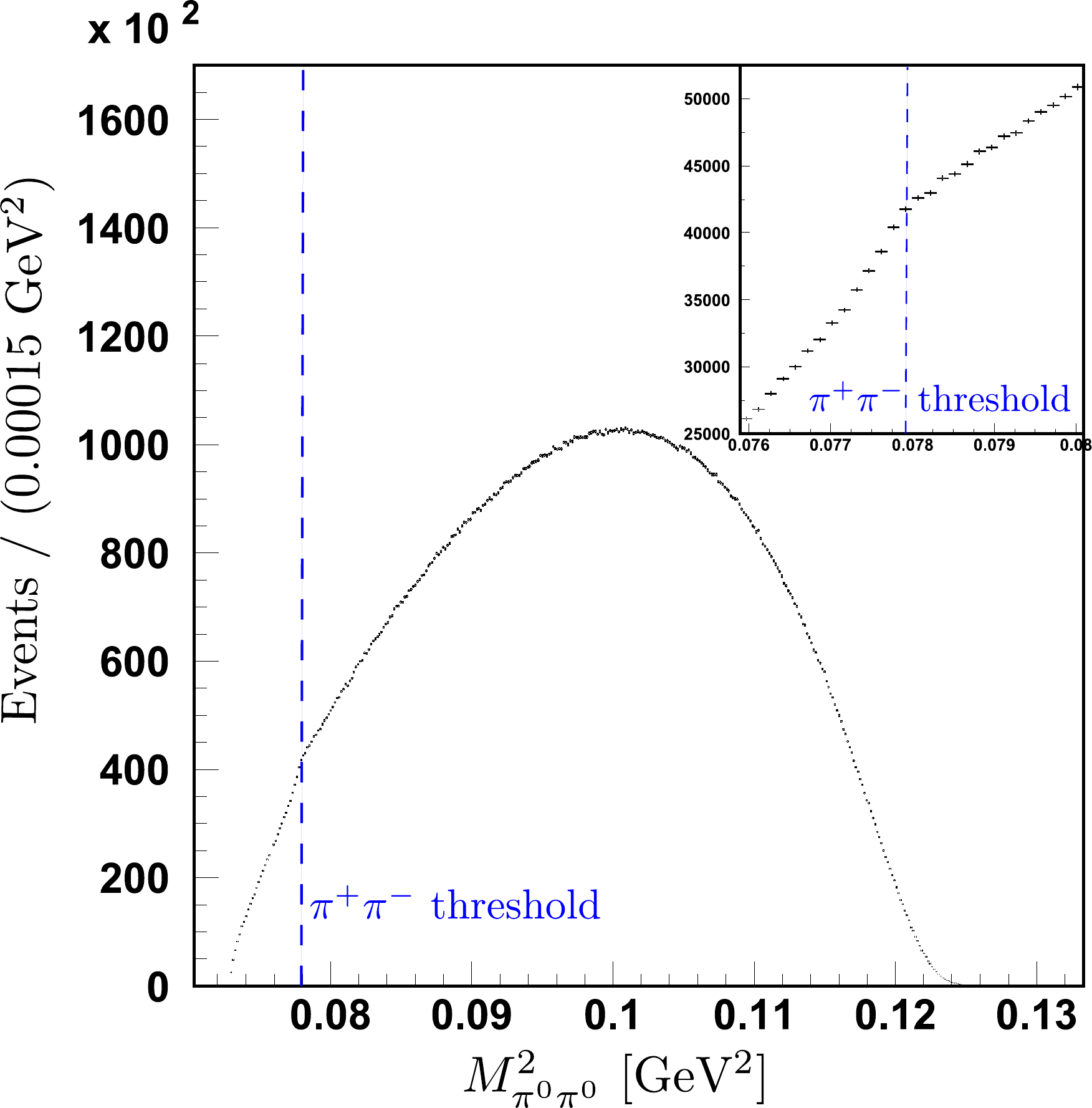}
    \begin{minipage}[t]{16.5 cm}
		\caption{The distribution of the square of the $\pi^0\pi^0$ invariant mass for the $K^+ \to \pi^+ \pi^0\pi^0$ decays, where a cusp structure appears at the $\pi^+\pi^-$ threshold. The figure is adapted from Ref.~\cite{Batley:2005ax}. 
		}
		\label{fig:NA48}
	\end{minipage}
	\end{center}
\end{figure}
%--------------------------------

%--------------------------------
\begin{figure}[t] 
	\begin{center}
		\includegraphics[height=3.5cm]{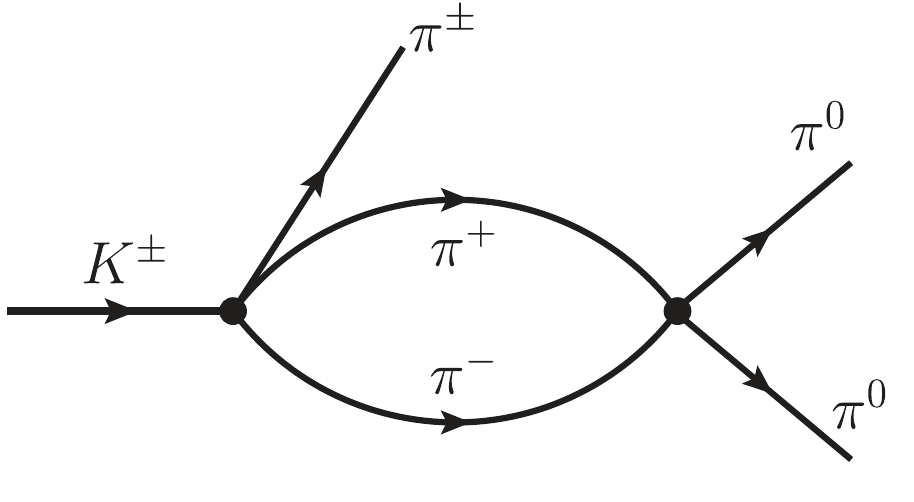}
    \begin{minipage}[t]{16.5 cm}
		\caption{The charge-exchange rescattering diagram which contributes to the $K^\pm\to \pi^\pm \pi^0\pi^0$ decays.}
		\label{fig:Kto3pi}
	\end{minipage}
	\end{center}
\end{figure}
%--------------------------------

A cusp-like structure in the $\pi^0\pi^0$ invariant mass distribution from the $K^\pm\to \pi^\pm \pi^0\pi^0$ decay was observed by the NA48/2 Collaboration in 2005~\cite{Batley:2005ax}, and the data are shown in Fig.~\ref{fig:NA48}. The presence of this structure can be ascribed to the
charge-exchange rescattering $\pi^+\pi^-\to
\pi^0\pi^0$ (see Fig.~\ref{fig:Kto3pi}), and the strength of the cusp is determined to the rescattering ampitude in the near-threshold region, which provides another method to measure the $\pi\pi$ scattering length.   
The method of using the threshold cusp singularity to measure the pion-pion interaction was first proposed by Budini and Fonda in 1961~\cite{Budini:1961bac}, and the process $K^+ \to \pi^+ \pi^0\pi^0$ was investigated in detail. However, limited by the experimental conditions it was impossible to do such a measurement at that time. This method was then forgotten and rediscovered by Cabibbo in 2004 \cite{Cabibbo:2004gq}. 
In fact, in 1997, Mei{\ss}ner {\it et al.} also pointed out that the mass difference between the charged and neutral pions generates a unitary cusp in the $\pi^0\pi^0\to \pi^0\pi^0$reaction, and the strength of the cusp is proportional to the scattering length combination $a_0-a_2$ which characterizes the strength for the charge-exchange $\pi^+\pi^-\to \pi^0\pi^0$ scattering at threshold~\cite{Meissner:1997fa}, where $a_0$ and $a_2$ represent the $S$-wave $\pi\pi$ scattering lengths in the $I=0$ and $I=2$ channels, respectively. The branching ratio of the $K^+\to \pi^+\pi^-\pi^+$,  $(5.59\pm 0.04)\%$, is much 
larger than that of the $K^+\to \pi^+ \pi^0\pi^0$, $(1.761\pm 0.022)\%$~\cite{Tanabashi:2018oca}. Therefore,  the
charge-exchange rescattering turns out to be important so that the cusp effect in the $\pi^0\pi^0$ invariant mass spectrum appears to be enhanced. 
This is one of the reasons why this measurement method is feasible. Based on $2.287\times 10^7$ events for the $K^\pm\to \pi^\pm \pi^0\pi^0$ decays recorded by the NA48/2 experiment at the CERN SPS, the best fit to the rescattering model proposed by Cabibbo and Isidori~\cite{Cabibbo:2005ez} gave~\cite{Batley:2005ax} 
\begin{eqnarray}
(a_0-a_2)M_{\pi^+} &=& 0.268\pm 0.010(\mbox{stat.})\pm 0.004(\mbox{syst.}) \pm 0.013(\mbox{ext.}), \nonumber \\
a_2 M_{\pi^+} &=& -0.041\pm 0.022(\mbox{stat.})\pm 0.014(\mbox{syst.}).
\end{eqnarray}

To precisely determine the $\pi\pi$ $S$-wave scattering lengths using the huge data sample of $K\to 3\pi$ decays, various theoretical frameworks have been developed based on the nonrelativistic effective theory, chiral perturbation theory, dispersion relation theory and so on~\cite{Colangelo:2006va,Bissegger:2008ff,Gasser:2011ju,Gamiz:2006km,Kampf:2008ts}.
In 2009, the updated results from the study of the full data sample of $6.031\times 10^7$ $K^\pm\to \pi^\pm \pi^0\pi^0$ decays collected by the NA48/2 experiment were reported~\cite{Batley:2000zz}. Using the Bern-Bonn nonrelativistic effective field theory formulation~\cite{Bissegger:2008ff,Colangelo:2006va}, which provided the most complete description of rescattering effects in the $K\to 3\pi$ process, the pion-pion scattering lengths were extracted to be  
\begin{eqnarray}
(a_0-a_2)M_{\pi^+} &=& 0.2571\pm 0.0048(\mbox{stat.})\pm 0.0025(\mbox{syst.}) \pm 0.0014(\mbox{ext.}), \nonumber \\
a_2 M_{\pi^+} &=& -0.024\pm 0.013(\mbox{stat.})\pm 0.009(\mbox{syst.})\pm 0.002(\mbox{ext.}).
\label{eq:apipi}
\end{eqnarray}
The $a_0-a_2$ combination was determined with a precision at the  $2\%$ level. This is a nice showcase that the kinematical singularities can be exploited to make precision measurements.

The $\pi\pi$ cusp structure was also discussed in some other processes including the $K_L\to
3\pi$, $\eta\to 3\pi$ and $\eta'\to
\eta\pi\pi$~\cite{Bissegger:2007yq,Ditsche:2008cq,Kubis:2009sb,Isken:2017dkw}.
However, it is difficult to accurately measure the cusp effects in  the $K_L\to 3\pi$ and $\eta\to 3\pi$ processes using the currently available experimental data. The process $\eta'\to \eta\pi\pi$ is a promising candidate, for which the cusp effect was predicted to have an effect of more than $8\%$ in the $\pi^0\pi^0$ energy spectrum below $\pi^+\pi^-$ threshold~\cite{Kubis:2009sb}. 
The $\eta^\prime$ meson can be produced in the $J/\psi\to \gamma \eta^\prime$ decay. Based on a sample of one billion $J/\psi$ events at the BESIII detector, no significant cusp structure was observed~\cite{Ablikim:2017irx}. However, with the much improved 10 billion $J/\psi$ events collected by the BESIII experiment, the cusp structure in  the $\eta'\to \eta\pi^0\pi^0$ decay is expected to be observed in the near future.

The charm and bottom factories, such as BESIII, CLEOc, Belle and Belle-II, have accumulated and/or will continue to accumulate huge data samples of heavy quarkonium dipion transitions. The feasibility of extracting the $\pi\pi$ scattering lengths using the cusp effect in heavy quarkonium dipion transitions was investigated in Ref.~\cite{Liu:2012dv}, and the $\Upsilon(3S)\to \Upsilon(2S)\pi^0\pi^0$ decay is the most promising among these decays to measure the threshold cusp. By using a Monte Carlo simulation, it was found that to reach the precision of $a_0-a_2$ as that given in Eq.~\eqref{eq:apipi}, $\mathcal{O}(2\times10^7)$ events need to be collected for the $\Upsilon(3S)\to \Upsilon(2S)\pi^0\pi^0$ decay.

The cusp phenomena in the pion photoproduction process $\gamma p \to \pi^0 p$ have been observed in experiments at Mainz~\cite{Schmidt:2001vg,Bernstein:1996vf,Fuchs:1996ja,Beck:1990da} and Saskatoon~\cite{Bergstrom:1996fq}. Taking into account that the electric dipole amplitude for the $\gamma p\to \pi^+ n$ reaction is much larger than that for the $\gamma p\to \pi^0 p$ reaction, the presence of a cusp at the $\pi^+ n$ threshold can be ascribed to the charge-exchange rescattering process $\gamma p \to \pi^+ n  \to \pi^0 p$, which has been studied long time ago~\cite{Faldt:1979fs,Laget:1981jq,Kamal:1989ps,Bernard:1992nc,Bernard:1994gm,Bernard:2001gz,Bernard:2005dj,Bernard:2006gx}. The strength of this cusp is related to the pion-nucleon scattering length. A more precise measurement of this cusp will give valuable information on the pion-nucleon scattering around threshold.

Analogous to the above method for measuring the $\pi\pi$ scattering length, Hyodo and Oka suggested that the $\pi\Sigma$ scattering lengths could be extracted from the cusp phenomena in the $\Lambda_c\to \pi\pi\Sigma$ decays~\cite{Hyodo:2011js}. One of the reasons for the interest in the $\pi\Sigma$ scattering lengths is that they are important quantities for understanding the nature of the $\Lambda(1405)$, which could be a dynamically generated resonance from the $\pi\Sigma-\bar{K}N$ coupled-channel interactions in the isospin-0 channel and might possess an intriguing two-pole structure~\cite{Oller:2000fj,Jido:2003cb} (see Ref.~\cite{Hyodo:2011ur} and the review focusing on the $\Lambda(1405)$ in the RPP~\cite{Tanabashi:2018oca} for reviews). The branching ratios of $\Lambda_c\to\pi\pi\Sigma$ decays are around a few percent~\cite{Tanabashi:2018oca}. The sizable branching ratios are beneficial for the extraction of the $\pi\Sigma$ scattering lengths, and the measurement on the cusp phenomena could be feasible in the high luminosity  Belle-II and BESIII experiments. 

At last, it is worthwhile to mention that the $K^- p$ correlation function in high-energy proton-proton collisions measured by the ALICE Collaboration at 13~TeV~\cite{Acharya:2019bsa} has a non-trivial behavior at the $\bar K^0 n$ threshold. It can be well described in the coupled-channel ($\bar K N$, $\pi\Sigma$, and $\pi\Lambda$) calculation of Ref.~\cite{Kamiya:2019uiw}, which has an evident cusp at the $\bar K^0 n$ threshold. The cusp was predicted earlier in Ref.~\cite{Haidenbauer:2018jvl}.

\subsection{Threshold cusps and new hadrons}
\label{sec:zc_cusp}

Resonances are normally searched for by seeking for peaks in invariant mass distributions (or bands in Dalitz plots).
As we discussed above, there should always be a cusp at the $S$-wave threshold of two particles as long as the two-body channel couples to final states that are measured.
Therefore, there exists the possibility that a peak in the energy distribution is caused by a threshold cusp. Then the situation of identifying resonance becomes complicated: the cusp might not require a pole to exist in the near-threshold region but be misidentified as a resonance of that mass; or if the production of the involved coupled channels is correlated so that the cusp can be employed to infer about the interaction strength, just like the pion-pion case discussed above, a strong cusp in this case may demand the existence of a near-threshold pole. In this regard, the discussion of threshold cusps is particularly relevant to the new hadrons because some of the $XYZ$ structures were suggested to be just threshold cusps, instead of resonances, by some authors. 
The suggestions and critiques will be reviewed in this subsection. 

%-------------------------------
\begin{table}[tb]
\begin{center}
\begin{minipage}[t]{16.5 cm}
\caption{The $X$ and $Z$ states that are close to the $S$-wave thresholds of a pair of ground state heavy mesons.  Their quantum numbers, masses and widths are given in the second to fourth columns. The nearest two-body channels and their thresholds are listed in the last two columns. The masses are taken from the RPP~\cite{Tanabashi:2018oca} ($X(3872)$ and $Z_c(4020)$ are named as $\chi_{c1}(3872)$ and $X(4020)$, respectively, therein).  }
\label{tab:thresholds}
\end{minipage}\\[2mm]
\begin{tabular}{l l c c c c c}
\hline\hline
Structure  & $I^G(J^{PC})$ & Mass [MeV] & Width [MeV] & Nearest channel & Threshold [MeV]  \\\hline
$\x$ &  $0^{+}(1^{++})$ & $3871.69\pm0.17$ & $<1.2$ &$D^0\bar D^{*0}$ & $3871.69\pm0.07$  \\
$Z_c(3900)^\pm$ &  $1^{+}(1^{+-})$ & $3887.2\pm2.3$ & $28.2\pm2.6$ & $D^+ \bar D^{*0}$ & $3876.50\pm0.07$ \\
$Z_c(4020)^\pm$ &  $1^{+}(?^{?-})$ & $4024.1\pm1.9$ & $13\pm5$ & $D^{*+} \bar D^{*0}$ &  $4017.11\pm0.07$\\
$Z_{b}(10610)^\pm$ & $1^{+}(1^{+})$ & $10607.2\pm2.0$ & $18.4\pm2.4$ & $B^0 B^{*-}$ & $10604.30\pm0.26$ \\
$Z_{b}(10650)^\pm$ & $1^{+}(1^{+})$ & $10652.2\pm1.5$ & $11.5\pm2.2$ & $B^{*0} B^{*-}$ & $10649.40\pm0.31$ \\
\hline\hline
\end{tabular}
\end{center}
\end{table}     
%-------------------------------

The most interesting ones of the new $XYZ$ states in the heavy-quarkonium mass region are located close to certain $S$-wave thresholds of a pair of heavy mesons. For instance, in the charmonium mass region, the $X(3872)$~\cite{Choi:2003ue} mass coincides with the $D^0\dzstarbar$ threshold, the charged $Z_c(3900)$~\cite{Ablikim:2013mio,Liu:2013dau} mass is also very close to the $D\bar D^*$ threshold, and the charged $Z_c(4020)$~\cite{Ablikim:2013xfr} is similarly close to the $D^*\bar D^*$ threshold; in the bottomonium mass region, the charged $Z_b(10610)$ and $Z_b(10650)$~\cite{Belle:2011aa} are nearby the $B\bar B^*$ and $B^*\bar B^*$ thresholds, respectively. 
All of them have a narrow width, and furthermore their quantum numbers are the same as the corresponding $S$-wave meson pairs except for the $Z_c(4020)$ whose quantum numbers have not been fully determined (see Table~\ref{tab:thresholds}). These properties and the nearest threshold for each of them are listed in Table~\ref{tab:thresholds}.

%--------------------------------
\begin{figure}[tb] 
\begin{center}
\includegraphics[height=3cm]{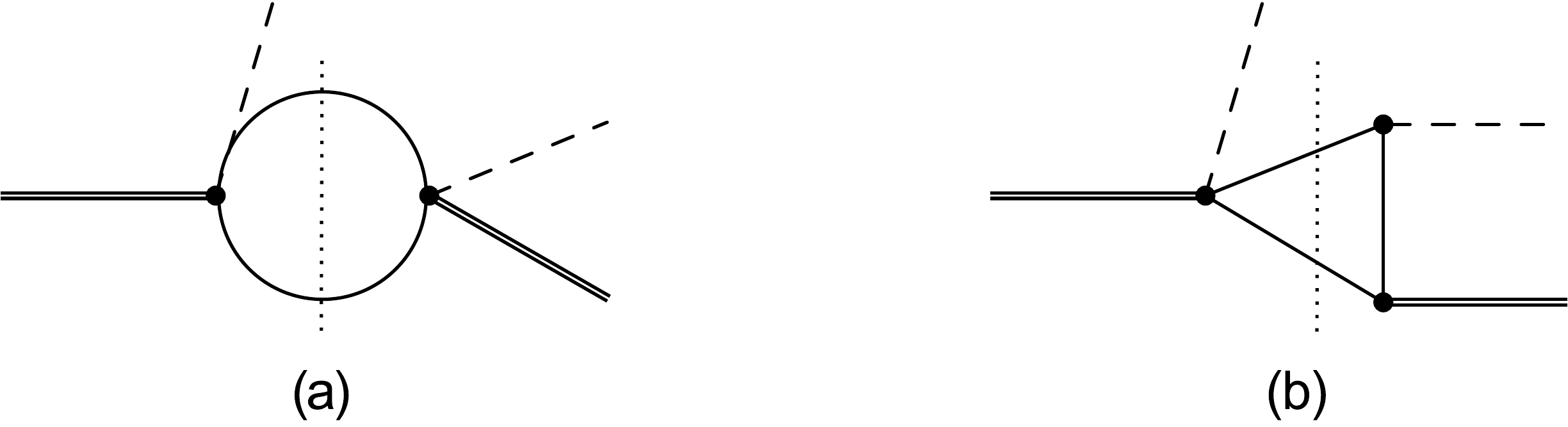} 
\begin{minipage}[t]{16.5 cm}
\caption{Cusp models for producing the $Z_c(Z_b)$ structures through intermediate open-flavor heavy meson loops. Diagram (a) depicts the model in Refs.~\cite{Bugg:2011jr,Swanson:2014tra}, and diagram (b) represents the model in Refs.~\cite{Chen:2011pv,Chen:2013coa}. The double lines, solid lines and dashed lines denote heavy quarkonia, charm (bottom) mesons and pions, respectively. The dotted vertical lines cross the intermediates particles contributing to the threshold cusp. For the $Z_c$, the diagrams correspond to $Y(4260)\to (\pi D\bar D^*+c.c.) \to \pi\pi J/\psi$.
\label{fig:cuspmodels}}
\end{minipage}
\end{center}
\end{figure}
%--------------------------------
These facts stimulated models speculating them to be not resonances but rather just threshold cusps due to coupled channels~\cite{Bugg:2004rk,Bugg:2011jr,Chen:2011pv,Chen:2013coa,Swanson:2014tra,Swanson:2015bsa}. Because of the coincidence of the $\x$ mass and the $D^0\dzstarbar$ threshold, the $\x$ was suspected to be a threshold cusp by Bugg in Ref.~\cite{Bugg:2004rk}\footnote{This paper also suggests the $p\bar p$ near-threshold enhancement observed by the BES Collaboration in $J/\psi\to \gamma p\bar p$~\cite{Bai:2003sw}, to be discussed below, a similar enhancement at the $\Lambda\bar \Lambda$ threshold in the $\Lambda\bar \Lambda\to p \bar p$ cross section measured by the PS185 Collaboration~\cite{Barnes:2000be}, and a peak of around the $\Sigma N$ threshold in the $\Lambda p$ energy spectrum of $K^-d\to \pi^- \Lambda p$~\cite{Eastwood:1971wj,Braun:1977ma} to be threshold cusps.}. However, he realized later on that the very narrow line shapes of the $\x$ in the $J/\psi\rho$ and $D^0\dzstarbar$ modes could not be fitted with only a threshold cusp, and a resonance or virtual state pole was necessary~\cite{Bugg:2008wu}. 
The cases of the $Z_c$ and $Z_b$ listed in the table were considered in Refs.~\cite{Bugg:2011jr,Chen:2011pv,Chen:2013coa,Swanson:2014tra}. These calculations focus on the processes where the $Z_c(Z_b)$ structures show up in the energy spectra of one pion and one heavy quarkonium, \ie, $Y(4260)\to J/\psi\pi^+\pi^-/h_c\pi^+\pi^-$ and $\Upsilon(5S)\to \Upsilon(1S,2S,3S)\pi^+\pi^-/h_b\pi^+\pi^-$. Such channels will be called ``inelastic" and those with the relevant open-flavor thresholds are to be denoted as ``elastic'' in the following. 
Thus, in these models, the final states were produced through the $D^{(*)}\bar D^* (B^{(*)}\bar B^*)$ rescattering at the one-loop level, as shown in Fig.~\ref{fig:cuspmodels}. 

In Ref.~\cite{Bugg:2011jr}, the rescattering amplitude is constructed using a dispersion integral of the following form:
\begin{equation}
	T(s) = \frac1{\pi} \int_{s_\text{th}}^\infty ds' \frac{g \,\rho(s') F_\Lambda(s')}{s'-s-i\epsilon}, \quad\text{with}\quad F_\Lambda(s)=e^{-2q^2(s)/\Lambda^2} ,
	\label{eq:bugg}
\end{equation}
where $g$ is the product of the couplings of the intermediate heavy meson pair to the initial and final states, $\rho(s)$ is the two-body phase space factor defined in  Eq.~\eqref{eq:rho}, and $F_\Lambda(s)$, with $q$ the magnitude of the c.m. momentum of the internal mesons, is a Gaussian form factor to tame the ultraviolet (UV) divergence of the integral. With $\Lambda=0.11$~GeV, narrow peaks similar to the $Z_b$ line shapes could be produced without the inclusion of a resonance pole. 
It is worthwhile to notice that such a small cutoff easily leads to narrow peaks, which however should not be understood as having explained the underlying dynamics.
The model in Ref.~\cite{Swanson:2014tra} is rather similar with the form factor chosen to be of the form $\exp(-s/\Lambda^2)$, and impressive agreement was achieved for the $Z_c$ and $Z_b$ line shapes by adjusting only two parameters (a coupling constant and a cutoff $\Lambda$) for each channel.

Unfolding the rescattering vertex in Fig.~\ref{fig:cuspmodels} to the exchange of a heavy meson gives the model in Refs.~\cite{Chen:2011pv,Chen:2013coa}, named as the initial single-pion emission mechanism by the authors. A dipole form factor was introduced to the exchanged heavy meson propagator. 
Taking the bottom sector as an example, in this way, their calculation produced cusps right around the $Z_b(10610)$ and $Z_b(10650)$ structures from the $B\bar B^*+c.c.$ and $B^*\bar B^*$ loops (here the two mesons refer to those connecting to the initial state), respectively, in the $\Upsilon(1S,2S,3S)\pi$ energy spectra, while no cusp was found at the $B\bar B$ threshold from the $B\bar B$ loop~\cite{Chen:2011pv}. Given that the $\Upsilon\pi$ can couple to the $S$-wave $B^{(*)}\bar B^*$ and couples to the $B\bar B$ pair only in higher partial waves, such a behavior is expected as discussed at the beginning of this section.

Do the impressive agreements achieved in these models for the observed $Z_c$ and $Z_b$ imply these structures to be simply due to coupled-channel threshold cusps, meaning that it is not necessary to introduce resonances? To answer this question, one has to analyze processes with the elastic channels in the final states. From the discussion in Section~\ref{sec:scatteringlength}, it is clear that the strength of a threshold cusp contains information about the rescattering causing the cusp. It could happen that the interaction required for producing these cusps is so strong that the one-loop approximation, which is the implicit assumption in the above mentioned models, becomes questionable. This is the critique raised in Ref.~\cite{Guo:2014iya}.

\begin{figure}[tb] 
\begin{center}
\includegraphics[height=5.4cm]{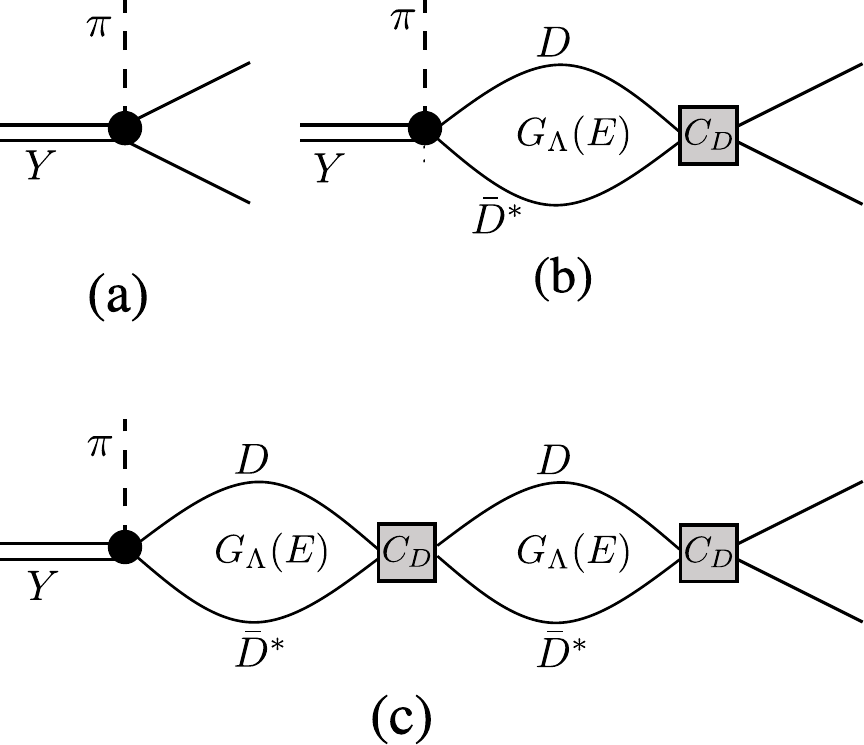} \qquad\quad
\includegraphics[height=5.4cm]{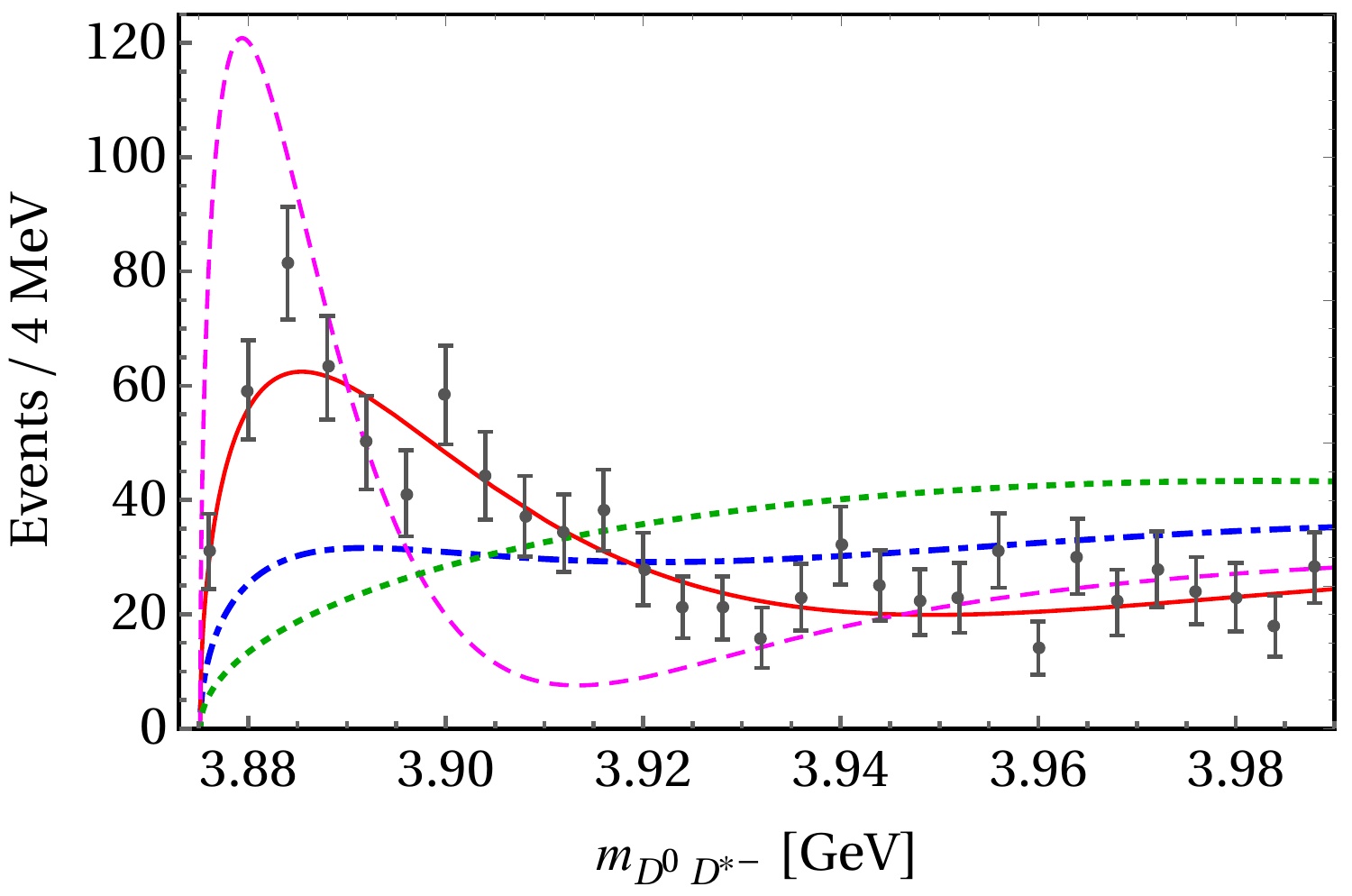}
\begin{minipage}[t]{16.5 cm}
\caption{Left: Tree-level, one-loop and two-loop diagrams for the decay $Y(4260)\to \pi D\bar D^*$. Right: Results in Ref.~\cite{Guo:2014iya} for the $D\bar D^*$ invariant mass spectrum of this process using parameters determined from the one-loop fitting to the BESIII data~\cite{Ablikim:2013xfr}. The results from the tree level, one-loop and two-loop
calculations are shown by the dotted (green), solid (red) and dashed (magenta) curves, respectively. 
The dot-dashed (blue) curve shows the one-loop result with the rescattering strength restricted to be small enough to justify a perturbative treatment. 
\label{fig:DDst}}
\end{minipage}
\end{center}
\end{figure}

Let us consider the $Z_c(3900)$ case with two coupled channels $J/\psi\pi$ and $D\bar D^*+c.c.$ We denote the direct production vertices for these two modes from the initial state as $g_\text{in}$ and $g_\text{el}$, respectively, and approximate the tree-level $S$-wave amplitudes for $J/\psi\pi\to D\bar D^*$ and $D\bar D^*\to D\bar D^*$ simply as constants $C_X$ and $C_D$, respectively.\footnote{The approximation works only  when no nearby singularity is present. The subtlety due to a TS from the $D_1 \bar D D^*$ triangle diagram will be addressed in Section~\ref{sec:6}. }  The direct $J/\psi\pi\to J/\psi\pi$ scattering may be neglected because it is Okubo--Zweig--Iizuka (OZI)~\cite{Okubo:1963fa,Zweig:1981pd,Iizuka:1966fk} suppressed, and lattice QCD calculations give a very small scattering length consistent with zero~\cite{Liu:2008rza}. Thus, the cusp models  may be expressed as the following one-loop amplitudes
\begin{eqnarray}
  g_\text{in} + g_\text{el}\, G_{\Lambda}(s)\, C_X, \qquad\text{and}\qquad 
  g_\text{el} \left[1 + G_{\Lambda}(s)\, C_D \right],
  \label{eq:1loop}
\end{eqnarray}
for the production of $J/\psi\pi$ and $D\bar D^*$ respectively, where 
\begin{equation}
G_\Lambda(s)=\int \frac{d^3 \vec{q}}{(2\pi)^3}\frac{F_\Lambda(\vec q\, ^2)}{\sqrt{s}-m_1
-m_2 -\vec q\, ^2/(2\mu) + i\epsilon} \ ,
\end{equation}
is the two-point loop function with $D\bar D^*$ as the intermediate states regularized using the same form factor as that in Eq.~\eqref{eq:bugg}. The formalism is equivalent to the one using the dispersive integral in Eq.~\eqref{eq:bugg}. The two terms in the second amplitude are represented as plots (a) and (b) in the left panel of Fig.~\ref{fig:DDst}. 
%--------------------------------
%--------------------------------
Because $g_\text{el}$ only serves as an overall normalization and does not affect the shape of the $D\bar D^*$ invariant mass distribution, the $G_{\Lambda}(E)\, C_D$ term can be fixed from fitting to the $D\bar D^*$ invariant mass distribution in the near-threshold region where the approximation of the rescattering as a constant contact term, $C_D$, is valid (which is the leading term in the nonrelativistic expansion). 
On the contrary, the $G_{\Lambda}(E)\, C_X$ term cannot be fixed because of its interference with $g_\text{in}$. Once $G_{\Lambda}(E)\, C_D$ is fixed from fitting to the data using the one-loop amplitude, one can compare the one-loop result with that from two loops shown as (c) in the left panel of Fig.~\ref{fig:DDst}, 
\begin{equation}
  g_\text{el} \left[1 + G_{\Lambda}(E)\, C_D + G_{\Lambda}(E)\, C_D\, G_{\Lambda}(E)\, C_D \right].
\label{eq:2loop}
\end{equation}
If the difference is small, the perturbative treatment using amplitude up to the one-loop level would be a valid approximation; otherwise, it implies that the one-loop coupled-channel model is not self-consistent. 

Such a comparison was performed in Ref.~\cite{Guo:2014iya} by fitting to the the BESIII data for both the $(D\bar D^*)^-$ invariant mass distribution of $e^+e^- \to \pi^+ (D\bar D^*)^-$~\cite{Ablikim:2013xfr} and the $J/\psi\pi^-$ invariant mass distribution of $e^+e^- \to \pi^+ \pi^-J/\psi$~\cite{Ablikim:2013mio},  both of which were measured at the $e^+e^-$ c.m. energy $E_\text{cm}=4.26$~GeV. The data can indeed be well described using Eq.~\eqref{eq:1loop}. The best fit to the data for the former process is shown as the solid curve in the right panel of Fig.~\ref{fig:DDst}, together with the BESIII data. 
Using the same parameters, the tree-level, which simply gives the shape of the phase space, and the two-loop, from Eq.~\eqref{eq:2loop}, results are depicted as the dotted  and dashed curves, respectively. The two-loop curve has a much sharper peak, largely deviating from the one-loop curve. This indicates that the $D\bar D^*\to D\bar D^*$ interaction determined in this way is nonperturbative, \ie, $ G_\Lambda(E)\,C_D = \order{1}$, in the near-threshold region. 
In fact, resumming the $D\bar D^*$ two-point bubbles up to infinite orders by $g_\text{el} /\left[1 -G_{\Lambda}(E)\, C_D \right]$ generates a pole in the vicinity of the $D\bar D^*$ threshold. 
That is to say, if the coupled-channel effects can be described by considering only the $D\bar D^*$ two-body rescattering, the narrowness of the near-threshold peak in the $D\bar D^*$ invariant mass distribution demands the $D\bar D^*$ interaction to be nonperturbative. 
If we force the interaction to be perturbative by hand, a narrow near-threshold peak cannot be produced in the $D\bar D^*$ channel. The result obtained by requiring $| G_\Lambda(E)\,C_D| = 1/2$ at the $D\bar D^*$ threshold is shown as the dot-dashed (blue) curve in the right panel of Fig.~\ref{fig:DDst}.
Notice that here the data in the inelastic channel, $e^+e^- \to \pi^+ \pi^-J/\psi$ for the $Z_c(3900)$, are not enough to determine the $D\bar D^*\to J/\psi\pi$ rescattering strength because it cannot be disentangled from the direct production $g_\text{in}$ in the first amplitude in Eq.~\eqref{eq:1loop}.\footnote{This is different from the case of $K^\pm\to \pi^\pm\pi^0\pi^0$ where the two channels $\pi^0\pi^0$ and $\pi^+\pi^-$ are related to each other via isospin symmetry. There is no symmetry connecting $g_\text{in}$ to $g_\text{el}$ in this case.}

In a modified version of the threshold cusp model of Refs.~\cite{Bugg:2011jr,Swanson:2014tra}, both the inelastic ($J/\psi\pi,h_c\pi$) and elastic ($D\bar D^*$, $D^*\bar D^*$) decay modes were considered for the $Z_c(3900)$ and $Z_c(4020)$~\cite{Swanson:2015bsa}. 
A Gaussian form factor as that in Eq.~\eqref{eq:bugg} was used for all the vertices, including the tree-level ones. It was found that the data of the $J/\psi\pi$ and $D\bar D^*$ energy spectra for the $Z_c(3900)$ and the $D^*\bar D^*$ and $h_c\pi$ energy spectra for the $Z_c(4020)$ could be well fitted. However, the fitting quality depends crucially on the cutoff parameter in the Gaussian form factor, and it was set to 0.2~GeV (for fitting to the $D\bar D^*$ data) and 0.3~GeV (for the other channels).\footnote{In Ref.~\cite{Bugg:2008sk}, a Gaussian form factor multiplied by the phase space was used to fit to the Belle data of the $D\bar D$ and $D^*\bar D^*$ energy spectra~\cite{Abe:2007sya}, and the $\Lambda_c\bar \Lambda_c$ energy spectrum~\cite{Pakhlova:2008vn}  was fitted using the $Y(4660)$ multiplied by a Gaussian form factor.}
Such a value is too small in the sense that the form factor already drops dramatically at an energy 20 to 30~MeV above the threshold ($\Lambda^2/(2\mu)\simeq 20$ to 45~MeV), and a peak with a width of this order would be produced with solely such a form factor without further dynamics. Were this true, one should expect similar peaks at all $S$-wave thresholds, in contradiction to observations.

Therefore, we conclude that, if there is no other nearby singularity, a narrow pronounced near-threshold peak cannot be produced just by a threshold cusp. It is more likely produced by a nearby pole in the unphysical Riemann sheet (with respect to the elastic channel) of the complex energy plane as a result of the involved strong interaction dynamics. 
It was suggested in Ref.~\cite{Guo:2013sya} that the $Z_c(3900)$ and $Z_c(4020)$ correspond to virtual state poles in the $I=1$ $D\bar D^*\to D\bar D^*$ amplitude, which may be located a few tens of MeV below the corresponding thresholds, and a multi-channel fit using a nonrelativistic formalism with unitarity built in Refs.~\cite{Hanhart:2015cua,Guo:2016bjq} to the Belle data suggests the $Z_b(10610)$ to be a virtual state and the  $Z_b(10650)$ to be a resonance. 

It is worthwhile to notice that in the above discussion, we have assumed that the production vertex (the $Y\to \pi D\bar D^*$ vertex $g_\text{el}$ in the considered example) does not have any nontrivial structure. In fact, for the $Z_c(3900)$ and $Z_c(4020)$ cases, the situation is more complicated because the production processes can proceed through triangle diagrams. The presence of TSs~\cite{Wang:2013cya,Wang:2013hga}, which are a few tens of MeV away, makes the problem more complicated. 
This tricky issue will be addressed in Section~\ref{sec:zc_ts}. 

Here we want to briefly comment on the lattice results by the HALQCD~\cite{Ikeda:2016zwx,Ikeda:2017mee} which suggest that the $Z_c(3900)$ is a threshold cusp due to strong channel coupling. In these calculations, the $\pi J/\psi$, $\rho\eta_c$ and $D\bar D^*$ coupled-channel potentials were derived from lattice QCD using the HALQCD method with unphysical pion masses between 410 and 700~MeV. The potential was then put into the Lippmann--Schwinger equation, and a virtual state pole far from the physical region was found.  We will not discuss their method, but only point out that the obtained $D\bar D^*$ invariant mass distribution is too broad to account for the BESIII double $D$-tagged data with little background at $E_\text{cm}=4.26$~GeV~\cite{Ablikim:2015swa}.\footnote{The $D\bar D^*$ data used for comparison in Refs.~\cite{Ikeda:2016zwx,Ikeda:2017mee} have a large background that were yet to be subtracted.} 
Improved results with the physical pion mass would be useful to shed light into the nature of the $Z_c(3900)$ structure.

More charged charmonium-like and bottomonium-like structures as threshold cusps at the thresholds of a pair of ground state heavy mesons were predicted in Refs.~\cite{Chen:2011xk,Chen:2011pu,Chen:2013wca}. 
From the above discussion, it becomes clear that it is unavoidable to have cusps at these thresholds as long as the intermediate particles can couple to the final states in an $S$ wave. However, how strong the cusps are depends on detailed dynamics, and the cusp could be rather dramatic if there is a nearby pole. In fact, a nice showcase of this point is provided by the well-established scalar meson $a_0(980)$. The recent high-statistics data from BESIII of the $\chi_{c1}\to\eta\pi^+\pi^-$ process have a peak with a prominent cusp structure at the $K\bar K$ threshold in the $\pi\eta$ distribution~\cite{Kornicer:2016axs}. This sharp cusp structure is a result of the nearby $a_0(980)$ resonance which strongly couples with the $K\bar K$ channel~\cite{Kornicer:2016axs,Liang:2016hmr}.

Let us also mention a near-threshold structure observed in the light meson sector.
The BES Collaboration observed a resonance called  $X(1835)$ in the $\pi^+\pi^-\eta'$ invariant mass distribution of the $J/\psi\to \gamma\pi^+\pi^-\eta'$ decay~\cite{Ablikim:2005um}, which was confirmed later at BESIII~\cite{Ablikim:2010au}. This structure is just below the $p\bar p$ threshold, and might be related to the threshold enhancement in the $p\bar p$ final states observed in Refs.~\cite{Bai:2003sw,BESIII:2010krt}. 
In fact, in the updated BESIII measurement, at the right shoulder of the $X(1835)$ peak an abrupt drop is seen around the $p\bar p$ threshold in the $\pi^+\pi^-\eta'$ invariant mass distribution~\cite{Ablikim:2016itz}. The drop is likely due to the opening of the $p\bar p$ threshold. If the events are divided into narrower bins, there should be a visible cusp exactly at the $p\bar p$ threshold, and a closer look at the immediate vicinity of the $p\bar p$ threshold can shed important light into the $p\bar p$ interaction (see Refs.~\cite{Zou:2003zn,Haidenbauer:2008qv,Chen:2010an,Haidenbauer:2012pu,Kang:2013uia,Dai:2017ont,Dai:2018tlc} in this context; for a review of earlier works on the low-energy $N\bar N$ interaction, we refer to Ref.~\cite{Klempt:2002ap}).

\subsection{Threshold cusps in the quark mass dependence}
\label{sec:mpidependence}

In lattice QCD calculations, the quark masses as parameters can be tuned. Although many calculations nowadays are being performed with light (up and down) quark masses around their physical values, there are still lots of calculations done with much larger light quark masses to reduce the computational cost. 
Such a feature should not be regarded as purely a disadvantage as varying the quark masses  can lead to new insights into the internal structure of hadrons, see, \eg, the discussions on the pion mass dependence of the $f_0(500)$ and the $\rho$ resonances (varying the up and down quark masses can be recast into varying the pion masses by virtue of the Gell-Mann--Oakes--Renner (GMOR) relation $M_\pi^2\propto m_u+m_d$~\cite{GellMann:1968rz}) in Ref.~\cite{Hanhart:2008mx}, and the suggestions that the pion and kaon mass dependence can be used to investigate the role of the $D^{(*)}K$ meson pairs in the $D_{s0}^*(2317)$ and $D_{s1}(2460)$~\cite{Cleven:2010aw,Du:2017ttu,Eichten:2019may}.

Pions and kaons are the pseudo-Nambu--Goldstone bosons of the spontaneous breaking of the approximate chiral symmetry in QCD, SU$(3)_L\times$SU$(3)_R\to $SU$(3)_V$. As a result, their masses squared are proportional to the light quark (up, down and strange) masses at leading order (LO) of the chiral expansion (GMOR relation). In contrast, the mass of a hadron other than these pseudo-Nambu--Goldstone bosons is linear in light quark masses at LO, \ie, $m_h = \mathring{m}_h + c\, m_q$, where $\mathring{m}_h$ is the hadron mass in the limit of $m_q=0$ and $c$ is a coefficient.  
In general, hadrons in different SU(3) multiplets have different values of $c$. As a result, when the quark masses are varied, the mass of a hadron ($A$) may coincide with the threshold of a hadron pair ($B$ and $C$) to which it couples. 
For instance, by equating the $\rho$ meson mass $m_\rho(M_\pi) = \mathring{m}_\rho + c_1\, M_\pi^2$ to $2M_\pi$ with the coefficient fixed from the mass difference between $\rho$ and $K^*$, $c_1 = (m_{K^*}-m_\rho)/(M_K^2-M_\pi^2) = 0.51 \text{GeV}^{-1}$, it can be easily estimated that the $\rho$ meson mass coincides with the $\pi\pi$ threshold at $M_\pi\simeq400$~MeV~\cite{Guo:2011gc}, which is consistent with the results from unitarized chiral perturbation theory~\cite{Hanhart:2008mx,Pelaez:2010fj}.
Let us start from the situation that $A$ is heavier than the $B+C$ threshold. Then $A$ can decay into $B+C$ and the c.m. momentum of the decay product is given by 
\begin{equation}
	\frac1{2m_A} \sqrt{ \left[m_A^2 - (m_B+m_C)^2 \right] \left[m_A^2 - (m_B-m_C)^2 \right]}.
\end{equation}
Then one can conclude that the $M_\pi$ value at which $m_A(M_\pi) = m_B(M_\pi) + m_C(M_\pi)$ is a branch point of the complex $M_\pi$ plane. Similar to the threshold cusp we discussed above in the energy distributions, there should also be a cusp in the quark mass dependence (or pion and/or kaon mass dependence) of physical quantities at the point when a hadron becomes unstable relative to a given channel.
To the best of our knowledge, such a singularity was first explored in Ref.~\cite{Guo:2011gc}, and cusps were found in the pion mass dependence of the squared pion charge radius and the derivative of the $\rho$ meson mass with respect to $M_\pi$.

\begin{figure}[tb] 
\begin{center}
\includegraphics[height=5.4cm]{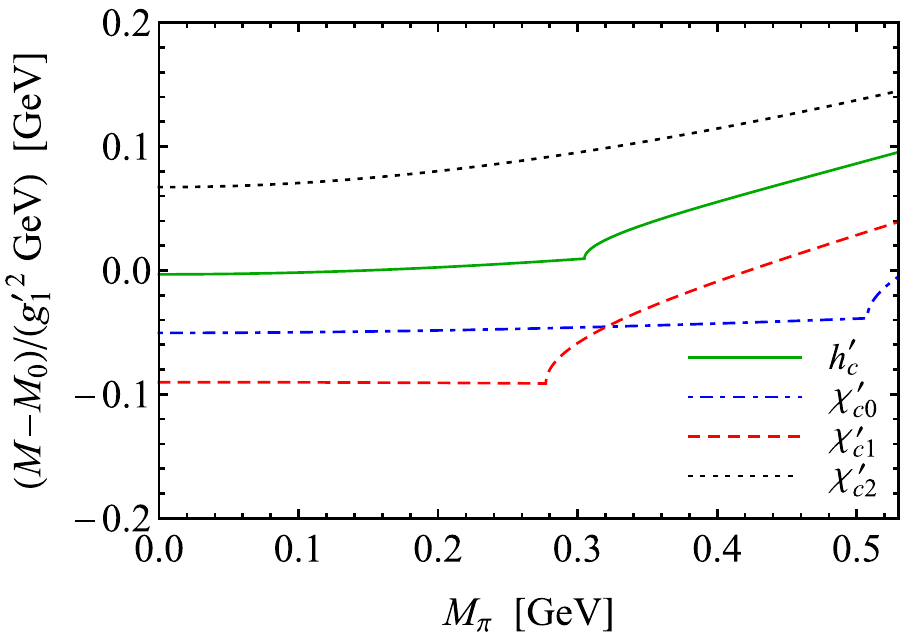} \qquad\quad
\includegraphics[height=5.4cm]{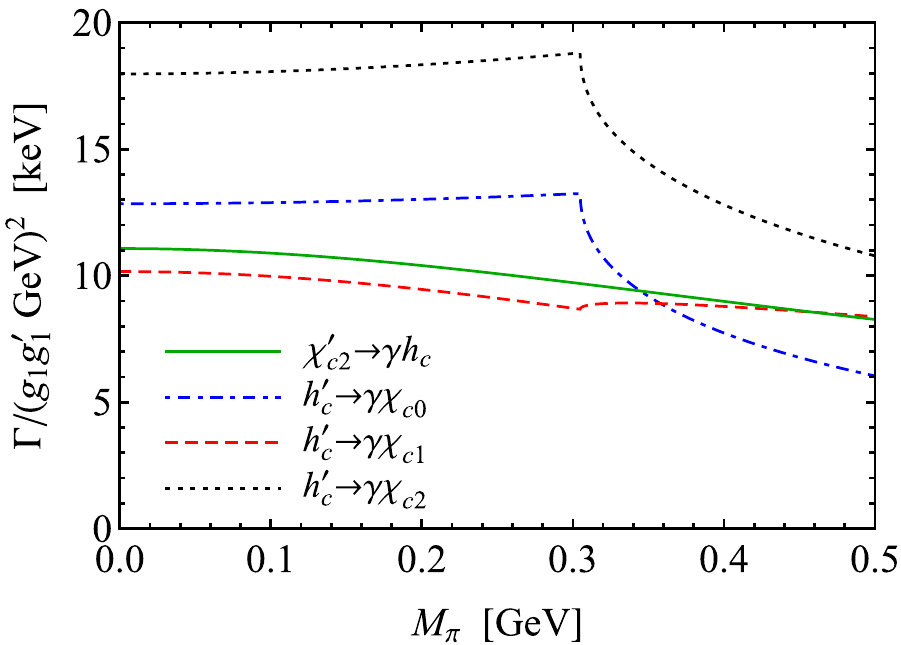}
\begin{minipage}[t]{16.5 cm}
\caption{Left: Pion mass dependence of the mass shifts of the $2P$ charmonia induced by charmed meson loops. Right: Pion mass dependence of the partial widths of the hindered M1 transitions between the $2P$ charmonia. Cusps happen when the $2P$ charmonium masses coincide with the thresholds of charmed meson pairs. The plots are adapted from Ref.~\cite{Guo:2012tg}.
\label{fig:mpidependence}}
\end{minipage}
\end{center}
\end{figure}

Such a nontrivial quark mass dependence needs to be considered in the chiral extrapolation of physical observables evaluated at large quark masses in lattice QCD. And it could provide new insights into the nature of the involved hadrons. 
The $2P$ charmonium states provide a good example. Among the four $2P$ states ($\chi_{cJ}(2P)$ with $J=0,1,2$ and $h_c(2P)$), only the candidates of $\chi_{c0}(2P)$ and $\chi_{c2}(2P)$ have been found, corresponding to the $\chi_{c0}(3860)$~\cite{Guo:2012tv,Chilikin:2017evr}\footnote{Note that the data statistics is rather low and the structure is broad, so that the observed $D\bar D$ invariant mass distribution can be easily fitted with different assumptions including the one without a resonance around this mass~\cite{Wang:2019evy}. Note also that the $\chi_{c2}(3930)$ was not taken into account in the analysis of Ref.~\cite{Chilikin:2017evr}, which might be reason that the central value of the extracted $\chi_{c0}(3860)$ mass is higher than that obtained  in Ref.~\cite{Guo:2012tv} from fitting to the Belle and \babar{} data for $\gamma\gamma\to D\bar D$~\cite{Uehara:2005qd,Aubert:2010ab}.} and the $\chi_{c2}(3930)$~\cite{Uehara:2005qd,Aubert:2010ab}. 
The $\chi_{c1}(2P)$ has the same quantum numbers as the $\x$. In addition, the puzzling $X(3915)$~\cite{Uehara:2009tx} might also be related to the $2P$ charmonia. There are suggestions that these observed structures originate from the $2P$ states whose mass spectrum gets shifted because of their coupling to the charm and anticharm meson pairs, see, \eg, Refs.~\cite{Zhou:2017dwj,Ortega:2017qmg,Cincioglu:2016fkm}. 
The quark mass dependence can be used to study this issue. For that, one needs to pay attention to the ``threshold'' cusps when the charmonia become unstable from a stable particle along decreasing the light quark mass.\footnote{Since a charm meson contains both charm and light quarks, the quark mass dependence of its mass should be stronger than that of a charmonium. Thus, at sufficiently large light quark mass, the open-charm meson masses would be large enough so that the charmonium would not be able to decay into a pair of charm mesons. Decreasing the light quark mass, the charmonium mass may become larger than the open-charm meson threshold.} The behavior has been predicted in Ref.~\cite{Guo:2012tg}, shown in Fig.~\ref{fig:mpidependence}. It is worthwhile to investigate whether the cusp behavior can be used to extract the corresponding coupling strength, similar to the extraction of scattering lengths discussed in Section~\ref{sec:scatteringlength}.

%--------------------------------
%--------------------------------

\bigskip

%% file: section4.tex
\section{Triangle singularities}
\label{sec:4}

Before we claim the experimentally observed resonance-like structures are genuine particles, such as multi-quark states or molecular states, it is necessary to investigate some other possibilities.  An intriguing character of the recently discovered $XYZ$ states is that many of them are located close to
two-particle thresholds. This is the reason why many of these structures are regarded as the candidates of hadronic molecules in many papers (for a review, see~\cite{Guo:2017jvc}). However, phenomenologically whether and which heavy flavor hadrons can form bound states is usually a model-dependent and fine-tuning problem. It is also a difficult problem for lattice QCD if the binding energy is small so that a large lattice size is needed to properly compute the system. 
On the other hand, the signal of some of these resonance-like structures may contain important (or even be dominated by) kinematic TS effects, which result from the rescattering processes with three particles in the intermediate state. 
The kinematical singularities of the rescattering amplitudes will behave themselves as bumps in the invariant mass distributions, and usually these singularities stay close to the pertinent thresholds. This implies the possibility of a non-resonance explanation for some peaking structures which would otherwise be due to exotic hadrons. 

\begin{figure}[tb]
  \begin{center}
  \begin{minipage}[t]{8 cm}
  \includegraphics[width=\linewidth]{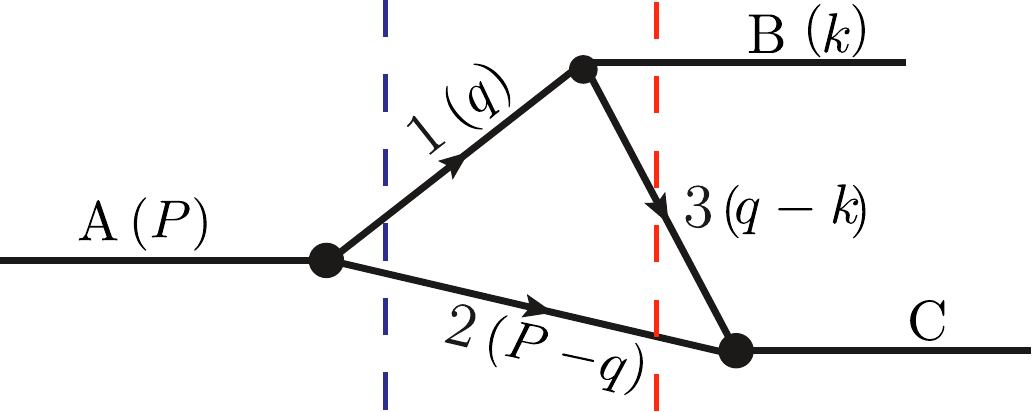}
  \end{minipage}
  \begin{minipage}[t]{16.5 cm}
   \caption{A triangle diagram for the reaction from A to B and C.  The two vertical dashed lines denote the two relevant cuts discussed in Section~\ref{sec:coleman-norton}.
   \label{fig:triangle}}
  \end{minipage}
  \end{center}
\end{figure}

The possible manifestation of the TS of $S$-matrix elements was already noticed in the 1960s and theoretical attempts were made to try to clarify whether some resonance-like structures were caused by the kinematic singularities or they were genuine resonance peaks. The so-called Peierls mechanism was proposed in 1961~\cite{Peierls:1961zz}, which suggested that  peaks could be produced from a triangle diagram without a genuine resonance. The original Peierls mechanism was proposed to study $\Delta\pi \to \Delta\pi$ reaction by exchanging a nucleon. Since the initial state must come from some short-distance source, the triangle diagram entered the game with the process corresponding to $m_1=m_C=m_{\Delta(1232)}$, $m_2=m_B=m_{\pi}$ and $m_3=m_N$ in Fig.~\ref{fig:triangle}. 
But in such a kinematic configuration the TS of the scattering amplitude in $m_A$ does not lie on the physical boundary, or in other words, the singularity is located on the wrong Riemann sheet which is far away from the physical region~\cite{Goebel:1964zz,Hwa:1963aa,Srivastava:1963aa,Aitchison:1964ak}. The location of the singularity is schematically shown as point $R'$ in Fig.~\ref{fig:riemannsheets}, whose path to the physical region is much longer than that of point $R$ just below the cut on the second Riemann sheet (a  detailed discussion on how one can test whether the singularity is on the physical boundary or not is given below in Section~\ref{sec:coleman-norton}). Therefore, the original Peierls mechanism actually does not work. 

The singularity can be located on the physical boundary in the modified and inverted Peierls mechanisms~\cite{Kacser:1964pl,anisovich1964logarithmic,month1965evidence,Schmid:1967ojm} (the paper by Schmid~\cite{Schmid:1967ojm} contains a clear review of various versions of the Peierls mechanism).
In the modified Peierls mechanism~\cite{Kacser:1964pl,Schmid:1967ojm}, the restriction of $m_1=m_C$ in the original Peierls mechanism is released. It focuses on the singularity effect in the distribution of $m_A$, \ie, the invariant mass of the initial state, with $m_C$ fixed around the threshold $(m_2+m_3)$. While the inverted Peierls mechanism~\cite{anisovich1964logarithmic} focuses on the singularity effect in the distribution of $m_C$, \ie, the invariant mass of the final state, with $m_A$ fixed near the threshold $(m_1+m_2)$. We will discuss the two cases separately in the following sections. 
However, Schmid in Ref.~\cite{Schmid:1967ojm} argued that for the single-channel case the rescattering diagrams cannot produce obvious peaks in the Dalitz plot projections, even if the rescattering amplitude possesses a TS on the physical boundary. For the elastic rescattering process, in addition to the triangle diagram, there must also be a corresponding resonance-production tree diagram. When it is added coherently to the triangle rescattering diagram, for which the TS dominates over the non-singular part, the effect of the triangle diagram is nothing more than multiplying a partial wave amplitude of the tree diagram by a phase factor. 
Therefore the singularities of the triangle diagram cannot produce obvious peaks in the angle integrated invariant mass distributions, though it can leave some footprint in the full Dalitz plot distribution~\cite{Schmid:1967ojm}.
This is the so-called Schmid theorem. But for the reactions involving inelastic rescattering processes~\cite{Anisovich:1995ab,Szczepaniak:2015hya,Debastiani:2018xoi}, the situation will be quite different from the single-channel case discussed in Ref.~\cite{Schmid:1967ojm}. A detailed discussion about the Schmid theorem can be found in Section~\ref{sec:Schmid}. 

Most of those proposed observable effects induced by kinematic singularities in 1960s were lacking experimental support at that time. Triggered by many new experimental discoveries in hadron spectroscopy, the importance of the kinematic singularity mechanism, especially the TS mechanism, was rediscovered in recent years and used to interpret some phenomena related to exotic hadron candidates.

\subsection{Physical picture of the triangle singularity }
\label{sec:coleman-norton}

\begin{figure}[tb]
  \begin{center}
  \includegraphics[width=9cm]{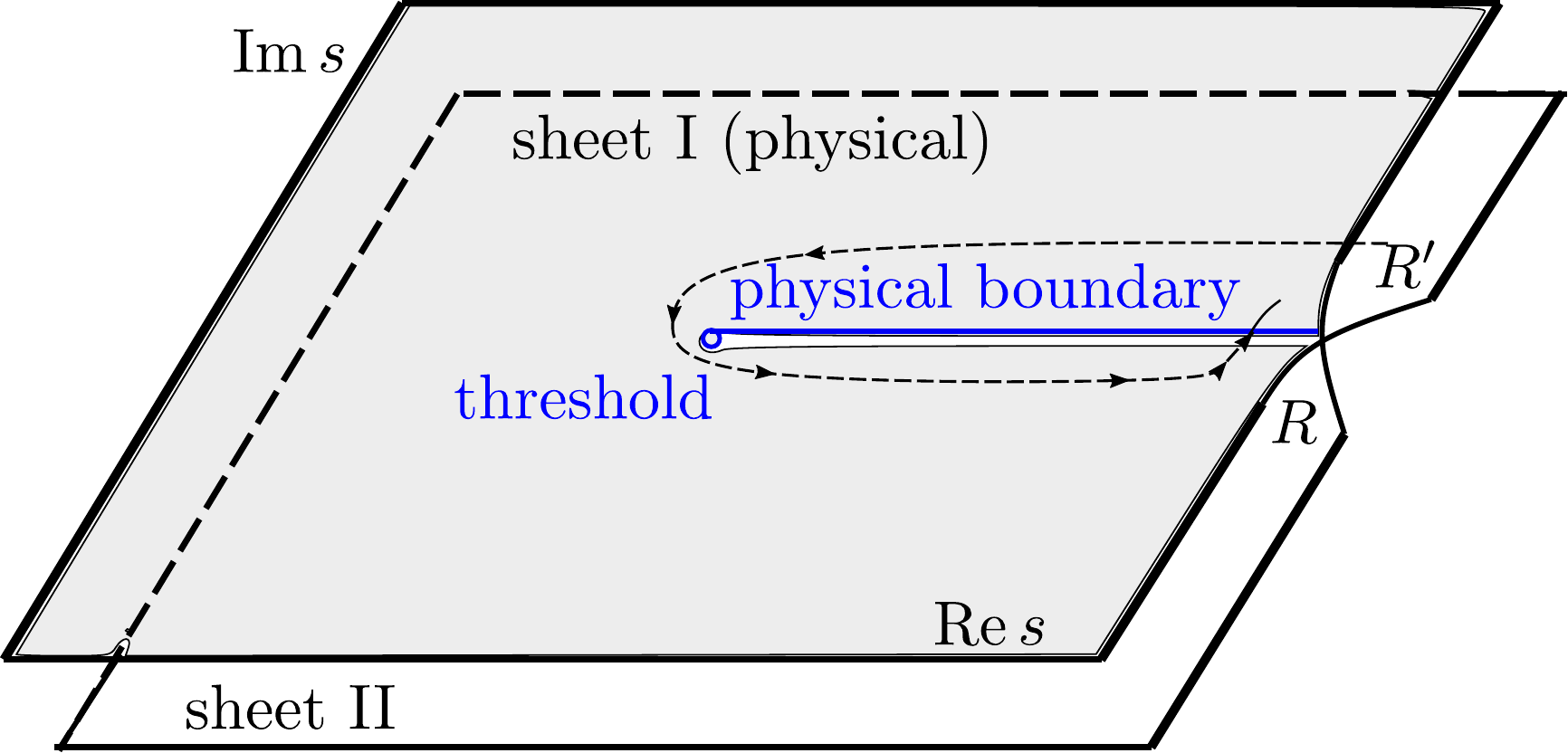}
  \begin{minipage}[t]{16.5 cm}
   \caption{
   \label{fig:riemannsheets}Structure of two Riemann sheets. The shaded area represents the first (physical) sheet. The upper (lower) edge of the cut on the first sheet is continuously connected to the lower (upper) edge of the cut on the second sheet. The dashed curve with arrows schematically shows the path for a point $R'$ in the upper half plane of the second Riemann sheet to affect the physical region.}
  \end{minipage}
  \end{center}
\end{figure}
The threshold of a two-body channel is a square-root branch point (see Eq.~\eqref{eq:qcm}), and a cut can be drawn from the threshold along the positive real $s$ axis until infinity to make the square-root function single valued in the whole Riemann surface. The cut divides the whole complex-$s$ plane into two Riemann sheets, as shown in Fig.~\ref{fig:riemannsheets}.
The physical region for an amplitude involving these two particles is the domain where the particles can go on shell, and is given by the upper edge along the cut on the first Riemann sheet, which is also called the physical Riemann sheet. It is continuously connected to the lower edge along the cut on the second Riemann sheet. 

It has been shown generally by Coleman and Norton that the Landau equations given in Eq.~\eqref{eq:landau}, together with the requirement that $\alpha_i \ge 0$ and all momenta are real, are the sufficient and necessary conditions for the Landau singularities to be on the physical boundary~\cite{Coleman:1965xm}, and the physical picture for these conditions are that for a given Feynman diagram, the interactions at all vertices can happen as classical processes in spacetime conserving energy and momentum, with internal particles on their mass shell and moving forward in time. 
This is called the Coleman--Norton theorem. The physical picture for the TS was discussed earlier in Ref.~\cite{Bronzan:1964zz}; see also Refs.~\cite{Eden:1960,Landshoff:1962} for discussions of the physical region singularities.

Let us consider the triangle diagram shown in Fig.~\ref{fig:triangle}, which represents a reaction from the initial state A to the final states B and C through a triangle diagram with intermediate particles 1, 2 and 3. The external A, B and C are not necessarily single particles as long as the interaction at each vertex is of short-distance. The physical picture for a TS to happen in the physical region is as follows.
Consider the rest frame of particle A.\footnote{See also Refs.~\cite{Schmid:1967ojm,Debastiani:2018xoi} for arguments in the $2+3$ c.m. frame.} Particle A decays into particles 1 and 2 flying back to back, then particle 1 decays into particles 3 and B. Particle 3 moves in the same direction as particle 2 with a larger velocity so that it can catch up with particle 2, and then particles 2 and 3 collide to form C in the final state. During this process,  all intermediate particles are on their mass shell so that they may propagate for an infinite time if they are stable. This means that all of the three vertices represent processes that can happen classically.
This picture corresponds to that of the leading Landau singularity for the decay region of $y_{12}<-1$, $y_{23}<-1$ and $y_{13}>1$ with $y_{ij}$ defined in Eq.~\eqref{eq:y}, which we shall discuss.

In the following, we review the recent formalism of Ref.~\cite{Bayar:2016ftu} where a physically intuitive equation for TSs in the physical region was derived, and the physical picture becomes quite apparent in this formalism.

Let us start from the scalar triangle loop integral
\begin{align}
    I(k) = i\int\!\frac{d^4q}{(2\pi)^4}
\frac{1}{\left(q^2-m_1^2+i\epsilon\right) \left[(P-q)^2-m_2^2+i\epsilon\right]
    \left[(q-k)^2-m_3^2+i\epsilon\right]},
    \label{eq:loop_triangle}
\end{align}
where the momenta have been labeled in Fig.~\ref{fig:triangle}. Since the TS happens when all particles are on shell, one may focus on the positive energy pole part of each propagator, and write
\begin{align}
    I(k) \simeq \frac{i}{N_m} \int \!\frac{d^4q}{(2\pi)^4}
\frac{1}{\left[q^0-\omega_1(q)+i\epsilon\right] \left[P^0-q^0-\omega_2(q)+i\epsilon\right]
    \left[q^0-k^0-\omega_3(\vec{q}-\vec 
k\,)+i\epsilon\right]} ,
    \label{eq:scalarIa}
\end{align}
where $N_m=8m_1m_2m_3$, $\omega^{}_{1,2}(q)=\sqrt{m_{1,2}^2+{q}^{2}}$ with 
$q\equiv|\vec{q}\,|$, $\omega^{}_3(\vec{q}-\vec 
k\,)=\sqrt{m_3^2+(\vec{q}-\vec k\,)^2}$. Here we have approximated $\omega_1\omega_2\omega_3$ by $m_1m_2m_3$ for simplicity without affecting the singularity structure (the expression without such an approximation can be found in Ref.~\cite{Bayar:2016ftu}).

Performing the contour integration over $q^0$, Eq.~\eqref{eq:scalarIa} becomes
\begin{eqnarray}
  I(k) &=& -\frac{1}{N_m}\int \frac{d^3\vec q}{(2\pi)^3}\,
   \bigg\{
   \left[ P^{0}-\omega_1(q)-\omega_{2}(q)+i\,\epsilon\right] 
   \left[E_{C}
   -\omega_{2}(q)-\omega_{3}(\vec{q} -\vec{k}\,)+i\,\epsilon\right] \bigg\}^{-1}
  \nonumber\\
  &=& -\frac{2}{\pi^2 N_m } \int_0^\infty d q\,\,
  \frac{q^2 f(q)}{P^{0}-\omega_1(q)-\omega_2(q)+i\,\epsilon} \,,
\label{eq:I}
\end{eqnarray}
where $E^{}_C=(m_A^2-m_B^2+m_C^2)/(2m_A)$ is the energy of particle C in the rest frame of A. The two terms in the curly brackets in the first line correspond to the two cuts shown as dashed lines in Fig.~\ref{fig:triangle}. The function $f(q)$ is given by
\begin{equation}
  f(q) = \int_{-1}^1 dz\,\frac1{ E_{C}-\omega_{2}(q) -
  \sqrt{m_3^2+q^2+k^2-2q\,k\,z} + i\,\epsilon} \, . \label{Eq:fq}
\end{equation}
The integral $I(k)$ is a function of various masses and external momenta involved in the triangle diagram. Here we choose to discuss the singularities in the variable $m_C$. 
From Eq.~\eqref{eq:I}, one needs to analyze the singularity structure of double integrals with one over the magnitude of loop momentum $l$ and the other over $z$, the cosine of the polar angle. The integrands of both 
integrals have singular points. However, a singularity of the integrand does not necessarily become a singularity of the integral. In the complex plane of the integration variable, if the integration contour can be deformed to avoid the 
singularity, the integral will be a regular function. There are cases that the contour cannot be deformed, and a singularity develops: endpoint singularity and pinch singularity, see Section~\ref{sec:landau_derivation}.

When particles 1 and 2 are on shell, the integrand of $I(q)$ is singular, and we have
\begin{equation}
  P^0 - \omega_1(q) - \omega_2(q) + i\,\epsilon = 0\, , 
\label{eq:cut1}
\end{equation}
corresponding to the left cut in Fig.~\ref{fig:triangle}, which has two solutions
\begin{equation}
  q_{{\rm on}\pm} \equiv \pm\left[\frac1{2 m_A} \sqrt{\lambda(m_A^2,m_1^2,m_2^2)} +
i\,\epsilon\right] . 
\label{eq:qon}
\end{equation}
The one with the positive sign, $q_{\text{on}-}$ is irrelevant since it is outside the integration region of $q$. Here the $i\,\epsilon$ is kept explicitly, and it is essential in order to determine how the integral contour is pinched in the complex-$q$  plane.

The function $f(q)$ has two endpoint singularities given by vanishing the denominator of its integrand for $z=\pm1$. That is when particles 2 and 3 are on shell and they move parallel or anti-parallel to each other,
\begin{equation}
  E_{C}-\omega_{2}(q) -
  \sqrt{m_3^2 + q^2+k^2\mp2q\,k} + i\,\epsilon = 0\, ,
  \label{eq:cut2}
\end{equation}
corresponding to the right cut in Fig.~\ref{fig:triangle}. The minus (positive) sign corresponds to 
$z=+1$ $(-1)$, and the situation that the 
momentum of particle 2 is parallel (anti-parallel) to the the momentum of particle C in the rest frame of A, respectively.
The two endpoint singularities of $f(q)$ then become singularities of the 
integrand of $I(q)$. 

Equation~\eqref{eq:cut2} has four solutions. The two solutions for $z=+1$ are given by
\begin{equation}
 q_{a+} = \gamma \left( \beta \, E_2^* + p_2^* \right) + i\,\epsilon\,, \quad
 q_{a-} = \gamma \left( \beta \, E_2^* - p_2^* \right) - i\,\epsilon\, ,
 \label{eq:qa}
\end{equation}
and the two solutions for $z=-1$ are given by
\begin{equation}
 q_{b+} = \gamma \left( - \beta \, E_2^* + p_2^* \right) + i\,\epsilon\,, \quad
 q_{b-} = - \gamma \left( \beta \, E_2^* + p_2^* \right) - i\,\epsilon\, ,
 \label{eq:qb}
\end{equation}
where
\begin{equation}
  E_2^* = \frac{m_{C}^2+m_2^2-m_3^2}{2 m_{C}},\quad
  p_2^* = \frac{\sqrt{\lambda(m_{C}^2,m_2^2,m_3^2)}}{2 m_{C}},
\end{equation}
are the energy and the magnitude of the three-momentum of particle 2 in the  c.m. frame of the $2+3$ 
system, respectively, $\beta=k/E_C$ is the magnitude of the velocity of the $2+3$ system in the rest frame of A, and $\gamma= 1/{\sqrt{1-\beta^2}} ={E_{C}}/{m_{C}}$ is the Lorentz boost factor. 
The four solutions differ from one another only by signs of individual terms,
and thus correspond to the momentum of the intermediate particle 2 in the rest 
frame of A in different kinematical regions.

\begin{figure}[tb]
  \begin{center}
    \includegraphics[width=0.65\linewidth]{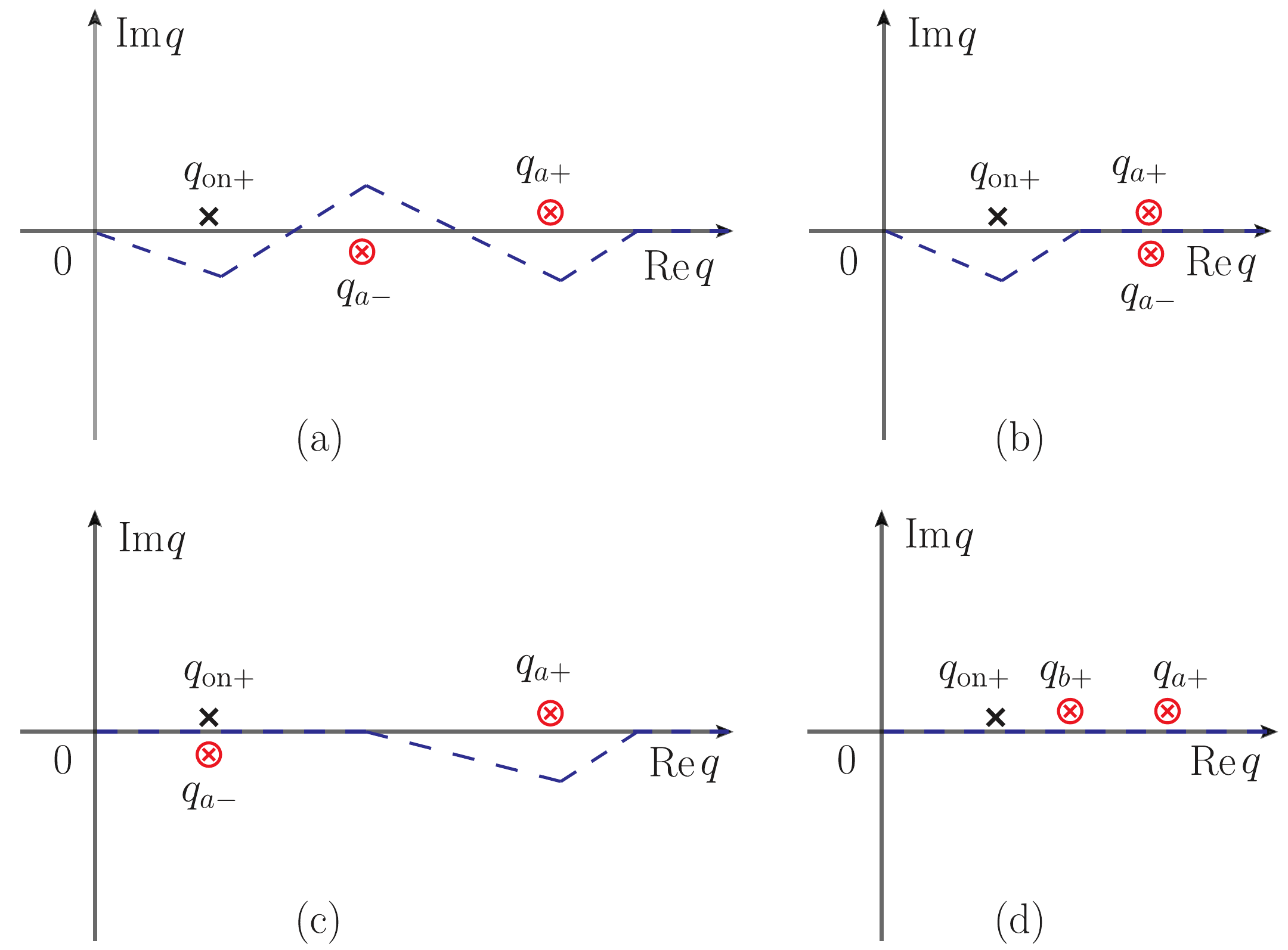}
  \begin{minipage}[t]{16.5 cm}
  \caption{Different situations for the locations of the pertinent singularities of the $I(k)$ integrand.  (a) and (d) are the cases without any pinching, (b) shows the case when $q_{a+}$ and $q_{a-}$ pinch the integration path, which gives
  the two-body threshold singularity at $m_C=m_2+m_3$, and (c) shows the case when the integration path is trapped between $q_{{\rm on}+}$ and $q_{a-}$, which gives the triangle   singularity. Possible integration paths are depicted as dashed lines.  $q_{\text{on}+}$ and $q_{b-}$, which are  outside the integration region, are not shown.
  \label{fig:sing1} }
  \end{minipage}
  \end{center}
\end{figure}

Among the four solutions of Eq.~\eqref{eq:cut2}, $q_{b-}$ is irrelevant for the integral over $q$ in $I(k)$ since it is always negative (when $\epsilon=0$). 
Furthermore, $q_{b+}=-q_{a-}$, and thus only one of them can be in the integration range from 0 to $+\infty$ of $q$. The locations of the pertinent singularities of the $I(k)$ integrand in Eqs.~\eqref{eq:qon},~\eqref{eq:qa} and 
\eqref{eq:qb} are shown in Fig.~\ref{fig:sing1}. The sub-diagrams correspond to different kinematical regions. When $q_{b+}>0$, there is no singularity in the lower half of the complex-$q$ plane as shown in (d), 
and $I(k)$ is regular in this region. The case for $q_{a-}>0$ is divided into 
three situations, shown as (a), (b) and (c) in the figure.
%%%%%
%%%%%
When the pole of the $I(k)$ integrand at $q_{{\rm on}+}$ and the two 
logarithmic branch points $q_{a\pm}$ take different values, the integration path can be deformed freely as long as it does not hit any of these points. Such a 
situation is shown in diagram (a), and $I(k)$ is analytic in the corresponding kinematical region. 
Since $q_{a-}$ and $q_{\rm on+}$, $q_{a+}$ are located on opposite sides of the real-$q$ axis, the integration path could be pinched between $q_{a-}$ and one of $q_{\rm on+}$ and $q_{a+}$ or even both of them simultaneously. Then the integration path cannot be deformed to avoid that 
point, and $I(k)$ gets a singularity. Diagram (b) shows the case for the 
pinching between $q_{a-}$ and $q_{a+}$, and it is easy to see that this can 
only happen when $p_2^*=0$ or $m_C=m_2+m_3$. Thus, this gives the 
two-body threshold cusp at the threshold of particles 2 and 3, which is a square-root 
branch point. When the pinching happens between $q_{a-}$ and $q_{\rm on+}$, as 
shown in diagram (c), the TS, which is a logarithmic branch point, develops. Therefore, the condition for $I(k)$ to have a TS in the physical region is given by~\cite{Bayar:2016ftu}
\begin{equation}
  \lim_{\epsilon\to 0} \left( q_{\rm on+} - q_{a-} \right) = 0 \, .
  \label{eq:trianglesing}
\end{equation}
When there is a real solution of this equation, there is a TS in the physical region, and the  solution gives the location of the singularity in a chosen variable. If both $q_{\rm on+}$ and $q_{a+}$ pinch the integration
path with $q_{a-}$ simultaneously, then the two-body threshold and the triangle
singularity coincide at $m_{C}=m_2+m_3$.

Now let us consider the kinematical region where $p_2 =
\lim_{\epsilon\to0}q_{a-} = \gamma (\beta\,E_2^*-p_2^*) >0$. The momentum of 
particle 3 in the rest frame of the initial particle $p_3=\gamma (\beta\,E_3^*+p_2^*)$, where $E_3^*$ is the energy of particle 3 in the c.m. frame of the $2+3$ system,
is positive as well.
This means that particles 2 and 3 move in the same direction in that frame (noticing $z=+1$). In addition, the velocities of these two particles in the same frame are given by
\begin{eqnarray}
  \beta_2 = \beta\, \frac{E_2^*-p_2^*/\beta}{E_2^*-\beta\, p_2^*}\,,\quad
  \beta_3 = \beta\, \frac{E_3^*+p_2^*/\beta}{E_3^*+\beta\, p_2^*}\,,
\end{eqnarray}
respectively. It is easy to see that $p_2>0$ leads to 
\begin{equation}
  \beta_3>\beta>\beta_2\,,
\end{equation}
which means that particle 3 moves faster than particle 2 and in the same direction in the rest frame of the initial particle A.
This, together with the requirement that all intermediate particles are on their mass shell, gives the condition for having a TS in the physical region. 
Thus, we get the physical picture implied by the Coleman--Norton theorem~\cite{Coleman:1965xm} for the TS: the singularity is on the physical boundary if and only if the diagram can be interpreted as a classical process in spacetime.

\subsection{Kinematic variables for  the physical region triangle singularity}
\label{sec:kinvars}

\begin{figure}[t]
 \centering
 \includegraphics[height=6cm]{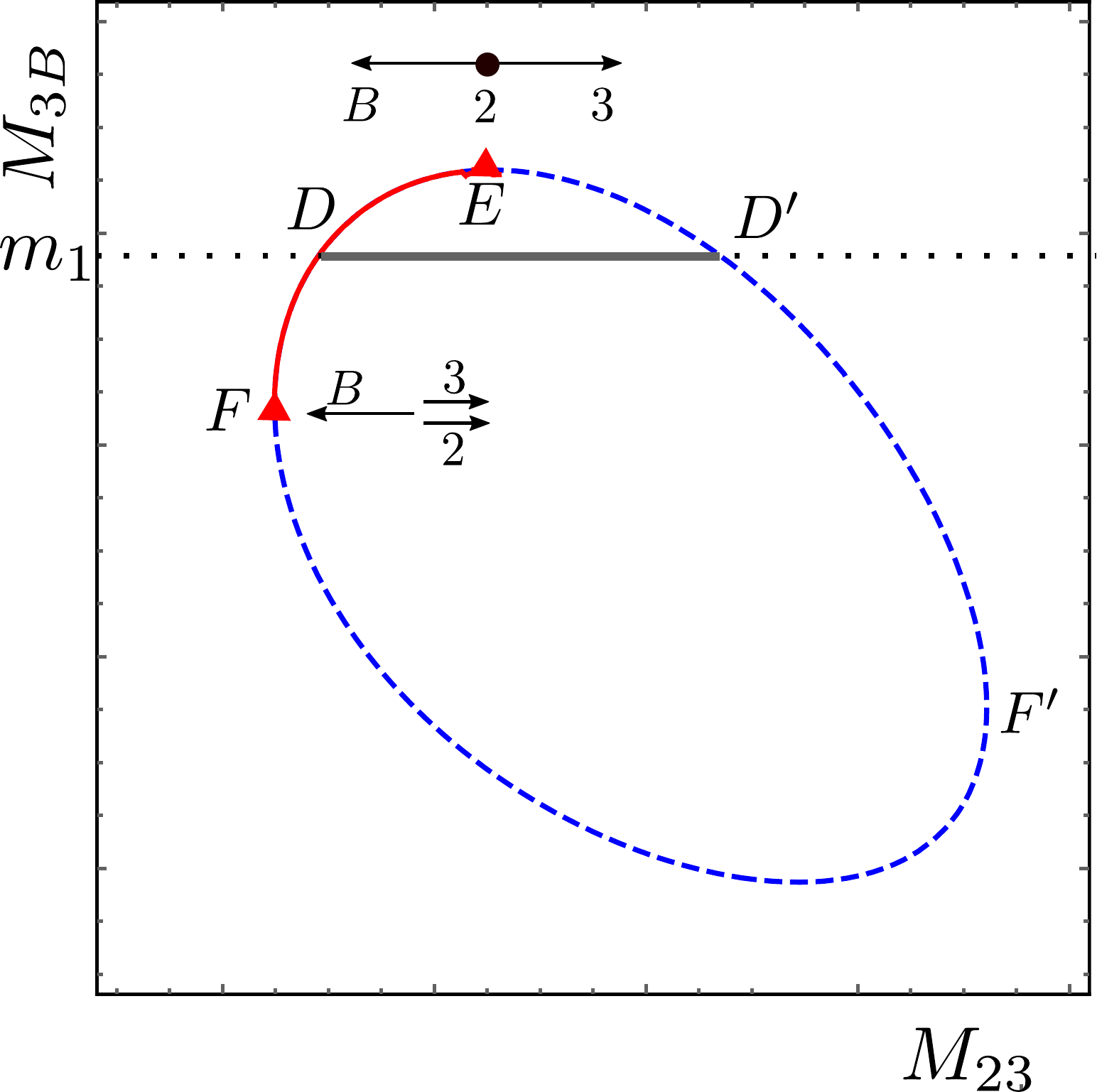} \qquad
 \includegraphics[height=6cm]{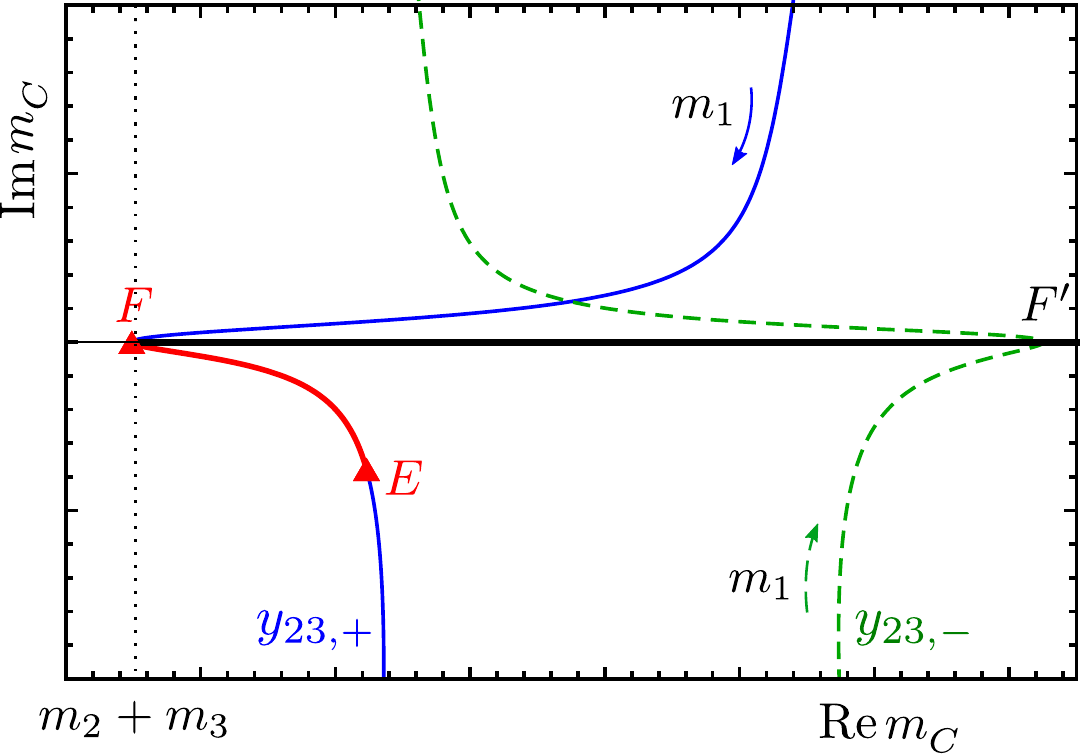}
  \begin{minipage}[t]{16.5 cm}
 \caption{Left: The Dalitz plot region for the process $A\to B +2+3$. The intersection of the resonance band for particle 1, with a mass $m_1$, with the Dalitz plot boundary (the dashed circle) at point $D$ produces singularities. The region when the singularity is on the physical boundary is shown as a solid segment along the Dalitz plot boundary. The moving directions of particles $2,3$ and B in the rest frame of A are also plotted.
 Right: Motion of the solutions of Eq.~\eqref{eq:landau_y}, given in Eq.~\eqref{eq:y23}, in the complex plane of $m_C=M_{23}$. In order to separate the trajectories from the real axis, a small negative imaginary part $(-i\,\Gamma_1/2)$ has been given to the $m_1$. Along the trajectories, only when located on the part between the points $E$ and $F$, the singularity is on the physical boundary (for $\Gamma_1$ approaching zero). Point $F'$ denotes the pseudo-threshold $m_A-m_B$ which is the end of the decay phase space. The solid straight line along the real axis is the cut from the $2+3$ threshold. The arrows show the moving directions of the singularities with increasing the mass of the internal particle 1. 
 }
 \label{fig:dalitz_traj}
  \end{minipage}
\end{figure}

From the above discussion, it is easy to find out the region of kinematical variables where a TS is on the physical boundary. Let us consider the process in Fig.~\ref{fig:triangle} with fixed values of $m_A$, $m_2$, $m_3$ and $m_B$, and try to find out in which region $m_1$ is for a singularity to be on the physical boundary.
We start from a large mass for particle 1 so that $m_1> m_A - m_2$. Apparently, particles 1 and 2 cannot go on shell. Decreasing $m_1$, particles 1 and 2 can be on shell when $m_1 = m_A - m_2$, labeled as point $E$ in the Dalitz plot for the process $A\to B+2+3$ in the left panel of Fig.~\ref{fig:dalitz_traj} (see also Chapter~4.13 in Ref.~\cite{Anisovich:2008zz} for a detailed discussion). At this point, particles 1 and 2 are produced at rest in the rest frame of A, and particle 3 from the decay of particle 1 can definitely interact with particle 2 classically.
Decreasing $m_1$ further, particles 1 and 2 move back to back. On the boundary of the Dalitz plot, particles 1, 2 and 3 always move collinearly. On the solid (red) segment, particles 2 and 3 are parallel to each other (along the same direction), and the velocity of particle 3 is still larger than that of particle 2 until point $F$ in the figure. At point $F$, they have the same velocity in the rest frame of A, and thus their total energy is at their threshold. If $m_1$ is decreased further, particle 3 would not be able to catch up with particle 2 any more.
At points $E$ and $F$, one has $y_{23}=-1$ and $y_{12}=-1$, respectively.
As a result, only when $m_1$ is within the following range:
\begin{equation}
 m_1^2 \in \left[ m_{1\text{low}}^2, m_{1,\text{up}}^2 \right],\quad\text{with} ~~m_{1\text{low}}^2= \frac{m_A^2 m^{}_3 + m_{B}^2 m^{}_2}{m^{}_2+m^{}_3} - m^{}_2 m^{}_3\,,~
 m_{1,\text{up}}^2 = \left(m^{}_A-m^{}_2 \right)^2 ,
 \label{eq:m1range}
\end{equation}
there can be a TS on the physical boundary. In terms of the mass of $C$ (the invariant mass of the $2+3$ system), it is within the range
\begin{equation}
 m_{C}^2 \in \left[ (m_2+m_3)^2,~ 
\frac1{m_1} \left[(m_1+m_2)(m_1m_2 + m_3^2) - m_2 m_B^2\right]
 \right],
 \label{eq:m23range}
\end{equation}
with the lower and upper limits corresponding to $m_1 = m_{1,\text{low}}$ and $m_{1,\text{up}}$, respectively. The range of the invariant mass squared for particles 1 and 2 for the TS to be in the physical region is 
\begin{equation}
  m_A^2 \in \left[(m_1 +m_2)^2, m_1^2 + m_2^2 + m_2 m_3 + \frac{m_2}{m_3} \left(m_1^2 -m_B^2\right) \right],
  \label{eq:mArange}
\end{equation}
with the lower and upper limits corresponding to $m_1 = m_{1,\text{up}}$ and $m_{1,\text{low}}$, respectively.
For discussions of such ranges, see, \eg, Refs.~\cite{Aitchison:1964zz,Anisovich:1966,Schmid:1967ojm,Szczepaniak:2015eza,Liu:2015taa,Guo:2015umn,Guo:2016bkl,Jing:2019cbw}. 

Noticing that the above discussion assumes that  the masses for all the particles are real, so if the singularity is really in the physical region, the amplitude would be logarithmically divergent, which is an infrared divergence because it happens when all particles are on-shell. This would not happen, because for all the intermediate particles being on-shell, particle 1 must be able to decay into particles 3 and B and gets a finite width. This width effectively adds a negative imaginary part to $m_1$, which moves the singularity off the real $m_C$ axis into the complex plane. 
As a result, the amplitude of the triangle diagram is safely finite, and has a peak in the invariant mass distribution of $M_{23}=m_C$ ($C$ can be regarded as more than one particle; this is the inverted Peierls mechanism with inelastic rescattering) because of the singularity.\footnote{It is more proper to take into account the width of particle 1 using the spectral function method. However, it has been shown in Ref.~\cite{Aitchison:1964ak} that using a complex mass for the resonance leads to an appropriate approximation for calculating the enhancement effects of the singularity near the physical region.}
Such phenomena have been extensively discussed in the context of new hadrons (see the rows labeled with ``F'' in Tables~\ref{tab:list-meson} and \ref{tab:list-baryon}), and will be the focus of Section~\ref{sec:6}.

In the right panel of Fig.~\ref{fig:dalitz_traj}, we show schematic trajectories of the two solutions of Eq.~\eqref{eq:landau_y} in the complex plane of $M_{23}=m_C$, corresponding to
\begin{equation}
  y_{23,\pm}^{} = y_{12}^{} y_{13}^{} \pm \sqrt{ (y_{12}^{2}-1) (y_{13}^{2}-1) } ,
  \label{eq:y23}
\end{equation}
with the external momenta in the $y_{ij}$ definitions being $p_1^2 = M_{23}^2$, $p_2^2 = m_B^2$ and $p_3^2 = m_A^2$.
Using the Coleman--Norton theorem, one knows on which Riemann sheet of the complex $m_C$ plane, which has a unitary cut starting from $m_2+m_3$, the two solutions are located. 
Since the physical boundary is the upper edge of the cut on the first Riemann sheet, which is continuously connected to the lower edge of the cut on the second Riemann sheet (see Fig.~\ref{fig:riemannsheets}), we can conclude that the $y_{23+}$ branch (solid curve in the plot) should always be on the second Riemann sheet. Then the  segment from $E$ to $F$, which is separated from the real axis because a complex mass $m_1-i\,\Gamma_1/2$ is used for particle 1, approaches the physical real $m_C$ axis from below by decreasing $\Gamma_1$.
On the contrary, the $y_{23-}$ branch, whose trajectory is shown as the dashed curve in the plot, is on the second Riemann sheet when it is above the real $m_C$ axis, and it crosses the cut at point $F'$, which is $m_A - m_B $, into the lower half-plane of the first Riemann sheet. Thus, it is always far away from the physical region and would not cause any visible impact on the physical amplitude. 

One may also fix the invariant masses in the final state, $m_B$ and $m_C$, and look for a TS in terms of $m_A$. 
Such a structure may mimic a resonance in the BC invariant mass distribution (the modified Peierls mechanism but with inelastic rescattering). 

\begin{figure}[tb]
  \centering
  \includegraphics[width=0.9\textwidth]{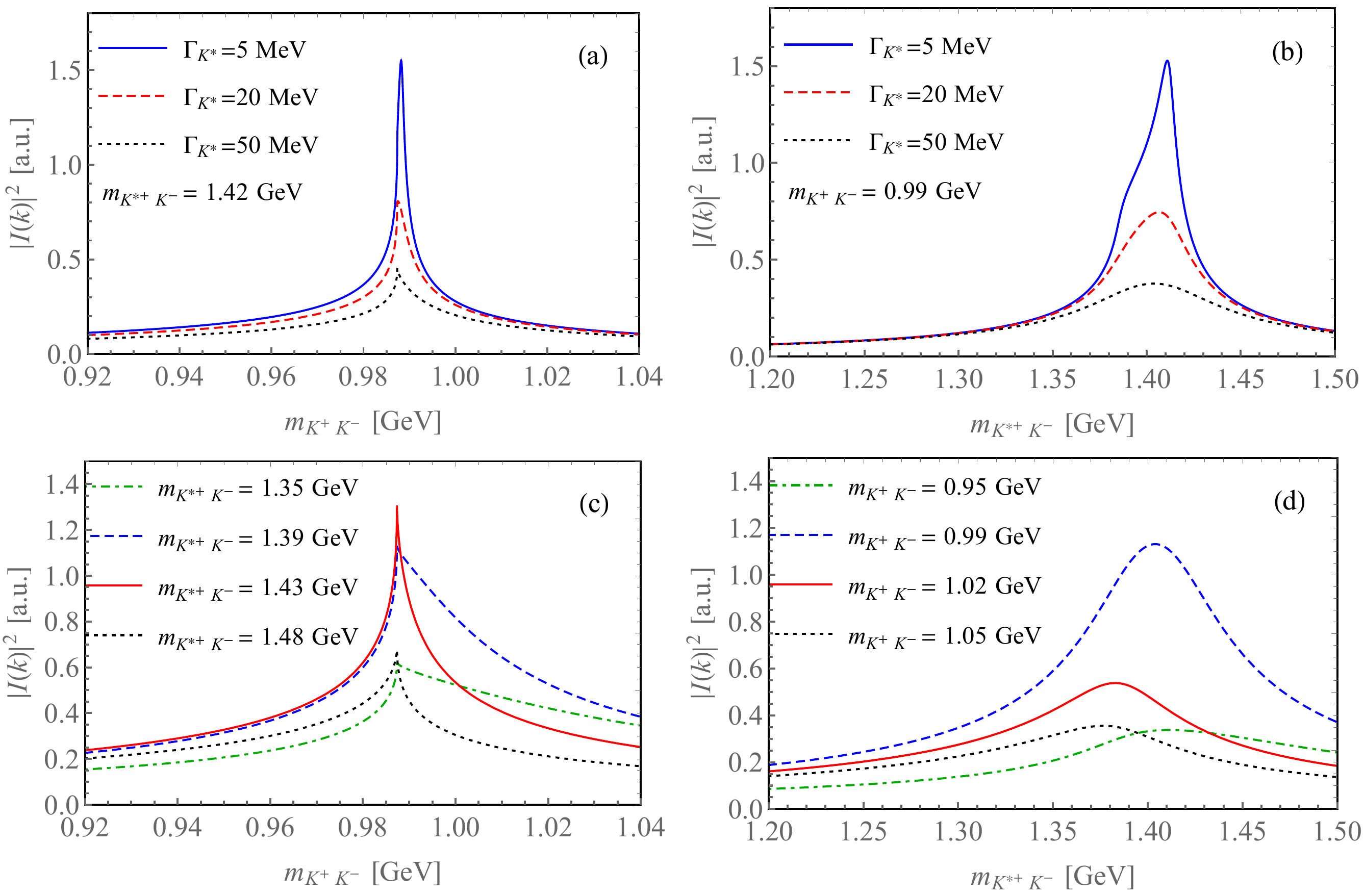}
  \begin{minipage}[t]{16.5 cm}
  \caption{The absolute value squared of the scalar triangle loop integral with intermediate particles 1, 2 and 3 being $K^{*+}$, $K^-$ and $K^+$, and the external particle $B$ being $\pi^0$ (in arbitrary units). For the $K^{*+}$ width taking three different values, (a) shows that structures in the $K^+K^-$ invariant mass distribution for $m_{K^{*+}K^-}=1.42$~GeV, and (b) shows the structures in the $K^{*+}K^-$ distribution for $m_{K^+K^-}=0.99$~GeV. Fixing the $K^{*+}$ width to 50~MeV, (c) shows the $K^+K^-$ distributions for $m_{K^{*+}K^-}$ fixed at a few chosen values, and (d)  shows the $K^{*+}K^-$ distributions for $m_{K^+K^-}$ fixed at different values.}
  \label{fig:kstarkk}
  \end{minipage}
\end{figure}
For example, let us consider a triangle diagram with internal particles 1, 2 and 3 being $K^{*+}$, $K^-$ and $K^+$, respectively, and particle $B$ being a neutral pion. Using Eq.~\eqref{eq:mArange}, one readily finds out that the TS is in the physical region for the $K^{*+}K^-$ invariant mass to be in the range $[1385.3, 1435.0 ]$~MeV; using Eq.~\eqref{eq:m23range}, the singular range for the $K^+K^-$ invariant mass is $[987.4, 1025.9]$~MeV. One sees that the singular range for the $m_{K^+K^-}$ covers the masses of the $f_0(980)$, $a_0(980)$ and $\phi$. 
In fact, a few structures around 1.42~GeV found in the $f_0(980)\pi$, $a_0(980)\pi$ and $\phi\pi$ final states have been proposed to be due to TSs. They are listed in Table~\ref{tab:list-meson}, and will be briefly reviewed in Section~\ref{sec:lightmesons}.

In order to see the TS effects, in Fig.~\ref{fig:kstarkk} we show the energy distributions induced by the scalar triangle integral, see Eq.~\eqref{eq:loop_triangle}. For all the plots, the units are arbitrary, but the relative heights in each plot are fixed. When the $K^{*+}K^-$ invariant mass is in the singular range given above, the $K^+K^-$ distribution has a sharp peak with a cusp at the threshold. 
Increasing the $K^*$ width reduces the height (notice that the coupling constants are not considered; for a discussion of the interplay of the coupling constant and the width of the internal particle, we refer to Ref.~\cite{Wu:2018xaa}), yet the peak keeps sharp. This is because kaons are stable, and the $K^*$ width cannot smear the threshold peak at the $K^+K^-$ threshold.
On the contrary, the TS induced peak in the $K^{*+}K^-$ distribution gets smeared by the $K^{*}$ width, see Fig.~\ref{fig:kstarkk} (b). The left shoulder of the solid (blue) curve in (b) has a $K^*\bar K$ threshold cusp smeared by the small width of 5~MeV, and such a structure becomes invisible when the $K^*$ width is much larger than the distance between the $K^*\bar K$ threshold and the TS location which is at about 1.412~GeV, see the dashed (red) and dotted (black) curves.  

Figure~\ref{fig:kstarkk} (c) and (d) show the sensitivity of the triangle diagram induced structures on kinematical variables. From plot (c), one sees that the cusp is more prominent when $m_{K^*\bar K}$ takes values in the range for the TS to be in the physical region, which is the case for $m_{K^*\bar K}=1.39$ and 1.43~GeV. 
Similarly, for the $K^*\bar K$ invariant mass distribution, the peaks in the dashed (blue) and solid (red) are more pronounced than those in the dotted (black) and dot-dashed (green) ones. This is because for the former ones, $m_{K\bar K}$ is in the singular range while it is not for the latter. 
Interestingly, one can see that the line shapes of the peaks are similar to that of a resonance parameterized in the Breit--Wigner form.
The sensitivity of the peaks on kinematical variables is a key feature for the TS, and can be used to distinguish it from a genuine resonance.

\subsection{Argand plot}
\label{sec:arganddalitz}

The Argand plot, the parametric plot of the real and imaginary parts of the reaction amplitude with the variation of energy, gives the phase motion of the amplitude around a resonant peak (see, \eg, Ref.~\cite{Taylor:1972} for an elementary textbook).
Normally, a circular rotation in the counterclockwise direction along with the energy increase is associated with a resonance, and is often regarded as its signature.\footnote{As pointed out in Ref.~\cite{Guo:2015umn}, such a counterclockwise motion of the Argand plot can also be reproduced with a simple two-body rescattering model like Eq.~\eqref{eq:1loop}, but with an unclosed circle.}
Argand plots have been used to extract resonances from the pion-nucleon scattering in the 1960s and 1970s~\cite{Hohler:1983}.
Recent developments from the experimental side enable us to access the Argand plots in various reactions in the heavy-hadron sector.
For example, the Argand plots of the $\psi'\pi$, $J/\psi p$, and $D^0p$ distributions were presented recently by the LHCb Collaboration in Refs.~\cite{Aaij:2014jqa,Aaij:2015tga,Aaij:2016fma}, respectively.

\begin{figure}[tb]
  \centering
  \includegraphics[width=0.9\textwidth]{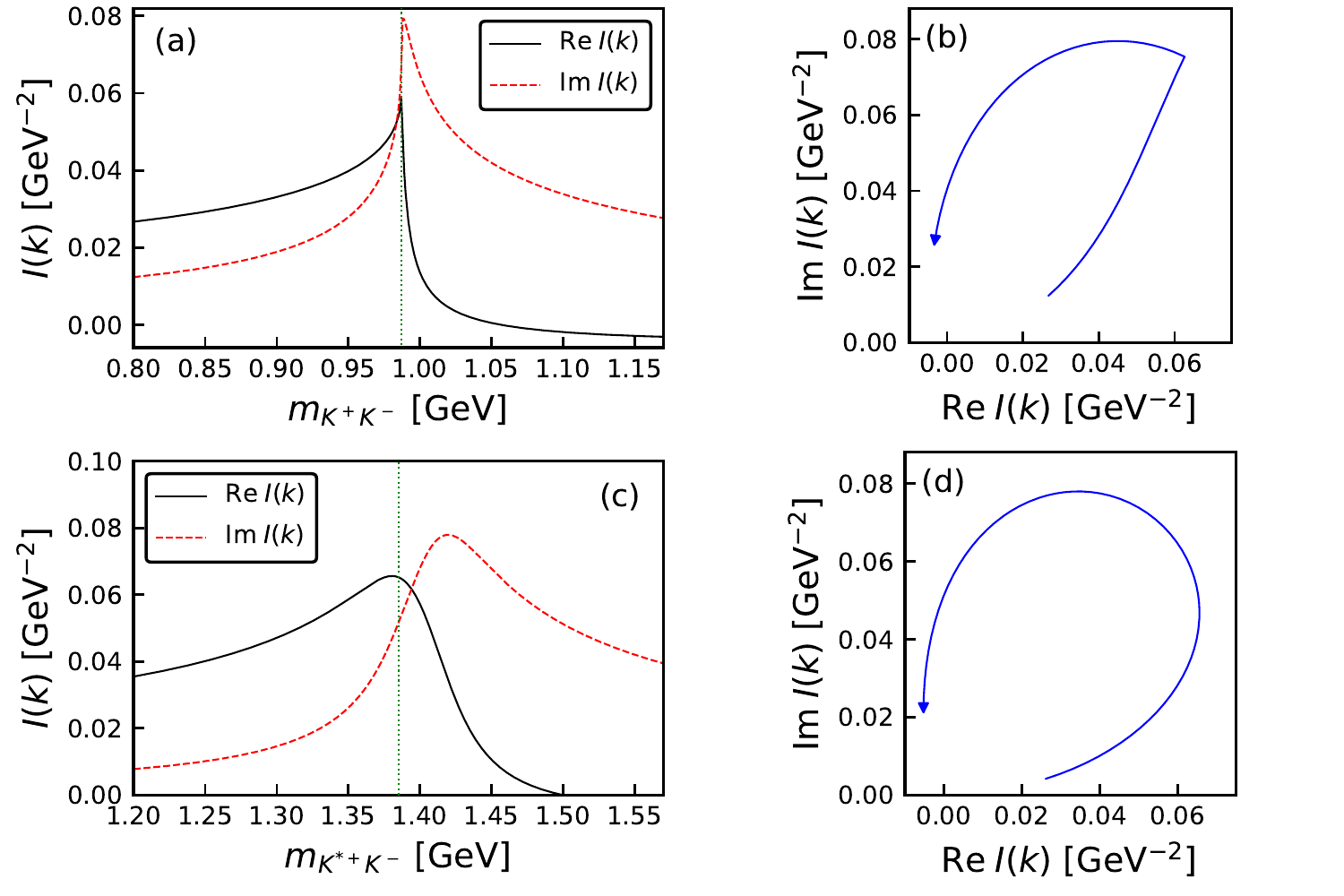}
  \begin{minipage}[t]{16.5 cm}
  \caption{Phase motion of the scalar three-point loop integral $I(k)$ for the intermediate states being $(1,2,3)=(K^{*+}, K^-,K^+)$ and the external particle B being the $\pi^0$. (a) shows the real and imaginary parts as functions of $m_{K^+K^-}$ for $m_{K^{*+}K^-}=1.42$~GeV; (c) shows the real and imaginary parts as functions of $m_{K^{*+}K^-}$ for $m_{K^{+}K^-}=0.99$~GeV; (b) and (d) are the Argand plots obtained by varying $m_{K^+K^-}$ (with $m_{K^{*+}K^-}$ fixed at 1.42~GeV) and $m_{K^{*+}K^-}$ (with $m_{K^{+}K^-}$ fixed at 0.99~GeV), respectively. 
  The two vertical dotted lines in (a) and (c) denote the $K^+K^-$ and $K^{*+}K^-$ thresholds, respectively.
  The $K^*$ width of 50~MeV is taken into account by replacing $m_{K^*}^2$ in the $K^*$ propagator with $m_{K^*}^2-i m_{K^*}\Gamma_{K^*}$.}
  \label{fig:Argand}
  \end{minipage}
\end{figure}
To see the Argand plot in the presence of a TS, we again take the example of the triangle diagram with the internal particles 1, 2 and 3 being $K^{*+}$, $K^-$ and $K^+$, respectively, and particle B being a neutral pion  (notice that only the masses of the particles are used here, and the other properties like the $P$-wave coupling of $K^*\to\pi K$, which are not important to see the TS properties below, are not considered). We take a value for each of the $K^+K^-$ and $K^{*+}K^-$ in their respective singular regions and vary the other one. For the case with $m_{K^{*+}K^-}=1.42$~GeV and varying $m_{K^+K^-}$, the real and imaginary parts of the scalar three-point loop integral $I(k)$ in Eq.~\eqref{eq:loop_triangle} and the Argand plot are shown in Fig.~\ref{fig:Argand} (a) and (b), respectively, and those for $m_{K^{+}K^-}$ fixed at 0.99~GeV with varying $m_{K^{*+}K^-}$ are shown in Fig.~\ref{fig:Argand} (c) and (d).

In plot (a), the energy of the prominent cusp in both the real and imaginary parts corresponds to the $K^+K^-$ threshold. In fact, the imaginary part reaches its maximum at an energy about 1~MeV higher than the threshold, corresponding to the location of the TS. The threshold cusp also shows up in the Argand plot in Fig.~\ref{fig:Argand}~(b).  As can be seen, the counterclockwise motion with increasing energy, which is normally associated with a resonant state, is captured by $I(k)$ while the shape looks deformed compared with the Breit--Wigner case. 
In the $K^{*+}K^-$ distribution in plot (c), the threshold cusp is smeared out due to the $K^*$ width, and consequently the corresponding Argand plot is smooth, mimicking the resonance behavior.

In Refs.~\cite{Mikhasenko:2015vca,Mikhasenko:2016mox,Nakamura:2019btl,Nakamura:2019emd,Liu:2019dqc,Braaten:2019gwc}, the Argand diagram of various amplitudes with triangle diagrams are considered.
In these works, it is found that the resonance-like behavior of the Argand diagram can be simulated by the specifically chosen triangle loop with a TS. 
In particular, the analysis of the COMPASS data on the $a_1(1420)$ structure~\cite{Adolph:2015pws} by introducing either a resonance or a triangle diagram with a nearby TS~\cite{Mikhasenko:2016mox,Ketzer:2019wmd} is a nice showcase for that the Argand plot is generally not able to distinguish a resonance from TS effects, as expected from plot (d) in Fig.~\ref{fig:Argand}. In the COMPASS amplitude analysis, it is found that these two scenarios can fit to the data with a similar quality, and the TS effects interfering with a background can perfectly reproduce the measured counterclockwise circular Argand plot, just like the one with a resonance plus a background~\cite{Akhunzyanov:2018lqa,Mikhasenko:2019talk}.

\subsection{Schmid theorem and Dalitz plot distribution}
\label{sec:Schmid}

\begin{figure}[tb]
  \centering
  \includegraphics[width=10cm]{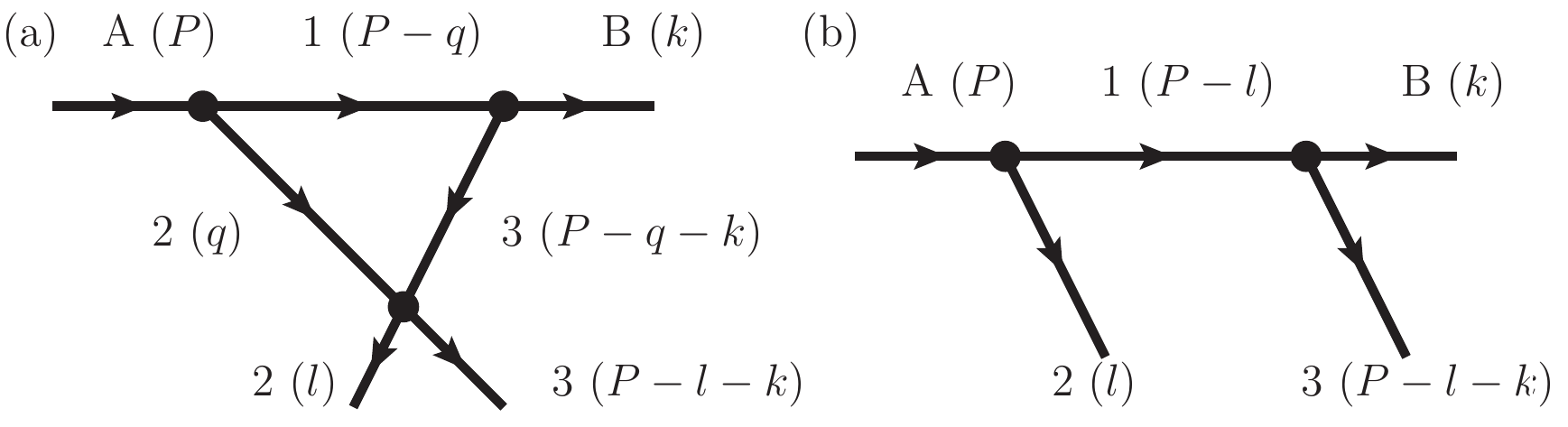}
  \begin{minipage}[t]{16.5 cm}
  \caption{(a) a triangle loop diagram with a rescattering of $23\rightarrow 23$, 
  and (b) a tree diagram with the same particles in the final state.}
  \label{fig:triangle_mod}
  \end{minipage}
\end{figure}

\begin{figure}[t]
 \centering
 \includegraphics[width=0.45\linewidth]{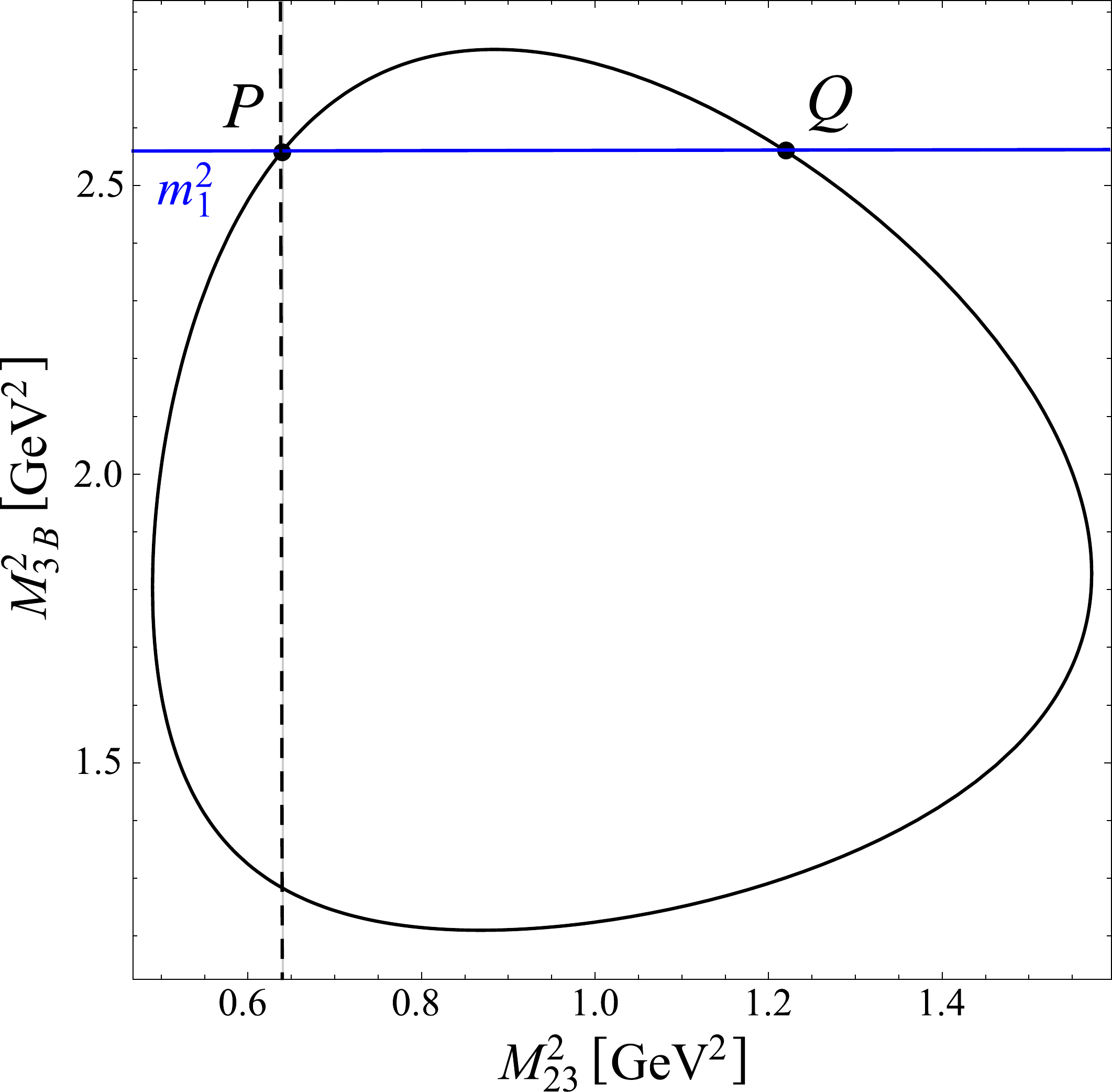}
  \begin{minipage}[t]{16.5 cm}
 \caption{Schematic picture of the Dalitz plot with a tree and TS contributions.
 The horizontal and vertical axes are $M_{23}^2$ and $M_{3B}^2$, respectively.
 The horizontal solid line corresponds to the band of particle $1$ from $T_{({\rm tree})}$,
 and the vertical dashed line to the band of the TS from $T_{({\rm loop})}$.}
 \label{fig:dalitz}
  \end{minipage}
\end{figure}

For the discussion on the Schmid theorem~\cite{Schmid:1967ojm}, let us consider the process shown in Fig.~\ref{fig:triangle_mod}~(a) where particles $2$ and $3$ involved in the triangle loop rescatter from each other and give the same particles in the final state.\footnote{We consider the case where all of the vertices are scalar, and the coupling constants $g_{A,12}$ and $g_{1,3B}$ are set to one for simplicity.}
In this case, the tree diagram with the same final state shown in Fig.~\ref{fig:triangle_mod}~(b) also contributes, and the contribution is not negligible in general.
The Schmid theorem claims that the logarithmic singularity of the triangle loop, $T_{({\rm loop})}$, is absorbed into the phase of $S$-wave projected part of the corresponding tree amplitude $T_{({\rm tree})}^{(l=0)}$ for particles 2 and 3 up to regular terms: 
\begin{equation}
    T_{({\rm tree})}^{(l=0)}+T_{(\rm loop)}\sim S_{23}\,T_{({\rm tree})}^{(l=0)},
\end{equation}
with $S_{23}$ being $S$-matrix element of the $2+3\to 2+3$ process. Here, we use $\sim$ to mean that this holds only when the logarithmic singular part dominates over the regular part for the loop contribution.
Hence, the loop contribution cannot be observed in the $2+3$ invariant mass distribution once the $2+3$ angle integration is performed.
One can understand it intuitively with the classical picture of the process:
without channel coupling, the amount of particles $2$ and $3$ produced at the tree level cannot be changed by the rescattering of $2+3\to 2+3$, and only their angular distribution is changed. 
Hence, particularly the angle independent $S$-wave part $T_{(\rm tree)}^{(l=0)}$ is not altered by such a rescattering up to an overall phase.
To see it more clearly, the phase space of the $\rmA\to \rmB+ 2+3$ process is useful (see Fig.~\ref{fig:dalitz} for a schematic figure).
While the three particles are on shell in the whole region, the collinear condition for the TS as discussed in Sec.~\ref{sec:coleman-norton} restricts the TS location on the phase space boundary.
With the angle $\theta$ between $\vec{k}$ and $\vec{q}$ in the $2+3$ c.m. frame, the invariant mass of the $3+\rmB$ pair, $M_{3B}$, is written as
\begin{align}
 M_{3B}^2=m_A^2+m_2^2-2\tilde{E}_A\tilde{E}_2+2\tilde{p}_A\tilde{p}_2\cos\theta,\label{eq_ct}
\end{align}
where the energies and momenta with tilde are the quantities in the $2+3$ c.m. frame.
Then, the TS condition of $\cos\theta=1$ in the $2+3$ c.m. frame~\cite{Debastiani:2018xoi} constrains the region of TS on the upper-half arc of the phase space boundary in Fig.~\ref{fig:dalitz};
the condition that the particles $1$ and B are parallel  in the particle A rest frame restricts the region to the upper-left arc of the phase space boundary in Fig.~\ref{fig:dalitz} (see also the left panel in Fig.~\ref{fig:dalitz_traj}).
The allowed ranges of $M_{23}^2$ and $M_{3B}^2$ in order that the TS appears are written as follows~\cite{Bayar:2016ftu} (see Eqs.~\eqref{eq:m23range} and \eqref{eq:m1range}, respectively):
\begin{align}
 M_{23}^2\in \left[ (m_2+m_3)^2,\;\frac{m^{}_A m_3^2-m_{B}^2m^{}_2}{m^{}_A-m^{}_2} + m^{}_A m^{}_2 \right],
 \label{Eq:m23limit}\\
  M_{3B}^2\in \left[ \frac{m^{2}_A m_3+m_{B}^2m^{}_2}{m^{}_2+m^{}_3} - m^{}_2 m^{}_3,\;(m_A-m_2)^2 \right].
 \label{Eq:m3Blimit}
\end{align}
The on-shell condition of particle $1$ is $M_{3B}=m_1$.
Then, the point of the TS is the upper-left arc of the boundary crossing with the line $M_{3B}=m_1$, which is point $P$ in Fig.~\ref{fig:dalitz}.
Then, the sequential decay of $\rmA\to 1+2$ with $1\to 3+\rmB$ following the rescattering of $2+3\to 2+3$ can proceed classically at point $P$, and the final rescattering redistributes the events at point $P$ along with the vertical line crossing it (the vertical dashed line in Fig.~\ref{fig:dalitz}).
The validity and limitation of Schmid theorem have been studied in Refs.~\cite{Aitchison:1969tq,Goebel:1982yb,Anisovich:1995ab,Szczepaniak:2015hya,Debastiani:2018xoi}.

Let us take a look at the triangle-loop amplitude in Fig.~\ref{fig:triangle_mod} (a).
With the Cutkosky cutting formula~\cite{Cutkosky:1960sp},
the discontinuity of $T_{\rm (loop)}$ across the $2+3$ cut is given by (here we define $s=M_{23}^2$)
\begin{align}
    T_{\rm (loop)}(s+i\epsilon)-T_{\rm (loop)}(s-i\epsilon)=2i\rho_{23}(s)T_{\rm (tree)}^{(l=0)}(s)T_{23}^*(s),\label{eq:disctloop}
\end{align}
where $\rho_{23}(s)$ is the phase space factor of the $2+3$ pair, $T_{23}$ is the $2+3$ scattering $T$-matrix in an $S$ wave, and $T_{\rm (tree)}^{(l=0)}(s)$ is the $S$-wave projection  of the tree-level amplitude for the $t$-channel exchange of particle 1, whose expression can be found in Refs.~\cite{Achasov:1989ma,Ketzer:2015tqa}.
The singularity of $T_{\rm (loop)}$ in the physical region corresponds to that of the $S$-wave projected tree-level amplitude $T_{\rm(tree)}^{(l=0)}$.
One should note that this discontinuity does not coincide with the imaginary part of $T_{\rm (loop)}$ multiplied by $2\,i$, and it is a complex function in general as stressed in, \eg, Ref.~\cite{Szczepaniak:2015eza}.
Now, we focus on the $S$-wave projected tree-level amplitude in the $2+3$ c.m. frame.
In this frame, taking only the positive energy pole, the tree-level amplitude for the $S$-wave projection $T_{({\rm tree})}^{(l=0)}$ is written as
\begin{align}
 T_{({\rm tree})}^{(l=0)}(s)= \frac{1}{2}\int_{-1}^1d\cos\theta\, T_{({\rm tree})}\sim\frac{1}{4kl}\log\left(\frac{\tilde{E}_A-\omega_1(k-q)-\omega_2(q)+i\epsilon}{\tilde{E}_A-\omega_1(k+q)-\omega_2(q)+i\epsilon}\right),\label{eq_tree}
\end{align}
where $\omega_1(k\pm q)=\sqrt{m_1^2+(k\pm q)^2}$ and $\omega_2(q)=\sqrt{m_2^2+{q}^2}$.
Hereinafter, the amplitude $T_{({\rm tree})}$ apart from its $S$-wave part is denoted by $T_{({\rm tree})}^{(l\neq0)}$.
Thus, in the $S$-wave part of the tree-level amplitude, Eq.~\eqref{eq_tree}, logarithmic singularities appear at $\tilde{E}_A-\omega_1(k\pm q)-\omega_2(q)+i\epsilon=0$.

Using the discontinuity relation in Eq.~\eqref{eq:disctloop} with a dispersion integral~\cite{Schmid:1967ojm,Szczepaniak:2015hya}, or evaluating the $q$ integral of Eq.~\eqref{eq:scalarIa} explicitly~\cite{Debastiani:2018xoi}, as given in, \eg, Ref.~\cite{Aitchison:1966zz} with the nonrelativistic reduction, the most singular part of $T_{({\rm loop})}$ is given by (note the sign difference from Ref.~\cite{Debastiani:2018xoi} due to the choice of convention)
\begin{align}
 T_{({\rm loop})}\sim \frac{i\,q_{\rm on}}{4\pi\sqrt{s} }T_{({\rm tree})}^{(l=0)}\,T_{23}(s),
\end{align}
with $q_{\rm on}=\sqrt{\lambda(s,m_2^2,m_3^2)}/(2\sqrt{s})$.
Then the sum of the $S$-wave projected tree-level amplitude $T_{({\rm tree})}^{(l=0)}$ with the triangle-loop contribution $T_{({\rm loop})}$ can be written as
\begin{align}
 T_{({\rm tree})}^{(l=0)}+T_{({\rm loop})} \sim &\; \left(1+\frac{i\,q_{\rm on}}{4\pi \sqrt{s}}T_{23}(s)\right)T_{({\rm tree})}^{(l=0)}
 \notag\\
 =&\; \eta e^{2\,i\,\delta_{23}}T_{({\rm tree})}^{(l=0)},\label{eq-schmid}
\end{align}
where the $2+3$ scattering $T$-matrix is written as 
\begin{equation}
  T_{23}(s)=\frac{8\pi \sqrt{s}}{2\,i\,q_{\rm on}}\left(\eta e^{2i\delta_{23}}-1\right),
\end{equation}
with $\eta$ and $\delta_{23}$ being the inelasticity and scattering phase shift for the $S$-wave $2+3\rightarrow 2+3$ scattering, respectively (see Eq.~\eqref{eq:TLtb} with $L=0$).
Then, the projection of the Dalitz plot onto the $M_{23}$ direction is, with Eq.~\eqref{eq-schmid}, written as follows:
\begin{align}
 \frac{d\Gamma_{A\to B23}}{dM_{23}}\propto&\int_{-1}^1d\cos\theta\, \left|T_{({\rm loop})}+T_{({\rm tree})}\right|^2
 =\int_{-1}^1d\cos\theta\left(\left|T_{({\rm loop})}+T_{({\rm tree})}^{(l=0)}\right|^2+\left|T_{({\rm tree})}^{(l\neq 0)}\right|^2\right)\label{eq_tsa}\\
 \sim &\int_{-1}^1d\cos\theta\left(\left|\eta e^{2i\delta_{23}}T_{({\rm tree})}^{(l=0)}\right|^2+\left|T_{({\rm tree})}^{(l\neq 0)}\right|^2\right)\notag\\
 =&\int_{-1}^1d\cos\theta\left(\left|T_{({\rm tree})}\right|^2-(1-\eta^2)\left|T_{({\rm tree})}^{(l=0)}\right|^2\right).
 \label{eq:schmid_cc_proj}
\end{align}
Then, there is no correction from the triangle loop up to the phase $\exp(2i\delta_{23})$ of $T_{({\rm tree})}^{(l\neq 0)}$ 
when the $2+3$ rescattering is completely elastic, \ie, $\eta=1$.
This is the original statement of the Schmid theorem in Ref.~\cite{Schmid:1967ojm}.
On the other hand, the effect of inelasticity $\eta\neq 1$ shows up with a strength of $1-\eta^2$~\cite{Szczepaniak:2015hya,Debastiani:2018xoi}.

Here, we note that the $M_{23}$ invariant mass distribution is given by the tree-level contribution, 
\begin{align}
 &\frac{d\Gamma_{A\to B23}}{dM_{23}}\propto\int_{-1}^1d\cos\theta \left|T_{({\rm tree})}\right|^2\\
 \sim&\frac{1}{\epsilon}\frac{(4kq)^{-1}}{\tilde{E}_A-\omega_2(q)}\left[\arctan\left(\frac{\tilde{E}_A-\omega_1(k-q)-\omega_2(q)}{\epsilon}\right)-\arctan\left(\frac{\tilde{E}_A-\omega_1(k+q)-\omega_2(q)}{\epsilon}\right)\right].
\end{align}
The terms in the square brackets give a finite value when the mass of particle~$1$ overlaps with the phase space of the $\rmA\rightarrow \rmB +2+3$ decay (from point $P$ to point $Q$ in Fig.~\ref{fig:dalitz}).
Although the coefficient contains a factor $1/\epsilon$ diverging more strongly than the logarithmic one in Eq.~\eqref{eq_tree}, the whole amplitude is in fact finite once the coupling of particle 1 to particles 3 and B is taken into account.\footnote{Evaluating the coupling $g_{1,3B}$ from the width of particle $1$ as $g_{1,3B}=\sqrt{8\pi m_1^2\Gamma_{1\to 3B}/p_{3}}$, the $1/\epsilon$ dependence is canceled by $\epsilon\propto \Gamma_{1\to 3B}$ from $g^2_{1,3B}$, then $d\Gamma_{A\to B23}/dM_{23}$ becomes finite.}
On the other hand, the Dalitz plot distribution is given by 
\begin{align}
 \frac{d^2\Gamma_{A\to B23}}{dM_{3B}^2dM_{23}^2}=&\;
 \frac{1}{(2\pi)^3}\frac{1}{32m_A^3} \left|T_{({\rm loop})}+T_{({\rm tree})}^{(l=0)}+T_{({\rm tree})}^{(l\neq 0)}\right|^2\notag\\
 \sim&\;\frac{1}{(2\pi)^3}\frac{1}{32m_A^3} \left|\eta e^{2i\delta_{23}}T_{({\rm tree})}^{(l=0)}+T_{({\rm tree})}^{(l\neq 0)} \right|^2.\label{eq:schmid1}
\end{align}
The interference term $2\re\left[\eta e^{2i\delta_{23}}T_{({\rm tree})}^{(l=0)}T_{({\rm tree})}^{(l\neq 0)*}\right]$ remains
in the absence of the $M_{3B}$, or equivalently the $\cos\theta$, integration.
Then, visible effects of the TS can be seen in the Dalitz plot even if $\eta=1$.
The amplitudes $T_{({\rm loop})}$ and $T_{({\rm tree})}^{(l=0)}$ are functions of $M_{23}$ and they are independent of $M_{3B}$, 
then the signal of the TS would be seen as a uniform band of $M_{23}$ in the Dalitz plot~\cite{Szczepaniak:2015hya}.
A schematic picture of the Dalitz plot in the presence of a TS is shown in Fig.~\ref{fig:dalitz}.
It is found in Ref.~\cite{Szczepaniak:2015hya} that even in the case with a channel coupling $2+3\to {2'}+{3'}$ in addition to the elastic channel,
the sum of the events of the Dalitz-plot projections (the sum of the events with the $2+3$ and ${2'}+{3'}$ pairs in the final state) does not exhibit the TS effect as a consequence of the probability conservation while the effect should be visible in each individual channel.
In Eq.~\eqref{eq:schmid_cc_proj}, the coefficient of $\left|T_{\rm (tree)}^{(l=0)}\right|^2$, $1-\eta^2$, coincides with $\left|S_{2+3\to 2'+3'}\right|^2$, the off-diagonal element of the $S$-matrix in the coupled-channel case.

In the derivation of the Schmid theorem, the energy region is restricted to that near the TS energy to ignore the terms that are regular and finite compared with the logarithmically singular terms.\footnote{As pointed out in Ref.~\cite{Debastiani:2018xoi}, the Schmid theorem does not matter even in the zero-width limit of the internal particles because the tree amplitude diverges as $1/\epsilon$ while the triangle amplitude does as $\log (\epsilon)$ when the coupling $g_{1,3B}$ is not considered:
the triangle-loop amplitude just gives a vanishing contribution in the zero-width limit compared with the tree amplitude. See also the discussion in Ref.~\cite{Goebel:1982yb}.}
However, due to the inevitable width of the internal particle (see the discussion in Sec.~\ref{sec:kinvars}), the singularity is smeared and gives just a finite contribution.
In such a case, the original Schmid theorem is not applicable, and all the terms that emerge from the tree-level and triangle loop diagrams should be taken into account in practice.

\begin{figure}[tb]
    \centering
    \includegraphics[height=5.1cm]{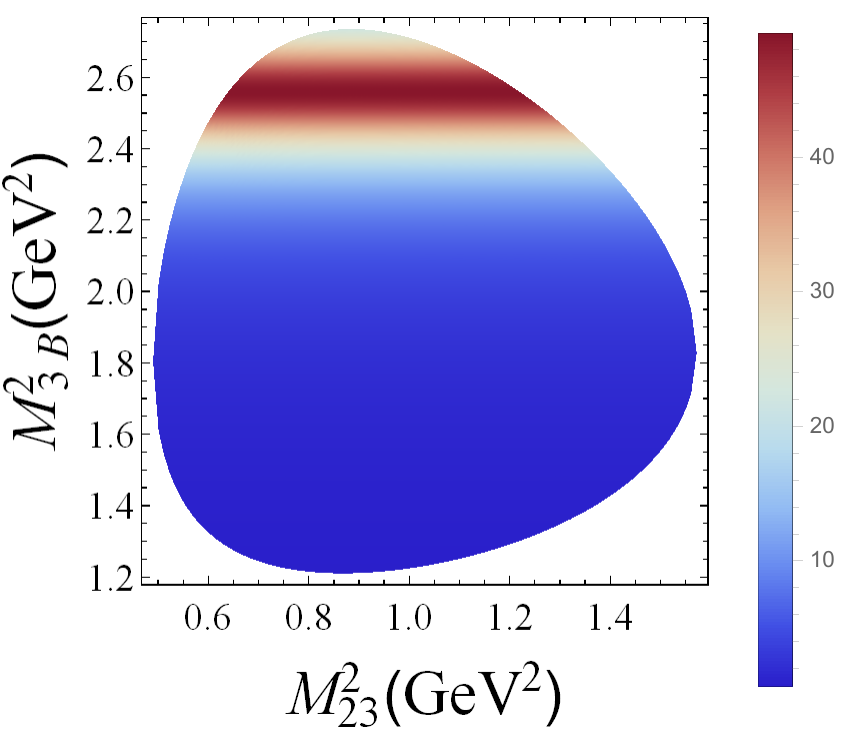}\hfill
    \includegraphics[height=5.1cm]{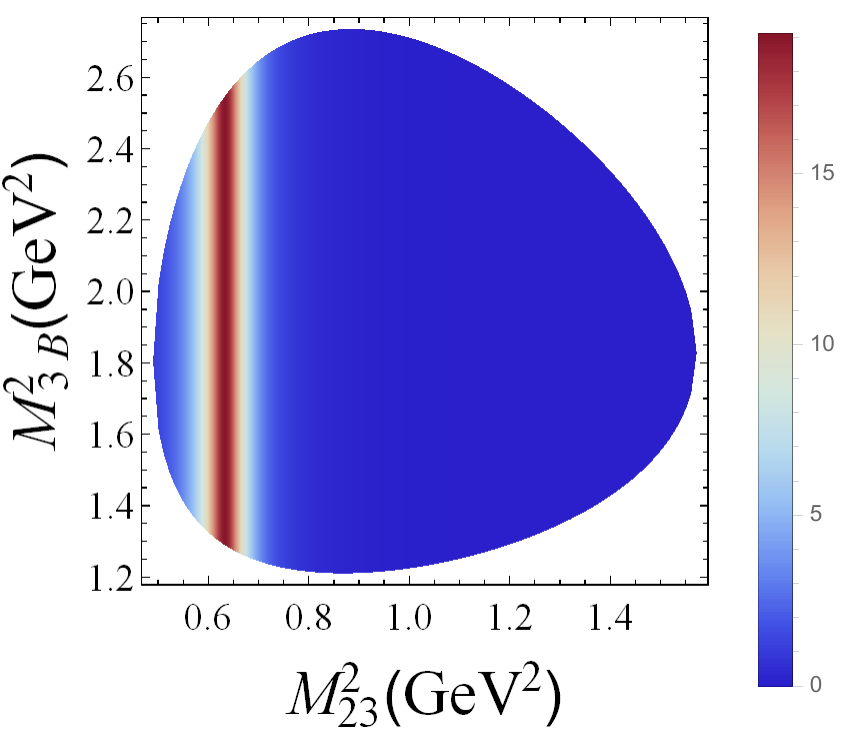}\hfill
    \includegraphics[height=5.1cm]{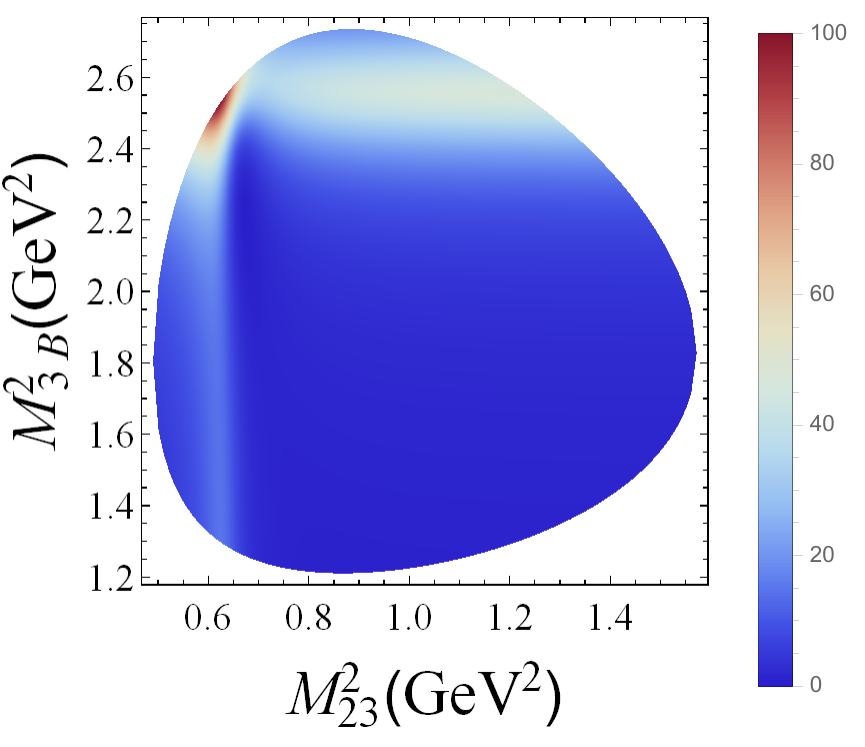}
   \begin{minipage}[t]{16.5 cm}
    \caption{The Dalitz plot in the $M_{23}^2$--$M_{3B}^2$ plane with the tree-level (left panel), triangle loop (middle panel), and tree+loop amplitudes (right panel).
    Since only the relative weight is important for Dalitz plots, we multiply each distribution by a factor of $10^5$ to make the values in the color bars not too small.
    }
    \label{fig:Dalitzftl}
   \end{minipage} 
\end{figure}

\begin{figure}[tb]
    \centering
    \includegraphics[width=7.5cm]{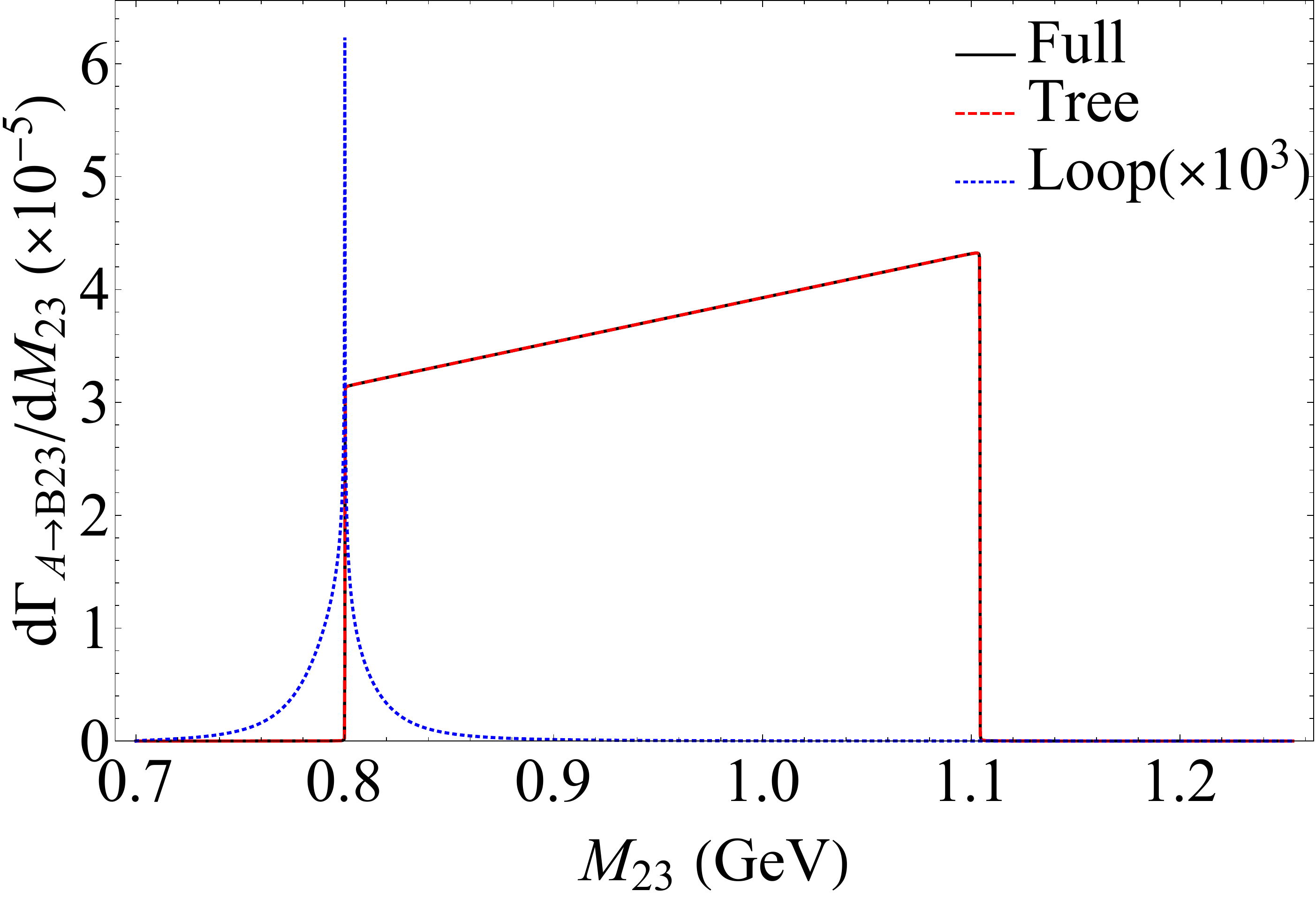}
    \hspace{1cm}
    \includegraphics[width=7.5cm]{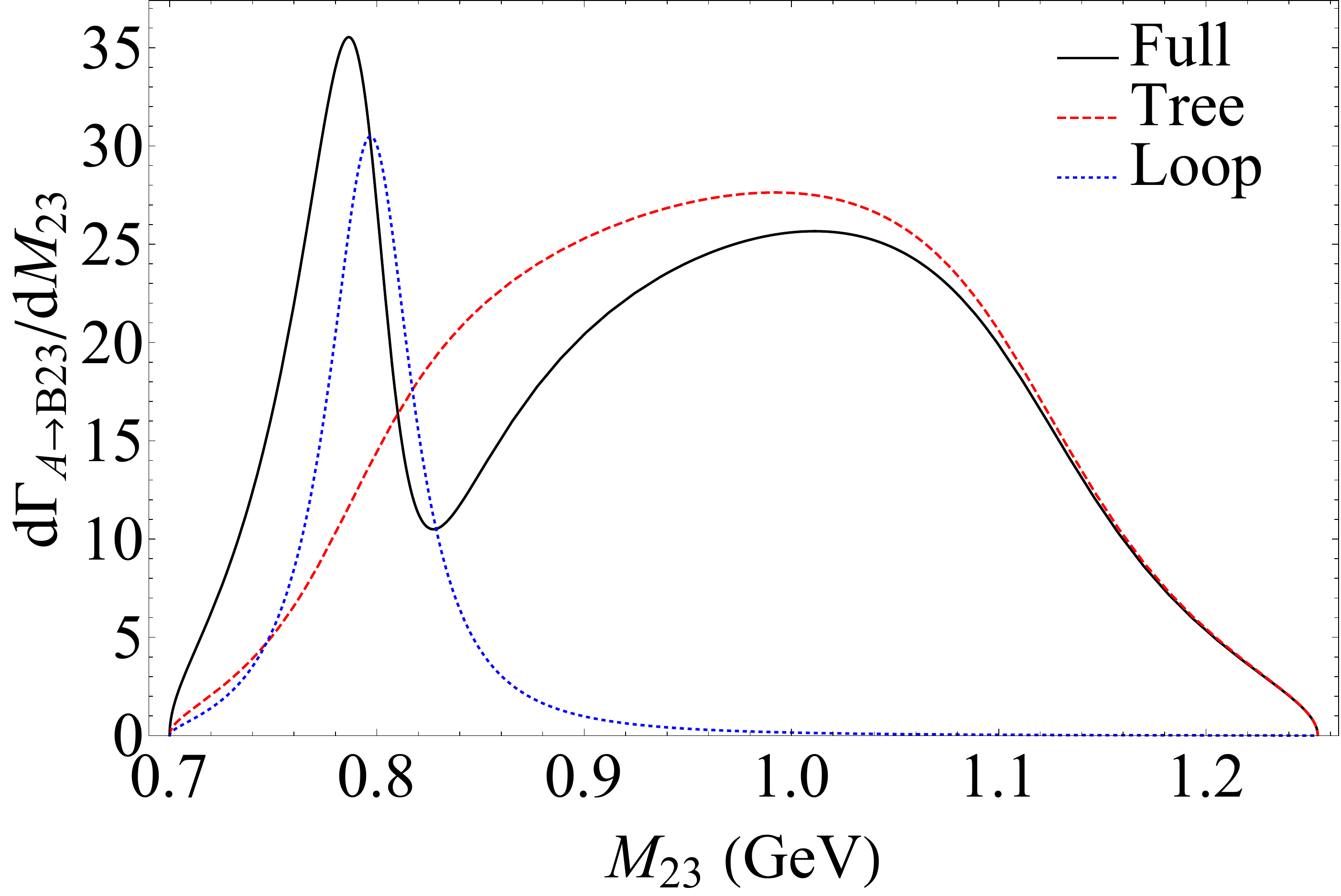}
    \begin{minipage}[t]{16.5 cm}
    \caption{The projection of the Dalitz plot on the invariant mass $M_{23}$ with a small width of particle $1$, $\Gamma_1=10^{-5}$~GeV (left panel), and with a finite width (right panel).
    In the plot of the left panel, the loop contribution is multiplied by a factor of $10^3$ for comparison. 
    }
    \label{fig:Dalitzprjwithoutwithwidth}
    \end{minipage}
\end{figure}

Here, as an example of the Dalitz plot and its projection, the case with $(m_1,m_2,m_3)=(1.6,0.5,0.2)$ and $(m_A,m_B,m_C)=(2.154,0.9,M_{23})$ (all units are in GeV) is considered.
In this case, a singularity appears at $0.8$~GeV in the $M_{23}$ distribution (see also Ref.~\cite{Debastiani:2018xoi}).
The width of particle $1$ is taken into account by replacing $m_1^2$ with $m_1^2-im_1\Gamma_1$ ($\Gamma_1=0.05$~GeV).
For the $2+3\to 2+3$ rescattering part, a Breit--Wigner amplitude with an energy-dependent width is used,
\begin{equation}
  T_{23}=\frac{g_{23}^2}{m_R^2-M_{23}^2-iM_{23}\Gamma_R(M_{23})},
\end{equation}
with $\Gamma_R(M_{23})=g_{23}^2q_{\rm on}/(8\pi M_{23}^2)$.
Here, $m_R=0.8$~GeV is taken to coincide with the TS energy, and the coupling constant $g_{23}$ is fixed with $\Gamma_R(m_R)=0.05$~GeV.
Now, not only the $S$-wave part, the full tree-level amplitude is included.
In Fig.~\ref{fig:Dalitzftl}, the Dalitz plot with the tree-level only, the loop only, and the sum of the  tree-level and loop amplitudes are shown from left to right in order, and  Fig.~\ref{fig:Dalitzprjwithoutwithwidth} shows the effect of a TS on the Dalitz-plot projection onto the $M_{23}$ distribution with a finite width of the internal particle~1.

In the left (middle) panel in Fig.~\ref{fig:Dalitzftl}, a horizontal (vertical) band of the tree-level (triangle-loop with a TS) contribution can be seen, and in the right panel the sum of the tree-level and loop contributions exhibits visibly two bands.
In the left panel in Fig.~\ref{fig:Dalitzprjwithoutwithwidth} for the Dalitz-plot projection to $M_{23}$ with a tiny $\Gamma_1$, the curves of the tree-level amplitude and the sum of the tree-level and the triangle loop amplitudes completely overlap with each other, and the effect of the triangle loop is invisible in this figure.
Note that the contribution from the triangle loop (the blue dotted line in the left panel in Fig.~\ref{fig:Dalitzprjwithoutwithwidth}) is multiplied by a factor of $10^3$ for comparison with that from the tree-level amplitude.
On the other hand, the inclusion of a finite width for the internal particle $1$ changes the situation drastically:
the tree-level and triangle loop contributions can be of the same order, and the effect of the rescattering can be visible in the $M_{23}$ distribution as shown in the right panel in Fig.~\ref{fig:Dalitzprjwithoutwithwidth}.

In addition to the width effect of the internal particles and the complexity of the $2+3$ rescattering, as pointed out in Ref.~\cite{Aitchison:1969tq},
additional production mechanisms, for example, particle B is directly produced from particle A, give another source of the violation of the Schmid theorem.
Furthermore, some specific features of the final-state rescattering amplitude can also provide corrections to the Schmid theorem as discussed in Ref.~\cite{Anisovich:1995ab} with examples.

Finally, we recall the Watson theorem for the final-state interaction~\cite{Watson:1952ji}, which states that the phase of the single-channel final-state interaction amplitude coincides with the scattering phase shift of the final-state particles. Therein, the phase factor is $\exp(i\delta_{23})$ instead of $\exp(2i\delta_{23})$ in the TS case from the Schmid theorem.
As pointed out in Ref.~\cite{Aitchison:1969tq}, the Watson theorem holds only when the primary production amplitude is real.
In the case of the triangle diagram, the corresponding production part is the tree-level amplitude from a resonance-exchange, which is complex and energy dependent.

\subsection{How to distinguish kinematic singularities and resonances}
\label{sec:distinguish}

Given that many of the newly observed candidates for exotic hadrons are beyond the expectation from the conventional quark model, and that the experimental signals for some of them may contain important contributions from, or even may be explained just by, kinematic singularities, it becomes crucial to distinguish kinematic singularities from resonances. 

Resonances are poles of the $S$ matrix. Therefore, their locations are due to the underlying strong interaction dynamics and fixed independently of the processes and channels. However, because of the different couplings to various channels and the interference with backgrounds, their line shapes may vary dramatically from case to case as already mentioned in the Introduction.

Kinematical Landau singularities because of the on-shell intermediate particles are normally branch points of the $S$ matrix.\footnote{One exception is given by the leading Landau singularity of the pentagon loop, which is a pole~\cite{Gribov:2009}, see Section~\ref{sec:character}.}
Therefore, their locations are not fixed by the dynamics but depend on the masses of the involved particles. As already discussed, the location of a threshold cusp is fixed, but that of a TS is determined by the masses of intermediate particles and the invariant masses of the external ones. Yet,  the strength of the singularities, \ie, how singular they behave in Dalitz plot distributions, is dictated by the dynamics. It is normally difficult to judge whether an invariant mass distribution peak is due to a resonance or due to a TS (if there is one in or close to the physical region). Nevertheless, the following key features of TSs  may be used to distinguish them from genuine resonances:
\begin{itemize}
    \item Strong sensitivity to kinematic variables: The dependence of the TS location on kinematic variables means that the singularity induced peak position as well as the shape of the peak changes along the variation of the external energies. For the study of a TS signal in the final state (in the distribution of $m_C$), one may vary $m_A$, \ie, the energy of the initial particles (\eg, the c.m. energy of the $e^+e^-$ pair); then the TS induced peak is sharper when $m_A$ is within the range given in Eq.~\eqref{eq:mArange}, and less sharp otherwise. 
    However, there should always be a cusp at the $m_2+m_3$ threshold in the distribution of $m_C$.\footnote{Notice that $C$ is normally more than one particle. Even if $C$ is a resonance formed in the $2+3$ rescattering, it must be able to decay into lighter final state particles. } One example is given in Section~\ref{sec:psi2Spipi}. For the study of a TS signal in the initial state energy $(m_A)$ distribution, one may divide the events into different bins of $m_C$. The peak in $m_A$ distribution should vary for $m_C$ in different bins, and it is more prominent when $m_C$ is within the range given in Eq.~\eqref{eq:m23range}.
    The dependence of the TS induced peaks on the initial state energy $m_A$ and on $m_C$ can be seen from plots (c) and (d) in Fig.~\ref{fig:kstarkk}, respectively.
    \item Quantum numbers: The existence of a TS is determined by the scalar triangle loop integral, and thus does not depend on the orbital angular momentum for each vertex. However, for the TS effects in the invariant mass distribution of a pair of internal particles, the logarithmic singularity in the amplitude gets multiplied by a c.m. momentum (to a positive power) factor if the particle pair is in a partial wave other than the $S$-wave. 
    This factor weakens the singular behavior and makes the peak much less sharp than the $S$-wave case, as can be seen from the explicit calculations in Ref.~\cite{Bayar:2016ftu} for the $P_c$. Consequently, the quantum numbers for a sharp TS peak are constrained to those of the $S$-wave pair of the corresponding internal particles.
    \item Schmid theorem for processes with interfering tree-level and triangle diagrams: As discussed above, for processes with both triangle and tree-level diagrams, see Fig.~\ref{fig:triangle_mod}, there is a subtle interference between the triangle and tree-level contributions. As a result, if the widths of all the internal particles are small such that the triangle diagram in the vicinity of the singular region is dominated by the TS term, the sum of TS and the tree-level diagram is just the tree-level diagram multiplied by a phase factor (for restrictions in the application of the Schmid theorem, see Section~\ref{sec:Schmid}).
    For such a case, there would be no TS induced peak in the projected invariant mass distributions though the interference may leave an imprint in the Dalitz plot, see Fig.~\ref{fig:Dalitzftl}. Thus, if a structure is presumed to receive important TS contribution, it would be valuable to measure the process involving elastic rescattering between the internal particles (2 and 3 in Fig.~\ref{fig:triangle_mod}).
\end{itemize}

Apart from these, it is also important to search for the resonant structures in processes free of the pertinent TSs. In this regard, reactions like the photoproduction and pion-induced production processes are as indispensable as the $e^+e^-$ and proton-(anti)proton collisions.

Next, we want to elaborate more on the first itemized point in the above.
In terms of Eq.~\eqref{eq:m23range}, one can define a quantity to reflect the size of the kinematic region where the TS can appear on the physical boundary (see the diagram in Fig.~\ref{fig:triangle} for the particle labels), \ie,
\begin{eqnarray}
	\Delta m_C \equiv \sqrt{\frac{m^{}_A m_3^2 - m_{B}^2
		m^{}_2}{m^{}_A-m^{}_2} + m^{}_A m^{}_2} - (m_2+m_3),
\end{eqnarray}
which is just the difference between the two bounds in of $m_C$ given in Eq.~\eqref{eq:m23range}, and describes the difference between the TS position and the normal threshold $m_2+m_3$.
With the invariant mass $m_B$ and the three internal masses fixed, when $m_A$ is equal to $m_1+m_2$, $\Delta m_C$ takes the maximal value
\begin{eqnarray}\label{eq:mCmax}
\Delta m_C^{\mathtt{max}} \approx \frac{m_2}{2 m_1(m_2+m_3)} \left[(m_1-m_3)^2-m_B^2\right].
\end{eqnarray} 
The above approximation is valid when $m_2[(m_1-m_3)^2-m_B^2] \ll m_1(m_2+m_3)^2$, which can usually be satisfied in the realistic rescattering processes. From Eq.~\eqref{eq:mCmax}, one can see that the $\Delta m_C^{\mathtt{max}}$ depends on two factors, the internal mass ratio ${m_2}/[{2 m_1(m_2+m_3)}]$ and the quantity $[(m_1-m_3)^2-m_B^2]$ which is correlated to the phase space of particle $1$ decaying into particles $3$ and B.

Similarly, one can define another kinematic variable internal $\Delta m_A$ for the TS in $m_A$. When $m_C=m_2+m_3$, the internal $\Delta m_A$ is maximal,
\begin{eqnarray}\label{eq:mAmax}
\Delta m_A^{\mathtt{max}} \approx \frac{m_2}{2 m_3(m_1+m_2)} \left[(m_1-m_3)^2-m_B^2\right].
\end{eqnarray}

\begin{figure}[tb]
	\centering
	\includegraphics[width=16.5cm]{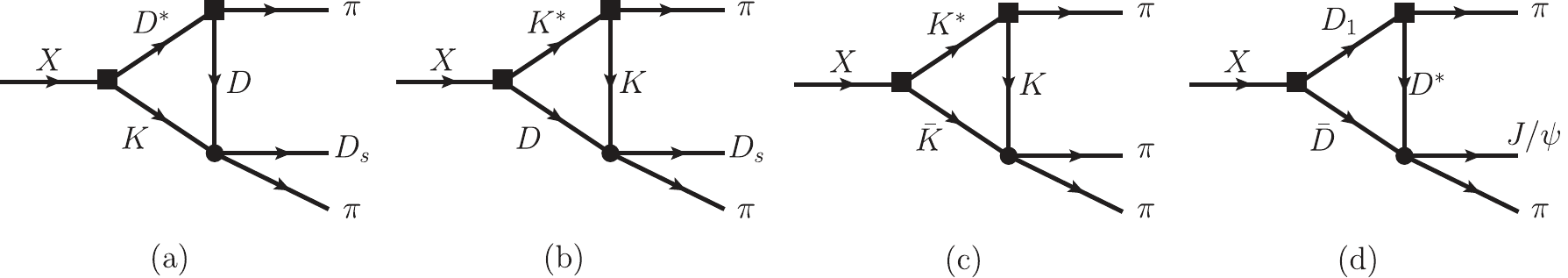}
  \begin{minipage}[t]{16.5 cm}
  \caption{Inelastic rescattering processes in 3-body decays via triangle diagrams. $X$ denotes some specified initial state with the proper quantum numbers. The upper pion lines correspond to particle B, and the lower external lines in the final states correspond to C in Fig.~\ref{fig:triangle}.
	}
  \label{fig:TSrange}
  \end{minipage}
\end{figure}

Larger $\Delta m_A^{\mathtt{max}}$ and $\Delta m_C^{\mathtt{max}}$ indicate larger kinematic regions where the TS can emerge, which implies that it would be easier to detect observable effects induced by the TS in experiments. Notice that as long as the invariant mass $m_B$ and the internal masses $m_{1,2,3}$ are fixed, $\Delta m_A^{\mathtt{max}}$ and $\Delta m_C^{\mathtt{max}}$ are determined. Some typical triangle diagrams are shown in Fig.~\ref{fig:TSrange}, and the corresponding $\Delta m_A^{\mathtt{max}}$ and $\Delta m_C^{\mathtt{max}}$ of these diagrams are listed in Table~\ref{tab:TSrange}. From Table~\ref{tab:TSrange}, one can see that $\Delta m_A^{\mathtt{max}}$ and $\Delta m_C^{\mathtt{max}}$ of Fig.~\ref{fig:TSrange}~(a) are quite small. This is because the $D^*$ mass  is very close to the $D\pi$ threshold, leading to tiny values for $[(m_1-m_3)^2-m_B^2]$ in Eq.~(\ref{eq:mCmax}) and Eq.~(\ref{eq:mAmax}). 
On the contrary, the phase space for $K^*$ decaying into $K\pi$ or $D_1$ decaying into $D^*\pi$ is much larger, therefore the $\Delta m_A^{\mathtt{max}}$ and $\Delta m_C^{\mathtt{max}}$ corresponding to Figs.~\ref{fig:TSrange}~(b), (c) and (d) are sizable. Among the four diagrams in Fig.~\ref{fig:TSrange}, (b) has the largest $\Delta m_A^{\mathtt{max}}$ and $\Delta m_C^{\mathtt{max}}$ because, besides the larger phase space factor, the internal mass ratios $m_2/m_3$ ($=m_D/m_{K}$) and $m_2/m_1$ ($=m_D/m_{K^*}$) are also relatively large. 

Larger $\Delta m_A^{\mathtt{max}}$ or $\Delta m_C^{\mathtt{max}}$ also implies the possibility of observing the movement of the TS peak in experiments by varying the kinematic configurations of the relevant rescattering process, which can serve as a criterion to distinguish the kinematic singularity from a genuine state as discussed above. 
For instance, it should be important to investigate the dependence of the $a_1(1420)$ peak~\cite{Adolph:2015pws} on the $\pi^+\pi^-$ invariant mass in the final state, which can be done by slicing the data into different bins of $m_{\pi^+\pi^-}$. 
A similar proposal was suggested in Ref.~\cite{Jing:2019cbw} to check whether the band around $m_{\phi\pi^0}=1.4$~GeV in the Dalitz plot for the $J/\psi\to \pi^0\phi \eta$ was due to a TS or a resonance~\cite{Ablikim:2018pik}.

One further example may be provided by the decay of $\bar B\to \bar K{K}^*{D}^{(*)}\to \bar K D_s^{(*)}\pi\pi$. This process has access to the ${K}^*{D}^{(*)}$ rescattering into $D_s^{(*)}\pi\pi$, which may be visualized by replacing $X$ in Fig.~\ref{fig:TSrange}~(b) by an incoming $\bar B$ and an outgoing $\bar K$. This triangle diagram has sizable $\Delta m_A^{\mathtt{max}}$ and $\Delta m_C^{\mathtt{max}}$. For this four-body decay process, one may observe the movement of a peak in the $D_s^{(*)}\pi$ distribution by varying the invariant mass $m_A$ =$\sqrt{(p_B-p_K)^2}$ in a certain range~\cite{Liu:2015taa}. 

\begin{table}
	\caption{Kinematic intervals $\Delta m_A^{\mathtt{max}}$ and $\Delta m_C^{\mathtt{max}}$ for the corresponding Feynman diagrams in Fig.~\ref{fig:TSrange}. All values are given in units of MeV. }\label{tab:TSrange}
	\begin{center}
		\begin{tabular}{l c c c c}
			\hline\hline  & Fig.~\ref{fig:TSrange}(a) & Fig.~\ref{fig:TSrange}(b) & Fig.~\ref{fig:TSrange}(c) & Fig.~\ref{fig:TSrange}(d) \\
			\hline $\Delta m_A^{\mathtt{max}}$ & 0.089 & 96 & 49  & 16 \\
			\hline $\Delta m_C^{\mathtt{max}}$ & 0.087  & 62  & 38 & 15  \\
			\hline\hline
		\end{tabular}
	\end{center}
\end{table}

\subsection{Amplitude analysis considering triangle singularities}

When peaks in the energy distributions can be imitated by the TS, the amplitude analysis considering the TS effects would be indispensable and necessary in order to extract reliable resonance parameters.
To the best of our knowledge, the pioneering work in the amplitude analysis with the TS effects taken into account is Ref.~\cite{Anisovich:1991qk}. In this work, the authors analyzed the data for the $p\bar p\to \pi^0\pi^0\pi^0$ and $\eta\eta\pi^0$ from the Crystal Barrel Collaboration~\cite{Aker:1991bk,Amsler:1992rx} considering both two- and three-body final state interactions in the dispersive framework. Although the extracted resonance parameters for the $f_0(980)$, $f_0(1335)$, $f_0(1505)$ and $f_2(1270)$ did not change much using parametrizations with and without considering the TSs, it was found that the production of the low-lying meson resonances, in particular the $f_0(980)$, was enhanced by the TSs.

Recently, some studies were devoted in this direction in relation with exotic hadrons. The TS effects may be tackled with by using an amplitude including the relevant triangle Feynman diagrams which can be computed directly in the conventional method (in the nonrelativistic case,  the expression for the triangle loop integral is given in Appendix~\ref{appendix}) or using the LoopTools library~\cite{Hahn:1998yk}. This is the approach employed in Ref.~\cite{Albaladejo:2015lob} which analyzed the BESIII data for the $Z_c(3900)$ by treating the charmed meson loops nonrelativistically. 
A dispersive formalism for amplitude analysis taking into account TSs with all $S$-wave couplings and spinless particles is discussed in Ref.~\cite{Szczepaniak:2015hya}, which has been applied in another analysis of the data for the $Z_c(3900)$ in Ref.~\cite{Pilloni:2016obd}. 
A coupled-channel formalism with the anomalous threshold included was proposed recently in Ref.~\cite{Lutz:2018kaz}.
For detailed discussions of the tricky evaluation of triangle diagrams in the dispersive formalism, we refer to Refs.~\cite{Lucha:2006vc,Melikhov:2019sob}.

To focus on seeing how the TS is implemented in the formulation in Refs.~\cite{Szczepaniak:2015hya,Pilloni:2016obd},
a simplified setup is considered here:
the $2+3\to 2+3$ amplitude is dominated by the $S$ wave and the left-hand cut of the amplitude is not considered.
We consider the amplitude projected to the $S$ wave, and the indices for the partial-wave decomposition are suppressed in the following.

\begin{figure}
    \centering
    \includegraphics[width=6cm]{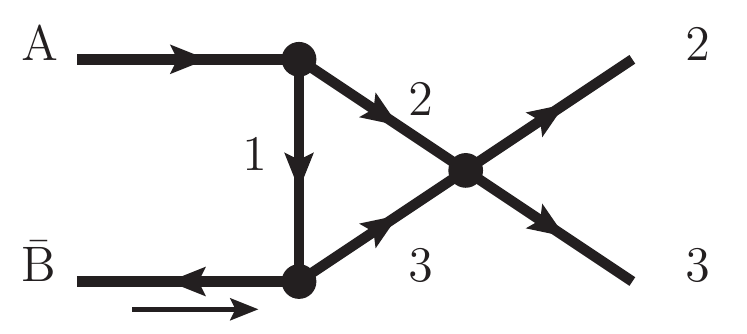}
    \caption{${\rmA}+\bar\rmB\to 2+3$ scattering via a triangle loop diagram.}
    \label{fig:diagCS}
\end{figure}

\begin{figure}[tb]
    \centering
    \includegraphics[width=9cm]{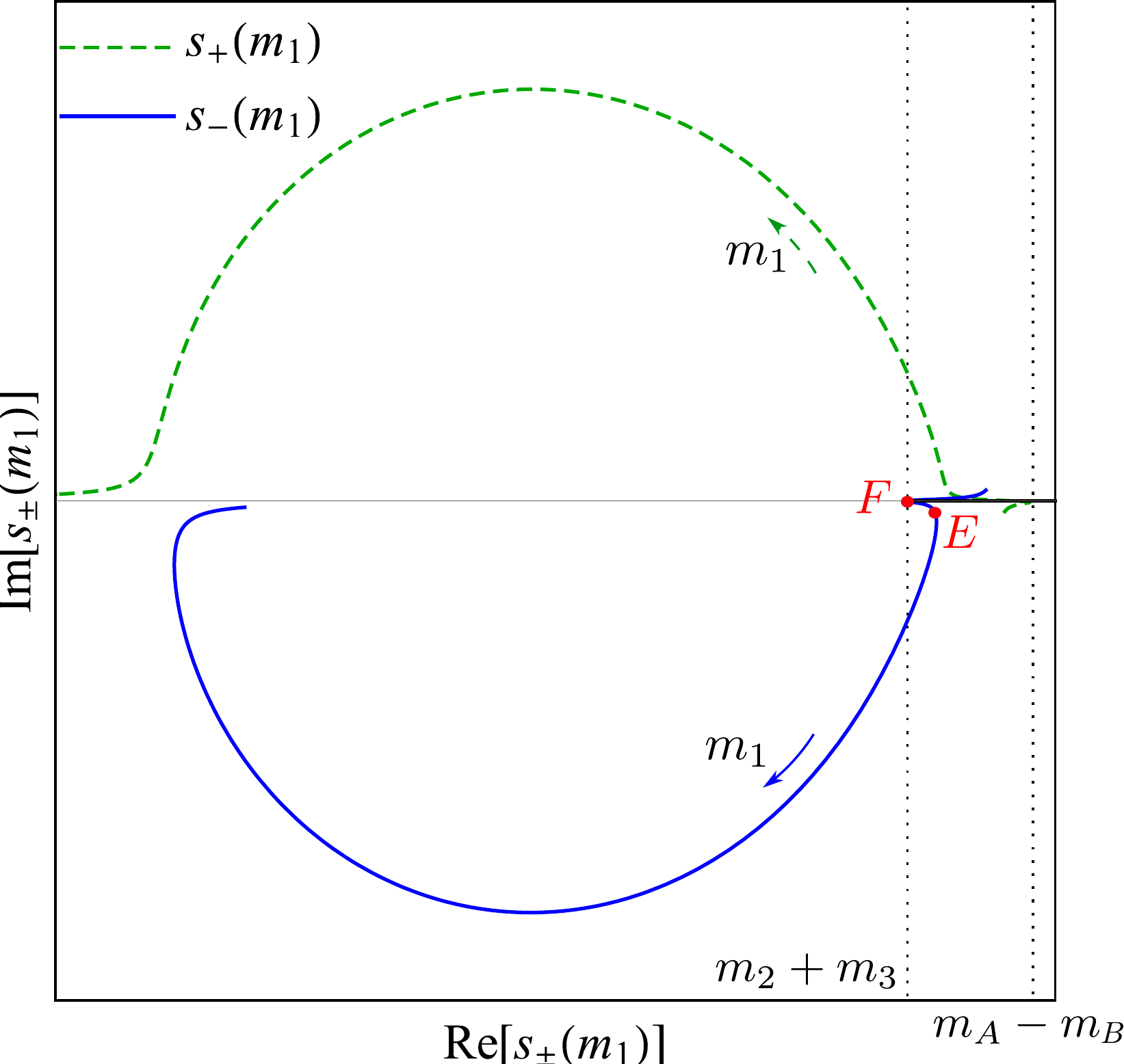}
    \caption{The motion of the branch point as a function of the particle-1 mass, $m_1$ with a small imaginary part being introduced for the plot.
    $m_1$ increases in the direction of the arrows.
    The point $F$ corresponds to the $2+3$ threshold.
    The solid (blue) curve $s_-$  crosses the real axis detouring around the point $F$ to avoid the integration contour, $(s_{\rm th},+\infty)$, and it is always on the second Riemann sheet with respect to this cut. Point $E$ corresponds to the upper bound of the range in Eq.~\eqref{eq:m23range}. The dashed (green) curve is the trajectory of $s_+$. When $m_1$ decreases, it crosses the unitary cut from the second Riemann sheet at $m_A-m_B$ to the lower half plane of the first Riemann sheet. The two dotted vertical lines label the threshold ($m_2+m_3$) and pseudo-threshold ($m_A-m_B$).
    } 
    \label{fig:bpmotion}
\end{figure}

Moving particle B in the final state to the initial state and relabelling it as $\bar\rmB$, we consider the two-body scattering of $\rmA+\bar\rmB\to 2+3$ shown in Fig.~\ref{fig:diagCS} [here $s=(p_2+p_3)^2$ and $t=(p_A-p_2)^2$].
Because all the particles in the loop should be on shell at the TS as we have learned in Sec.~\ref{sec:coleman-norton}, the singularity from the $t$-channel particle exchange in the ${\rm A}+\bar{\rm B}\to 2+3$ scattering process needs to be taken into account in addition to the unitary cut in $s$-channel.
With the discontinuity of the $2+3$ cut, the dispersion integral gives an expression of the $S$-wave ${\rm A}+\bar{\rm B}\to 2+3$ amplitude, $\mathcal{M}_s$, with a manifest $2+3\to 2+3$ elastic cut~\cite{Aitchison:1969tq,Szczepaniak:2015eza}: 
\begin{align}
\mathcal{M}_s(s)=\int_{\sth}^\infty\frac{ds'}{\pi} \frac{C(s')\rho_{23}(s') T_{23}^*(s')}{s'-s - i\epsilon},\label{eq:ABbto23pre}
\end{align}
with $\sth=(m_2+m_3)^2$ and $\rho_{23}(s)$ the two-body phase-space factor for particles 2 and 3. Here, the $i\epsilon$ in the denominator is included to denote that the physical boundary is defined on the upper edge ($s+i\epsilon$) of the cut on the first Riemann sheet.
In this dispersion integral, the analytic property of $C(s)$ representing the primary $\rmA+\bar \rmB\to 2+3$ transition is important.
In the simple case in Fig.~\ref{fig:diagCS}, this $C(s)$ is given by the $S$-wave projection of the $t$-channel exchange of particle 1 which has logarithmic branch points:
\begin{align}
    C(s)=\frac{1}{4\tilde{p}_A\tilde{p}_2}\log\left[\frac{m_1^2-t(s,\cos\theta=-1)}{m_1^2-t(s,\cos\theta=+1)}\right],
    \label{eq:treeprojection}
\end{align}
with $t(s,\cos\theta)=m_A^2+m_2^2-2\tilde{E}_A\tilde{E}_2+2\tilde{p}_A\tilde{p}_2\cos\theta$.
In terms of the $s$ variable, they are given by
\begin{equation}
 s_\pm = m_A^2+m_B^2+\frac{-(m_A^2+m_1^2-m_2^2)(m_1^2+m_B^2-m_3^2)\pm\sqrt{\lambda(m_A^2,m_1^2,m_2^2)\lambda(m_1^2,m_B^2,m_3^2)}}{2m_1^2}.
\end{equation}
The motion of the branch points as a function of the mass of the exchanged particle, $m_1$, is shown in Fig.~\ref{fig:bpmotion} (see also Refs.~\cite{Szczepaniak:2015eza,Bronzan:1963mby}).
In the figure, $s_\pm$ corresponds to the branch point with $\cos\theta=\pm 1$ in Eq.~\eqref{eq:treeprojection}.
The part of the figure around the threshold $(m_2+m_3)$ and pseudo-threshold $(m_A-m_B)$ is the same as that given in the right panel in Fig.~\ref{fig:dalitz_traj}.
One sees a peculiar motion of the $s_-$ branch point near point $F$ corresponding to the $2+3$ threshold. When $m_1$ decreases,  $s_-$ circles around the threshold from the lower half plane to the upper half one without crossing the right-hand unitary cut. 
Between point $F$ and point $E$, which is the upper bound of the range in Eq.~\eqref{eq:m23range}, the $s_-$ branch point under the cut on the second Riemann sheet pinches the integration contour in Eq.~\eqref{eq:ABbto23pre} with $s+i\epsilon$ from the above, leading to the TS on the physical boundary. 
This subtle interplay of the left- and right-hand cuts is the origin of the singularity.
In addition to the $2+3\to 2+3$ rescattering, from the crossing symmetry, the other three-body final state interactions where the particles B rescatters with particle 2 or 3 may also need to be considered.
Further generalization of the scattering amplitude $\mathcal{M}$ with a dispersive integral gives~\cite{Szczepaniak:2015hya,Pilloni:2016obd}, 
\begin{align}
\mathcal{M}(s)=c(s)+T_{23}(s)\int_{\sth}^\infty\frac{ds'}{\pi}\frac{c(s')\rho_{23}(s') }{s'-s - i\epsilon},\label{eq:disp2}
\end{align}
where $c(s)$ comes from the projection of the $t$- and $u$-channel particle exchanges which contain the left-hand-cut information as $C(s)$ in Eq.~(\ref{eq:ABbto23pre}).
Thus, a general $\rmA\to \rmB +2+3$ decay amplitude that contains TS effects can be constructed by demanding crossing symmetry, analyticity and unitarity~\cite{Anisovich:1991qk,Szczepaniak:2015hya,Pilloni:2016obd}.
In practice, the two-body transition amplitude, \eg, $T_{23}$ here, needs to be parametrized. Depending on the specific system, it can be done by using the Omn{\`e}s dispersive representation (\eg, Refs.~\cite{Molnar:2019uos,Chen:2019mgp}), the isobar model, the $K$-matrix formalism (\eg, Ref.~\cite{Pilloni:2016obd}), or effective field theories (\eg, Ref.~\cite{Albaladejo:2015lob}).

Works to include the TS effects to the coupled-channel scattering can be found, \eg, in Refs.~\cite{Roca:2017bvy,Samart:2017scf} expecting that the triangle loop provides a characteristic energy dependence to the interaction kernel.
In these works, a particular attention was paid to the potential role of the TS in building up hadronic molecules. 
More recently, a coupled-channel formalism including the effects of the anomalous threshold was proposed in Ref.~\cite{Lutz:2018kaz}.
Therein, a generalized potential developed in Ref.~\cite{Lutz:2015lca}, which is basically constructed by partial-wave decomposition of the $t$- and $u$-channel exchange processes, is used.
Because the information of the left-hand cut is contained in the generalized potential and the right-hand cut is implemented by the dispersion integral with the generalized potential in the driving term, 
the use of the generalized potential makes it possible to take into account the anomalous-threshold effects in this formalism.
While, actually, this approach faces a problem that the Schwarz reflection principle cannot be satisfied in general, a minimal prescription to maintain the reflection principle and the coupled-channel unitarity is proposed in the paper.

\bigskip

%% file: section5.tex
\section{Triangle singularities in the initial state energy spectrum}
\label{sec:5}

In the following, we will review some processes among those listed in Tables~\ref{tab:list-meson} and \ref{tab:list-baryon} that have been proposed to have significant contributions from TSs.
Triangle singularities can be manifested as observable effects in both the initial ($A$ in Fig.~\ref{fig:triangle}) and final ($C$ in Fig.~\ref{fig:triangle}) states. From the above discussed sensitivity of the singular location on the kinematical variables, it is evident that the energies of the initial and final states are intertwined in this consideration. 
There can be a peak in both the initial and final state invariant mass distributions. In this section, we discuss the structures in the initial state energy spectrum with the final state invariant mass fixed to a specific region (or value), which is normally the mass of  a resonance in the vicinity of the threshold of two intermediate particles.

\subsection{Light mesons}
\label{sec:lightmesons}

The TS produced from the $K^*\bar K K$ loops has been pointed out in 1960s, which was used as an example for the modified Peierls mechanism in the paper by Schmid~\cite{Schmid:1967ojm}. Lots of discussions emerged in recent years relating such TS effects to the signals for meson resonance candidates around 1.4~GeV, including the $\eta(1405/1475)$, the $a_1(1420)$, and the $f_1(1420)$, with the first coupled to $K^*\bar K$ in the $P$ wave and the latter two in the $S$ wave. 
The works about the $\eta(1405/1475)$ and the $a_1(1420)$ will be summarized in the following subsections, and some other processes related to TSs in light-meson sector are summarized afterwards.

\subsubsection{\texorpdfstring{$\eta(1405/1475)$}{eta1405}}

The BESIII collaboration reported their measurement of the radiative decay $J/\psi\to\gamma \eta(1405/1475)$ in the exclusive decay channel $\eta(1405/1475)\to f_0(980)\pi\to  3\pi$ in 2012~\cite{BESIII:2012aa}. It was found that the isospin violation in this  $\eta(1405/1475)$ decay was anomalously large and could not be explained by the $a_0(980)$--$f_0(980)$ mixing. This abnormal phenomenon is explained in Ref.~\cite{Wu:2011yx} which proposes that the TSs play a crucial role to enhance the isospin-violating effects and the $\eta(1405/1475)$ signal around 1.4~GeV in the $3\pi$ energy spectrum could be due to TSs. The kinematical conditions for the presence of TSs in the physical region are well satisfied in the rescattering process, 
There are two kinds of triangle loops: the ones with charged $K^{*+}K^- K^+$ and $K^{*-}K^+ K^-$ and those with the neutral strange mesons. 
The difference between the charged and neutral loops contributes to this isospin breaking process. When the $\pi^+\pi^-$ invariant mass is outside the region between the $K^+K^-$ and $K^0\bar K^0$ thresholds, these two contributions almost cancel with each other exactly. While between the thresholds, they do not; and since the TS locations differ slightly, the isospin breaking receives a significant enhancement.
Moreover, as a result, the peak for the $f_0(980)$ receives contributions from both the $f_0(980)$ resonance and the TSs (see Fig.~\ref{fig:kstarkk}~(a) for the TS peak generated by the charged $K^*\bar KK$ loop). It is sandwiched between the charged and neutral $K\bar K$ thresholds and is much narrower than that in isospin symmetry preserving reactions.

A later detailed analysis suggests that the BESIII data for the $\eta(1405/1475)$ structure may contain a small contribution from the $f_1(1420)$ in the $3\pi$ decay channel which can be disentangled by the angular distributions of the neutral pion and the recoiled photon~\cite{Wu:2012pg}. Similar analysis can be found in Ref.~\cite{Aceti:2012dj} where the two-body coupled-channel ($\pi\pi$, $K\bar K$ and $\pi\eta$) final state interactions were considered using the chiral unitary approach. 

Another puzzle concerning the $\eta(1405/1475)$ is that the experimentally observed invariant mass spectra for the $\eta(1405/1475)\to a_0(980)\pi$ and the $\eta(1405/1475)\to K^*\bar{K}+c.c.$ are different. The PDG actually lists the $\eta(1405)$ and the $\eta(1475)$  as two individual states. However, considering that their masses are close to each other and their decay modes are similar, whether they can be identified as a single resonance is a long standing puzzle~\cite{Klempt:2007cp}. The TS mechanism suggests a solution that there is only one state.
Specifically, the decays of the resonance $\eta(1405/1475)$ into the $a_0(980)\pi$ or the $f_0(980)\pi$ can proceed via the $K^*\bar KK$ loops, whose TSs in the invariant mass of the initial state shift the peak position of the genuine resonance.

Although the TS mechanism seems to be promising for understanding the puzzles related to the $\eta(1405/1475)$, it was argued in Ref.~\cite{Achasov:2015uua} that the non-zero width of $K^*$ in the loop integral could lead to a significant suppression of the decay rate (see the comparison of the curves in Fig.~\ref{fig:kstarkk}~(a) assuming various $K^*$ widths), and therefore the dominance of triangle diagrams in the isospin violating channel might become questionable. 
The decay pattern of the $\eta(1405/1475)$ with the $K^*$ width effects included in computing the triangle diagrams was carefully reanalyzed in Ref.~\cite{Du:2019idk}. 
It was found that a self-consistent description of the $K\bar{K}\pi$, $\eta\pi\pi$ and $3\pi$ decay modes for the $\eta(1405/1475)$ could be provided in the TS mechanism with the $a_0(980)$--$f_0(980)$ mixing taken into account, leading to the conclusion that the TS mechanism still played a decisive role in understanding the $\eta(1405)$ and $\eta(1475)$ puzzles. 
For a detailed discussion on this issue and the relation between the $\eta(1405/1475)$ with pseudoscalar glueball candidates, we refer to Refs.~\cite{Du:2019idk,Qin:2017qes}.

\subsubsection{\texorpdfstring{$a_1(1420)$}{a11420}}

A further support  of the importance of TSs in the  $K^*\bar KK$ triangle diagrams comes from the observation of the $a_1(1420)$ by the COMPASS Collaboration~\cite{Adolph:2015pws} and the detailed analyses following that in Refs.~\cite{Ketzer:2015tqa,Mikhasenko:2016mox,Akhunzyanov:2018lqa,Ketzer:2019wmd,Mikhasenko:2019talk}.
The $a_1(1420)$ was reported in the $P$-wave (from the freed-isobar partial-wave analysis) $\pi f_0(980)$, $f_0(980)\to \pi\pi$ channel in the $\pi^-p\to \pi^+\pi^-\pi^-p$ reaction ($p_\pi=190$~GeV$/c$)~\cite{Adolph:2015pws,Akhunzyanov:2018lqa}.
Various models were proposed for the nature of the $a_1(1420)$. The authors of Ref.~\cite{Basdevant:2015wma} explain the observation using the Drell--Hiida--Deck mechanism~\cite{Drell:1961zza,Deck:1964hm} assisted with resonant final state interactions, which can lead to one $a_1$ resonance to peak at different energies in the $S$-wave $\rho\pi$ and $P$-wave $f_0(980)\pi$ final states. There are also works proposing the  $a_1(1420)$~\cite{Wang:2014bua,Chen:2015fwa,Gutsche:2017oro,Sundu:2017xct,Murakami:2018spb,Gutsche:2017twh,Nielsen:2018uyn} to be a tetraquark resonance.

The possible understanding of the $a_1(1420)$ peak as a TS peak was discussed in Ref.~\cite{Ketzer:2015tqa},\footnote{The connection of the $a_1(1420)$ to the TS effect was  first pointed out by Qiang Zhao at the HADRON 2013 Conference in Nara when the COMPASS data was first reported~\cite{Zhao:private}. }
followed by Ref.~\cite{Aceti:2016yeb}.
As already discussed,   the $K^*\bar{K}K$ triangle loops can produce a peak  at about $1.4$~GeV in the $\pi f_0(980)$ channel.
In Refs.~\cite{Ketzer:2015tqa,Aceti:2016yeb}, the $K^*\bar{K}$ pair is produced from the broad $a_1(1260)$ in the $S$ wave, and the triangle loop is formed with a $P$-wave $K^*$ decay into $\pi K$ and the $f_0(980)$ formation from the $K\bar{K}$ pair that finally give a $P$-wave $\pi f_0(980)$ in the final state.
The decay of $a_1^-(1260)\to \pi^-\rho^0\to\pi^-\pi^+\pi^-$, the dominant decay channel of $a_1(1260)$, was also calculated, and a good agreement  with the experimental result was found for the production strength relative to that via the $f_0(980)\pi^-$.
More information about the $a_1(1420)$ can be found in the recent review~\cite{Ketzer:2019wmd}.

To clarify the nature of $a_1(1420)$, two models for the $P$-wave $f_0(980)\pi$ production mechanism are compared~\cite{Mikhasenko:2016mox,Ketzer:2019wmd,Mikhasenko:2019talk}:
the production via a Breit--Wigner resonance with a mass of about 1.4~GeV for the $a_1(1420)$, or via $K^*\bar KK$ triangle diagrams with an $a_1(1260)$ resonance producing the intermediate $K^*\bar K$ pair.
From these analyses, the COMPASS data can be fitted with comparably good quality with these two models, and no significant difference could be claimed. 
In particular, the two models give very similar Argand plots after considering the interference with backgrounds.
Thus, further efforts are required to reveal the nature of the $a_1(1420)$ peak. One possibility would be to check the variation of the peak against varying the $\pi^+\pi^-$ invariant mass. That is, if the $a_1(1420)$ peak is dominantly due to TS effects, when the $\pi^+\pi^-$ invariant mass moves away from the TS region (from the $K\bar K$ threshold to about 1.03~GeV~\cite{Jing:2019cbw} computed using Eq.~\eqref{eq:m23range}), which contains the  $f_0(980)$ location, the peak should decrease with a pace following the expectation from the $K^*\bar KK$ triangle loops (after considering background contributions).

The study carried out in Ref.~\cite{Dai:2018rra} suggests a possible emergence of the $a_1(1420)$ peak in the $\pi^- f_0(980)$ distribution of the decay $\tau^-\to\nu_\tau\pi^-f_0(980)$, and the branching fraction was predicted.
Similarly, in Ref.~\cite{Pavao:2017kcr}, the TS effect of the $K^*\bar{K}K$ loops in the decay of $B\to \bar{D}^*\pi f_0(980)$ was studied.
In these works~\cite{Dai:2018rra,Pavao:2017kcr}, the production of the $\pi a_0(980)$ mode was also calculated.
The predictions in these works can be confronted with future experimental data to clarify the nature of the $a_1(1420)$ peak.

\subsubsection{Other reactions}
\label{sec:lighthadronsothers}

The $J/\psi\to\eta\pi^0\phi$ was studied in Ref.~\cite{Jing:2019cbw}.
In the mechanism considered in this work,
the decay process is driven by the isospin violation from the mass splittings between the charged and neutral internal $K$ and $K^*$ mesons, similar to the above discussed $\eta(1405/1475)\to f_0(980)\pi$.
The isospin violation is enhanced by the $K^*\bar{K}K$ TSs, which can also produce a peak around $1.4$~GeV in the $\phi\pi^0$ energy spectrum.
A band structure around $m_{\pi^0\phi}^2=1.9$~GeV$^2$ is visible in the Dalitz plot, in the $m_{\pi^0\phi}^2$--$m_{\eta\pi^0}^2$ plane, measured by the BESIII Collaboration~\cite{Ablikim:2018pik}.
The theoretically calculated Dalitz-plot distribution with the TS mechanism captures a feature found experimentally, \ie, the accumulation of events at the phase-space boundary of the TS band lying in the $\pi^0\phi$ direction. In the TS mechanism, that is because the outgoing particles in the data are all moving collinearly when the TS is on the physical boundary.
One promising method to check if the band structure comes from TS effects or not was proposed.
The band around $1.4$~GeV in the $\pi^0K^+K^-$ (the $\phi$ meson was reconstructed from $K^+ K^-$) invariant mass distribution should be largely reduced when the $K^+K^-$  invariant mass is far away from the $\phi$-meson mass region.
On the other hand,  in general the band should be independent of the invariant mass of $K^+K^-$ if it is from a resonance unless the resonance is of hadronic molecular type with one of the constituent being the $f_0(980)$ which also couples strongly to the $K\bar K$. 
The $S$-wave $\pi^0\phi$ channel has quantum numbers $J^{PC}=1^{+-}$. The $h_1(1380)$ or an isovector resonance at 1.5~GeV claimed in Ref.~\cite{Bityukov:1986yd} can couple to this channel, and a structure around $1.4$~GeV in the $K^*\bar{K}$ distribution was seen in some charmonia decays~\cite{Ablikim:2015lnn,Ablikim:2018ctf}. Further efforts from both theoretical and experimental sides are needed to clarify the nature of this Dalitz plot band.
Now, we move into a curious feature of the TS which becomes very clear by comparing the isospin-forbidden $J/\psi\to\eta\pi^0\phi$ and isospin-allowed $J/\psi\to\pi^0\pi^0\phi$ processes:
the TS peak in the $\pi^0\phi$ distribution in the $J/\psi\to\eta\pi^0\phi$ is stable against changing the subtraction constant for the regularization of the UV divergence.
This is because this $J/\psi\to\eta\pi^0\phi$ process is isospin forbidden, and mainly driven by the difference between the charged and neutral $K^*\bar KK$ loops, the TSs of which lead to an enhancement of isospin violation as discussed above.
As one can see, the scalar triangle loop is UV convergent, and the TS appears as a kinematic effect when the internal particles are on shell (meaning that the TS is an infrared effect), which should not be affected by the UV part.
Thus, a large cancellation of the non-singular part irrelevant to the TSs occurs due to the isospin symmetry (a similar cancellation of the UV divergent part due to the isospin symmetry was found in Ref.~\cite{Guo:2010ak}). 
However, in the $J/\psi\to\pi^0\pi^0\phi$ case, while the TS peak position in the $\pi^0\phi$ distribution, which is much broader than the isospin breaking case, does not depend on how the UV divergence in the loops is treated, a sizable modification could happen to the width of the peak when different subtractions for the UV divergence are used~\cite{Jing:2019cbw}.

Enhanced isospin violation by TSs was also proposed for  the $D_s^+\to \pi^+\pi^0a_0(980)$~$[f_0(980)]$ and $\bar{B}_s^0\to J/\psi \pi^0a_0(980)$~$[f_0(980)]$~\cite{Sakai:2017iqs,Liang:2017ijf}.
In the isospin-forbidden $\pi^0f_0(980)$ mode, a narrow $f_0(980)$ line shape, of several MeV, is obtained as a result of the cancellation between charge and neutral kaon loops as above.
Due to the emission of one additional particle ($\pi^+$ and $J/\psi$ in the $D_s^+$ and $\bar B_s^0$ decays, respectively), the $\pi^0f_0(980)$ invariant mass can be varied, and the magnitude of the isospin violation, the ratio of the differential width of  the isospin-forbidden $\pi^0f_0(980)$ to that of the isospin-allowed $\pi^0a_0(980)$ modes, also vary depending on the $\pi^0f_0/a_0$ invariant mass.
By checking the energy dependence of the $\pi^0f_0(980)$ and $\pi^0a_0(980)$ productions, the role of the TSs in the isospin-violating processes would become more clear.

The amplitude of the $K^*\bar{K}^*K$ loop has a singularity at $1810$~MeV in the invariant mass of the $K^*\bar K^*$.
In Ref.~\cite{Xie:2016lvs}, by considering the production of $K^*\bar{K}^*$ from the $f_2(1640)$ and the formation of the $a_1(1260)$ resonance from  the intermediate $K^*\bar{K}$ and its subsequent decay into $\pi\rho$, an unavoidable peak at $1.8$~GeV in the $\pi\pi\rho$ spectrum was obtained.
The process $\tau\to\nu_\tau\pi f_1(1285)$ was also studied in a similar way: the $a_1(1260)$ resonance in the final state is replaced with the $f_1(1285)$~\cite{Oset:2018zgc}.
In the $\tau$-decay case, since the $\tau$ mass is smaller than the $K^*\bar K^*$ threshold, the TS of the $K^*\bar{K}^*K$ loop does not lie in the phase space. Nevertheless, the TS located in the complex plane still plays a sizeable role and leads to a significant distortion of the $f_1\pi$ invariant mass spectrum from a plain phase space distribution.

\subsection{Light baryons}

In this subsection, recent studies containing light baryons are summarized.\footnote{In fact, most of the old discussions in the 1960s on the TSs are related to processes with light baryons involved due to the limited experimental facilities. For instance, some processes were proposed to study the TS with light baryons in Ref.~\cite{Chang:1964pr}.}

First, let us summarize studies related to the $\Lambda(1405)$ resonance.
The mass of the $\Lambda(1405)$ resonance is slightly below the $\bar KN$ threshold, and couples to the $\bar KN$ channel strongly (and thus was proposed to be a $\bar K N$ molecule whose existence was predicted in 1959 by Dalitz and Tuan~\cite{Dalitz:1959dn,Dalitz:1960du} prior to the discovery of the $\Lambda(1405)$ in 1961~\cite{Alston:1961zzd}).
One reason that this resonance has been paid much attention is that it was proposed to have a special feature known as a two-pole structure~\cite{Oller:2000fj,Jido:2003cb,Hyodo:2011ur,Mai:2014xna}:
the higher pole is located close to the $\bar{K}N$ threshold, and the lower one is rather close to the $\pi\Sigma$ threshold with a large imaginary part on the same Riemann sheet; the peak of the $\Lambda(1405)$ is produced as a manifestation of these two poles.
It is found in Ref.~\cite{Jido:2003cb} that the higher (lower) pole strongly couples to the $\bar{K}N$ $(\pi\Sigma)$ channel (see, \eg, Refs.~\cite{Hyodo:2011ur,Kamiya:2016jqc} and the topical review on the $\Lambda(1405)$ in the RPP~\cite{Tanabashi:2018oca} for more discussions).
Then, one can expect sizable contributions from triangle loops with the $\bar KN$ or $\pi\Sigma$ merging into the $\Lambda(1405)$ in certain processes.

The effects of the TS on the $\gamma p\to K^+\Lambda(1405)$ reaction were investigated in Ref.~\cite{Wang:2016dtb} with a particular interest in the angular distribution.
The total cross section of the $\Lambda(1405)$ photoproduction shows a peak around $1.9$~GeV in the photon energy, and the angular distribution increases rapidly around $\cos\theta=0$~\cite{Moriya:2013hwg}, with $\theta$ the angle between the photon and the kaon in the $\gamma p$ c.m. frame,  which was not reproduced well theoretically (here we refer to Ref.~\cite{Nakamura:2013boa} for a paper containing a detailed analysis of this reaction).
It was proposed in Ref.~\cite{Wang:2016dtb} that the reaction receives important contributions from the  $K^*\Sigma\pi$ loops with the $K^*\Sigma $ coupled to the $\gamma p$ via the $N^*(2030)$ resonance. The triangle loops can produce a peak around $2.1$~GeV 
in the $K^+\Lambda(1405)$ distribution (1.9~GeV of the photon energy in the proton rest frame).
By including additional phenomenological terms, the peak around $1.9$~GeV and the rapid rise of the cross section at $\cos\theta\sim 0$ around $W=2$~GeV can be reproduced fairly well~\cite{Wang:2016dtb}.

In Ref.~\cite{Bayar:2017svj}, the processes $\pi^-p\to K^0\Lambda(1405)$ and $pp\to pK^+\Lambda(1405)$ with the subsequent decay of $\Lambda(1405)\to\pi\Sigma$ are considered.
Similar to the previous case, the $K^*\Sigma\pi$ triangle diagram with the $\pi\Sigma$ pair turning into the $\Lambda(1405)$ develops a singularity at about $2.1$~GeV.
Experimentally,  the $\pi^-p\to K^0\Lambda(1405)$ was reported in Ref.~\cite{Thomas:1973uh}, and the process $pp\to pK^+\Lambda(1405)$ was studied in Refs.~\cite{Zychor:2007gf,Agakishiev:2012xk,Adamczewski-Musch:2016vrc}.
It was found that in these reactions the $\Lambda(1405)$ peak appeared slightly lower than $1.4$~GeV, \ie, lower than the location of the $\Lambda(1405)$ peak in other reactions. For instance, in the photoproduction case, the peak is above $1.4$~GeV in the experimental data~\cite{Moriya:2013eb}, which can be described with a reasonable agreement using unitarized chiral models~\cite{Roca:2013cca,Nakamura:2013boa,Mai:2014xna}.
In the studies of the $pp$ and $\pi^- p$ reactions~\cite{Geng:2007vm,Hyodo:2003jw} using similar unitarized chiral models to account for the $\pi\Sigma$--$\bar K N$ coupled-channel final state interactions, which generates the $\Lambda(1405)$~\cite{Kaiser:1995eg,Oset:1997it,Oller:2000fj}, the $\pi\Sigma$ energy spectra were also found to peak above $1.4$~GeV. It is because that the amplitude is dominated by the higher one of the two $\Lambda(1405)$ poles.
In Ref.~\cite{Bayar:2017svj}, it is found that the  $\bar{K}^*\Sigma\pi$ triangle loops enhance the contribution from the $\pi\Sigma\to\pi\Sigma$ rescattering. As a result, the peak position is lowered due to the lower pole of $\Lambda(1405)$.\footnote{In the two-pole picture of the $\Lambda(1405)$, the higher pole couples dominantly to the $\bar K N$ while the lower one couples mainly to the $\pi\Sigma$.}
Thus, the TS effects and the two-pole structure of the $\Lambda(1405)$ resonance provide one plausible explanation of the peaking locations in different reactions.

In Ref.~\cite{Dai:2018hqb}, the $\Lambda_c\to\pi\pi\pi\Sigma$ decay, with the $\pi\Sigma$ produced from the $\Lambda(1405)$, is studied considering a $K^*N\bar{K}$ loop.
With the formula in Eq.~\eqref{eq:trianglesing}, the energy where a TS appears on the physical boundary is found to be $1.88$~GeV in the $\pi\Lambda(1405)$ ($ \pi\pi\Sigma$) invariant mass.
The weak decay is assumed to proceed through the $W^+$ emission which becomes the $\pi^+$ in the final state. Then the $\pi^0\Lambda(1405)$ production from the isoscalar $sud$ source breaks isospin symmetry~\cite{Dai:2018hqb}. 
However, similar to the case of the $\eta(1405/1475)\to\pi f_0(980)$, the isospin violation is expected to get enhanced by the TS effects.
The obtained $\pi\Sigma$ invariant mass distribution has a rather narrow peak around the $\Lambda(1405)$ region. The width is only of several MeV, which is of the order of the isospin mass splitting between the $K^- p$ and 
$\bar K^0 n$ thresholds. This special prediction needs to be examined experimentally.

The case with the same final-state particles, but with the $\Lambda(1405)$ replaced by the $\Sigma(1430)$, was studied in Ref.~\cite{Xie:2018gbi}. In this case, the process is isospin symmetry allowed.
The $\Sigma(1430)$ is an isospin-vector structure which couples strongly to the $\bar K N$ hypothesized in Ref.~\cite{Oller:2000fj} as a bound state and in Ref.~\cite{Jido:2003cb} as a strong cusp effect (see Refs.~\cite{Oller:2006jw,Guo:2012vv,Roca:2013cca} for further investigations of this structure).
In Ref.~\cite{Xie:2018gbi}, the $\bar{K}N$-$\pi\Sigma$ coupled-channel amplitude constructed in Ref.~\cite{Oset:1997it}, which does not have a pole in the proper Riemann sheet in the $I=1$ channel but has a pronounced $\bar{K}N$ cusp, was used.
The TS of the $K^*N\bar{K}$ loop produces a peak structure in the $\pi\pi\Sigma$ distribution around $1.9$~GeV, and in the $\pi\Sigma$ distribution a narrow cusp peak around the $\bar{K}N$ threshold is also clearly visible.
The corresponding measurements may be done at BESIII and Belle-II, and an observation of the predicted structures will be able to provide valuable information to our limited knowledge on the $I=1$ part of the $\bar{K}N$ interaction.

The analyses of recent high-statistics experimental data provide fruitful information on hadron resonances.
An enhancement of the near-threshold $\pi^0N(1535)$ production in the $\gamma p\to\eta\pi^0p$ process was found in Ref.~\cite{Gutz:2014wit} though it is naively expected to be suppressed because of the $P$-wave near the threshold.
This problem was addressed by taking into account the TS effects in Ref.~\cite{Debastiani:2017dlz}.
The $\pi^0N(1535)$ can be produced via the $\Delta\,\eta\, p$ loop with the $\eta\, p$ merging into the $N(1535)$.
The large coupling of the $N(1535)$ to the $\eta N$ channel, which is a salient feature of the $N(1535)$,\footnote{The large branching fraction of the $N(1535)$ to $\eta N$ reported by the PDG~\cite{Tanabashi:2018oca} is not compatible with the naive quark-model expectation, suggesting a large $s\bar s$ component~\cite{Liu:2005pm},
and a description of it as a hadronic molecule in a coupled-channel analysis including the $\eta N$, $K\Lambda$, and $K\Sigma$ channels, which contain hidden-strange contents, was proposed to resolve this problem in Ref.~\cite{Kaiser:1995cy}. Many further studies were devoted on this topic, see, \eg, Refs.~\cite{Nieves:2001wt,Inoue:2001ip,Arndt:2006bf,Khemchandani:2013nma,Garzon:2014ida,Sekihara:2015gvw}.} makes the role of the TS important.
The $\eta\Delta$ production is initiated by the intermediate $\Delta(1700)$ ($J^P=3/2^-$) resonance, which gives the dominant contribution in the near-threshold energy as seen experimentally~\cite{Ajaka:2008zz}.
A bump-like structure of the $\pi^0N(1535)$ production cross section around $1.2$~GeV of the initial-photon energy in the laboratory frame (about $1.78$~GeV of the $\gamma p$ energy in their c.m. frame) coincides with the TS energy of the $\Delta\eta p$ loop, indicating the importance of the TS effects.

In Ref.~\cite{Roca:2017bvy}, the TS effects on the properties of the $N(1700)$ with $J^P=3/2^-$ was studied.
A relatively large branching fraction of the $N(1700)$ to the $\pi\Delta$ channel was reported by some experimental analyses though the results of these analyses do not converge
as can be seen, \eg, from Table~I of Ref.~\cite{Roca:2017bvy}.
The authors considered the scattering amplitude from a pair of vector meson and an octet ground state baryon into the $\pi \Delta$, which goes through a $\rho N\pi$ triangle loop with the $\rho N$ rescattering by the exchange of a pion into the $\pi\Delta$. It was found that the triangle loop produced a bump around 1.7~GeV (the $\rho N$ threshold), and enhanced the contribution of the $\pi\Delta$ channel to the dynamics of the $N(1700)$ resonance.
The nucleon resonance $N(1875)$ $(J^P=3/2^-)$, the existence of which is supported by the partial-wave analysis in Ref.~\cite{Anisovich:2011fc} and by the recent high-statistics experimental data~\cite{Sokhoyan:2015fra}, was studied in Ref.~\cite{Samart:2017scf} along the same line. 
Building upon the meson-baryon interaction framework in Ref.~\cite{Sarkar:2004jh}, the $\Sigma^*K\Lambda$ and $\Delta\pi\pi$ loops were considered to take account of the decays of the resonance into the $\pi N(1535)$ and $\sigma N$, respectively. It was found that the TS from the former, which is slightly off the physical region because the mass of the $N^*(1535)$ is smaller than the $K\Lambda$ threshold, could produce a peak around the $N(1875)$ mass (also around the $\Sigma^* K$ threshold).

\subsection{\texorpdfstring{${XYZ}$}{XYZ}  states}

\subsubsection{Use of TS: measuring the \texorpdfstring{${X(3872)}$}{X(3872)} binding energy}

\begin{figure}[tb]
    \centering
    \includegraphics[width=0.435\textwidth]{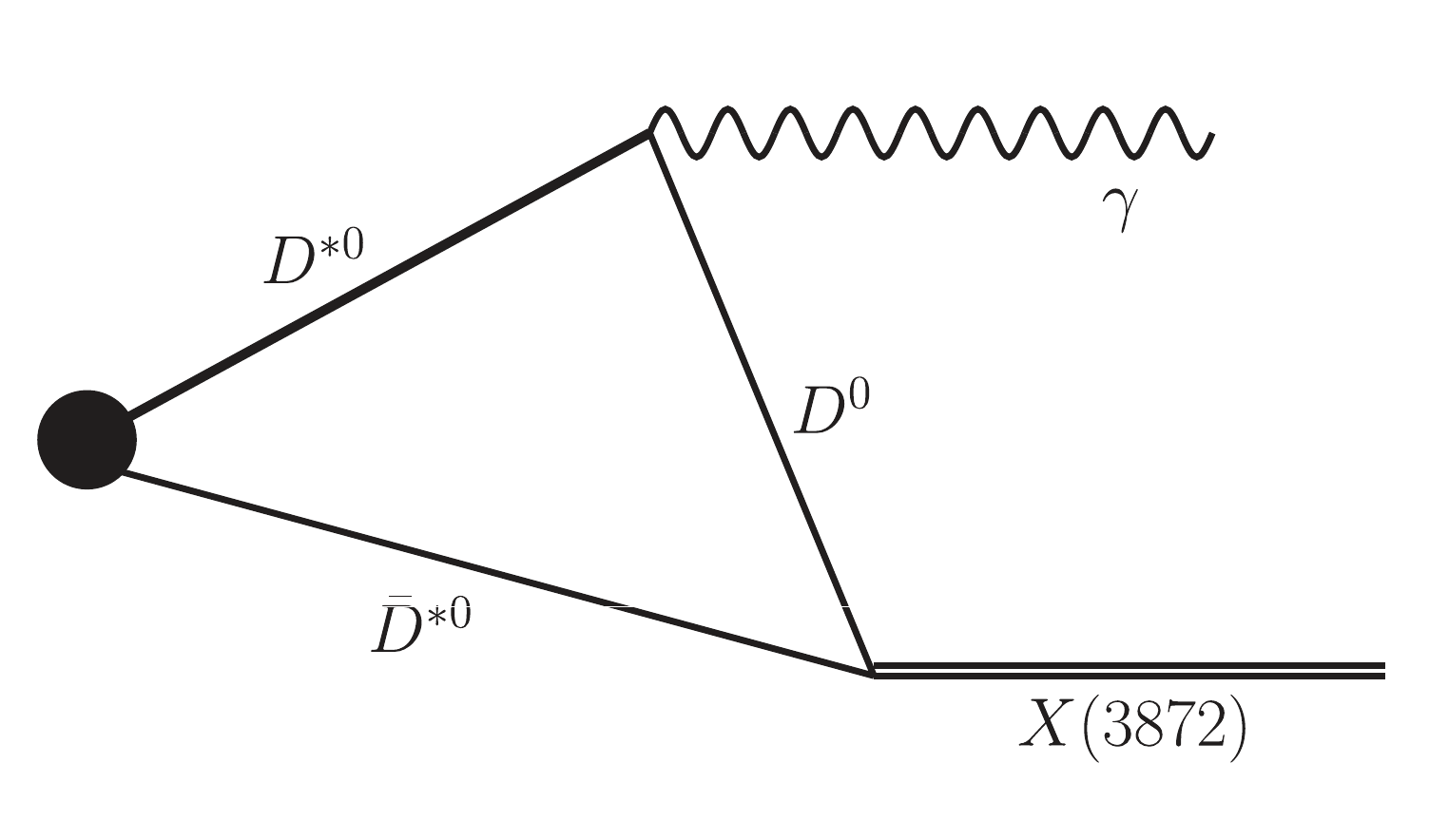} \quad
    \includegraphics[width=0.435\textwidth]{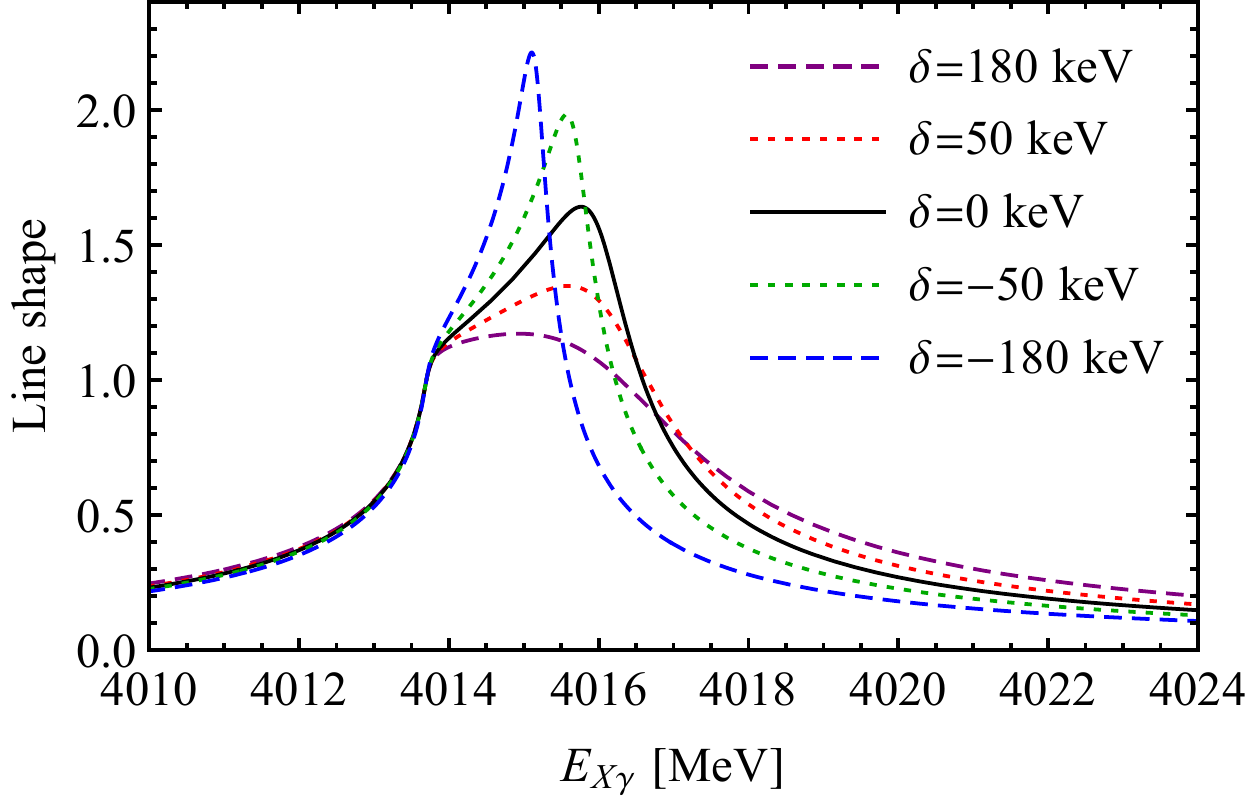}
    \begin{minipage}[t]{16.5 cm}
    \caption{Left panel: the mechanism for producing the $\x$ associated with a photon from a short-distance $\dzstar\dzstarbar$ source; right panel: dependence of the $\gamma\x$ line shape on the $\x$ binding energy.} 
    \label{fig:Xmass}
    \end{minipage}
\end{figure}

The triangle singularity does not only lay traps in identifying hadronic resonances, it can also be used to make precision measurements thanking the sensitivity of the singularity to kinematical variables. In Ref.~\cite{Guo:2019qcn}, it was suggested that the $\x$ binding energy, that is $\delta = M_{D^0} + M_{D^{*0}} - M_{\x}$, can be measured by the mechanism shown in the left panel of Fig.~\ref{fig:Xmass}, and a precision of one-order-of-magnitude smaller uncertainty compared with the current one ($\delta=0\pm180$~keV using the PDG average values for the involved masses~\cite{Tanabashi:2018oca}) may be achieved.

This mechanism requires the $\dzsdzsbar$ pairs be produced  at short distances in some high energy experiments, which means that one can neglect any other impact from the $\dzsdzsbar$ production vertex and cares only about the production strength. Then one of the $\dzsdzsbar$ decays into a photon and a $D^0$ (or $\bar D^{0}$), which then coalesces with the other $\bar D^{*0}$ (or $D^{*0}$) into the $\x$. The $\x$ needs to be constructed using decay modes other than the $D^0\bar D^0\pi^0$ or $D^0\bar D^{*0}+c.c.$ in order to avoid the intricate interference between the triangle diagram and tree-level ones, \ie, $\dzsdzsbar\to \gamma D^0\bar D^{*0} +c.c.$ without the rescattering between the  $D^0$ (or $\bar D^{0}$) and the $\bar D^{*0}$ (or $D^{*0}$), that would also contribute in such a case.
The mechanism is based on the extreme sensitivity of the TS location to the $\x$ binding energy. When the initial energy of the $\dzsdzsbar$ pair is within the range given by Eq.~\eqref{eq:mArange}, the logarithmic TS is located in the physical region and is shifted to the complex plane due to the $D^{*0}$ width. Because the $D^{*0}$ width, which was predicted to be $(55.3\pm1.4)$~keV~\cite{Guo:2019qcn,Rosner:2013sha} from the width of the charged $D^*$ using isospin symmetry, is tiny, the singularity is in the immediate vicinity of the physical region, resulting in the extreme sensitivity to the binding energy mentioned above.
This can be clearly seen from the right panel of Fig.~\ref{fig:Xmass} which shows the $\gamma\x$ line shapes for $\delta$ ranging from $-180$~keV to 180~keV.
The behavior can be easily understood. The cusp at the left shoulder of the peak is because of the opening of the $\dzsdzsbar$ threshold, whose location is thus fixed, and the following peak arises from the TS. The two singularities together lead to the particular line shape.
Using Eq.~\eqref{eq:trianglesing}, it is easy to see that the singularity in the total energy in the c.m. frame is located at
\begin{equation}
    E_{X\gamma}^\text{TS} = 2 m_{D^{*0}} + \frac{x^2}{4m_{D^0}} + \mathcal{O}\left(\frac{x^3}{m_{D^0}^2}\right),
     \label{eq:ts}
\end{equation}
with $x=m_{D^{*0}}-m_{D^0}-2\sqrt{-m_{D^0}\delta}+\delta$. From this expression, one sees that when the $\x$ is below the $D^0\bar D^{*0}$ threshold, $\delta>0$, $x$ is complex, meaning that the singularity is off the real energy axis and produces a smooth peak. When the $\x$ is above threshold, $\delta<0$, the singularity is in the physical region and becomes complex only when the $D^{*0}$ width is taken into account; consequently, it produces a much sharper peak. 
Notice that the curves on the right panel of Fig.~\ref{fig:Xmass} were computed without considering the width effect of the $\x$. The $\x$ needs to be reconstructed from the decay modes $J/\psi\pi^+\pi^-(\pi^0)$ and so on, so that in principal the line shape needs to be convolved with the $\x$ spectral function. Were the $\x$ width is as large as 1~MeV, the TS structure would get significantly smeared; for  the $\x$ width to be $\lesssim100$~keV, which is very likely given that the partial width of the $\x\to D^0\pi^0\pi^0$ is only about 30--50~keV in the hadronic molecular picture~\cite{Fleming:2007rp,Guo:2014hqa,Dai:2019hrf},  the impact on the line shapes is small as was checked. It needs yet to be taken into account for a more thorough analysis. 

This method measures the binding energy for the $\x$ directly, and thus surpasses the restriction for its precision set by the uncertainties of the $\x$ (currently 170~keV~\cite{Tanabashi:2018oca}) and even the $D^{(*)0}$ (50~keV~\cite{Tanabashi:2018oca}) masses.
By generating synthetic data for the $\gamma\x$ line shape using the Monte Carlo method, it was found that the precision could reach the level of about 10~keV to a few tens of keV, depending on the actual binding energy, with $\mathcal{O}(10^3)$ events. We refer to Ref.~\cite{Guo:2019qcn} for more details about the simulation.

To make the full use of the sensitivity to the binding energy and have the threshold cusp, the $\dzsdzsbar$ needs to be in an $S$ wave, so that the quantum numbers for the $\gamma\x$ should be $J^{PC}=1^{+-}$.
Several reactions may be applicable for such a proposal: 
\begin{itemize}
    \item[(1)] $e^+e^-\to \pi^0\gamma\x$ at high luminosity $e^+e^-$ colliders with the c.m. energy at $\sim 4.4$~GeV: The cross sections for the open-charm productions  $e^+e^-\to \pi^+ D^0 D^{*-}$~\cite{Ablikim:2018vxx} and $D^+D^-\pi^+\pi^-$~\cite{Ablikim:2019okg}  measured by the BESIII Collaboration reached their maxima around 4.4~GeV. The cross section needs to be estimated in order to see whether this is achievable at future super tau-charm factories under discussion.
    \item[(2)] $B\to K\gamma \x$: Given that the production of the $h_c$ with the same $J^{PC}$ quantum numbers in the $B$ decays is much smaller than that for the $J/\psi$, the rate for this reaction might be too small to be feasible at the current $B$ factories.
    \item[(3)] $p\bar p\to \gamma \x$: This should be the most promising reaction. With the c.m. energy tuned to measure the line shape between 4010 and 4020~MeV, a high precision measurement of the $\x$ binding energy is foreseen at \panda~which has a brilliant energy resolution. Together with the possible high-precision measurements of (the upper limit of) the $\x$ width at \panda~\cite{PANDA:2018zjt} and Belle-II~\cite{Hirata:2019gqg}, a deeper understanding of the puzzling $\x$ is foreseen.
\end{itemize}

\subsubsection{Use of TS: enhancing production of hadronic molecules}

One of the most salient features of the hadronic molecules is that they couple strongly to their constituents. 
The long-distance part of their production in various processes happens through producing their constituents first (for a discussion, see Ref.~\cite{Guo:2017jvc}).
Furthermore, prime candidates of hadronic molecules are close to the relevant thresholds and couple in $S$-wave.
Therefore, it is natural to expect that the production of hadronic molecules can get enhanced by triangle singularities for suitably chosen intermediate particles. For a recent review dedicated to hadronic molecules, see Ref.~\cite{Guo:2017jvc}. The concept of compositeness for hadronic molecules is also nicely discussed in Ref.~\cite{Hyodo:2013nka}.

In the mechanism discussed in the above subsection, the $\dzsdzsbar$ pair needs to be in an $S$-wave in order to have the threshold cusp, giving rise to the special line shape shown in Fig.~\ref{fig:Xmass}. The triangle diagram has a TS even if the $\dzsdzsbar$ pair is in a $P$-wave. But in this case, there is no threshold cusp as discussed in Section~\ref{sec:3}, and thus a single smooth peak due to the TS is expected to be seen with a suppression at the $\dzsdzsbar$ threshold due to the $P$-wave factor. This is exactly what was found by Braaten {\it et al.} in Refs.~\cite{Braaten:2019gfj,Braaten:2019gwc}, where the TS effects were  studied in the $e^+e^-\to \gamma\x$ reaction. 
The same reaction was first studied in Ref.~\cite{Dubynskiy:2006cj} where, however, the existence of a TS was not realized.
Similar enhancement effects in the $\x$ production associated with a pion were computed for the $B$-meson decays~\cite{Braaten:2019yua} and hadro-productions~\cite{Braaten:2019sxh}.

\begin{figure}[tb]
	\centering
	\includegraphics[width=0.3\textwidth]{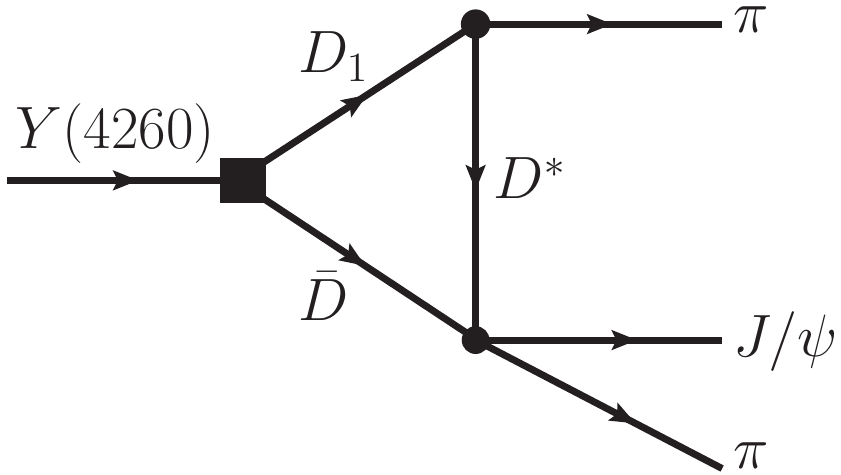} 
  \begin{minipage}[t]{16.5 cm}
	\caption{The $Y(4260)\to J/\psi\pi\pi$ decay via the $D_1(2420)\bar D D^*$ triangle rescattering diagram.} 
  \label{fig:YtoJpsipipi}
  \end{minipage}
\end{figure}
So far in all the observations, the $Z_c(3900)$ was always observed together with a pion, and their production was found to be highly correlated to the initial energy. In the original discoveries, the c.m. energies for producing the $Z_c(3900)^\pm\pi^\mp$ at both BESIII and Belle experiments are around 4.26~GeV~\cite{Ablikim:2013mio,Liu:2013dau}. Confirmations and higher-statistics observations of the $Z_c(3900)$ by BESIII was done with $e^+e^-$ c.m. energies of 4.23 and 4.26~GeV~\cite{Ablikim:2013xfr,Ablikim:2015tbp,Ablikim:2015swa,Collaboration:2017njt}. The D0 Collaboration also reported signals for the $Z_c(3900)^\pm\to J/\psi\pi^\pm$ from the semi-inclusive $b$-flavored hadron decays~\cite{Abazov:2018cyu,D0:2019zpb}, and found that the $Z_c(3900)$ only existed in the data with the $J/\psi\pi^+\pi^-$ constrained in the energy region between 4.2 and 4.3~GeV, suggesting that the observed production happened only through the intermediate process $Y(4260)\to Z_c\pi$.  
One possible explanation for these experimental facts is that the $Z_c(3900)$ production is enhanced by the nearby TSs. The triangle diagram can have $D_1(2420)\bar D D^*$ as the intermediate particles~\cite{Wang:2013cya,Wang:2013hga}, as shown in Fig.~\ref{fig:YtoJpsipipi}, and the $Z_c$, which couples strongly to the $D\bar D^*$ and is an isovector $D\bar D^*$ molecular candidate~\cite{Wang:2013cya,Guo:2013sya}, is contained as a pole in the $\bar D D^*\to J/\psi\pi$ rescattering amplitude.
More discussions about the TS effects related to the $Z_c(3900)$ will be given in Section~\ref{sec:zc_ts}.

Similarly, for the $\x$ being a prominent $D\bar D^*$ hadronic molecular candidate, one can expect its production in $Y(4260)\to\gamma\x$ should also be enhanced through the same $D_1(2420)\bar D D^*$ triangle diagrams. Thus, it was predicted in Ref.~\cite{Guo:2013nza} that the $Y(4260)\to\gamma\x$ transition rate should be large.
The $\x$ was then observed at BESIII in the suggested reaction~\cite{Ablikim:2013dyn}, which has been the source for BESIII to produce the $\x$ and study its properties~\cite{Ablikim:2019zio}.

For the production of the $Z_b(10610)$ and $Z_b(10650)$ states, to be denoted as $Z_b^{(\prime)}$, in the $\Upsilon(11020)$ decays, since the $\Upsilon(11020)$ is very close to the thresholds of $B_1(5271) \bar{B}$, it was pointed out in Refs.~\cite{Wang:2013hga,Bondar:2016pox} that TSs  could be important to enhance the production rates.
However, the narrow $B_1(5721)$ is mainly a meson with $s_\ell^P=3/2^+$, where $P$ denotes the parity and $s_\ell$ is the total angular momentum of the light quark system, including the light quark spin and the orbital angular momentum, which becomes a good quantum number in the heavy quark limit~\cite{ManoharWise}, and it has been shown that the $S$-wave production of a pair of $3/2^+$ and $1/2^-$ (\ie, ground state $S$-wave heavy mesons) mesons in $e^+e^-$ collisions is suppressed in the heavy quark limit~\cite{Li:2013yka}. Thus, a mixing between the $3/2^+$ $B_1$ and $1/2^+$ $B_1'$ axial-vector bottom mesons, though suppressed in the heavy quark limit as well, is introduced in Ref.~\cite{Bondar:2016pox}.

It is argued in Ref.~\cite{Wu:2018xaa} that the $B_1' \bar B^{(*)} B^{(*)}$ and $B_0^* \bar B^{*} B^{(*)}$ loops, where $B_1^\prime$ and $B^\ast_0$ are the lowest-lying $s_\ell^P=1/2^+$ bottom mesons, play an important role in the copious production of the $Z_b$ states in the $\Upsilon(10860)\to Z_b^{(\prime)}\pi$ processes. To date, the $B_1^\prime$ and $B^\ast_0$ have not been discovered yet. Using the values $m_{B_1^{\prime}} =5584$~MeV and $m_{B_0^{\ast}} =5535$~MeV, which were predicted in Ref.~\cite{Du:2017zvv} using heavy quark flavor symmetry in a framework which can describe both the lattice~\cite{Liu:2012zya,Moir:2016srx} and experimental data~\cite{Aaij:2016fma} for the $D\pi$ $S$-wave systems~\cite{Albaladejo:2016lbb,Du:2017zvv}, the nominal thresholds of the $B_0^*\bar B$ and $B_1'\bar B^*$ are close to the mass of the $\Upsilon(10860)$, and they can couple in an $S$-wave.
The $B^*\bar B^{(*)}$ also couple to the $Z_b^{(\prime)}$ in an $S$-wave. As a result, the triangle loops are potentially important and could be the reason why the two $Z_b$ states can be easily produced in the $\Upsilon(10860)$ decays. It is also shown in  Ref.~\cite{Wu:2018xaa} that the large widths of the $B_1^\prime$ and $B^\ast_0$ do not suppress such a contribution like one would naively guess. This is because their widths are correlated with the pionic couplings for the dominant decay channels into the $B^{(*)}\pi$, which enter the triangle amplitude as a multiplicative factor.

The productions of the $D_{s0}^*(2317)$ and $D_{s1}(2460)$ via triangle diagrams are considered in Ref.~\cite{Sakai:2017hpg}.
The decay processes $B^-\to K^-\pi^-D_{s0}^{*+}$ and $B^-\to K^-\pi^-D_{s1}^+$ are considered.
The triangle loops of $K^*DK$ and $K^*D^*K$ (the external $K^-$ and the internal $K^*D^{(*)}$ pair are produced from $B^-$ and the $D^{(*)}K$ fuse to produce the $D_{s0}^*$ ($D_{s1}$) in the final state) have singularities around 2.85 and 3~GeV in the $D_{s0}^*\pi$ and $D_{s1}\pi$ invariant mass distributions by putting the $DK$ and $D^*K$ invariant masses slightly above the thresholds.
One promising model for the $D_{s0}^*$ and $D_{s1}$ mesons is that they are molecular states composed of $KD$ and $KD^*$, respectively~\cite{Barnes:2003dj,vanBeveren:2003kd,Kolomeitsev:2003ac,Chen:2004dy,Guo:2006fu,Guo:2006rp,Gamermann:2006nm,Gamermann:2007fi,Liu:2012zya,Albaladejo:2016lbb,Du:2017zvv} (for a recent review, see Ref.~\cite{Guo:2019dpg}).
While these mesons are in the bound region of these channels, due to their large couplings to their constituent hadrons, the TSs slightly off the physical boundary in the complex plane are still found to be relevant, producing characteristic peaks in the $\pi^- D_{s0}^{*+}$ and $\pi^- D_{s1}^+$ spectra~\cite{Sakai:2017hpg}.

Other than productions, the effect of the TS on the hadron decays of hadronic molecular candidates was also discussed.
In Ref.~\cite{Achasov:2019wvw}, the $\pi^+\pi^-$ mass spectrum in the $X(3872)\to\pi^+\pi^-\pi^0$ decay was studied with the $\x$ spectral function taken into account.
In the $\pi^+\pi^-$ spectrum, the triangle loops of $D^{*0}\bar D^0D^0 + c.c.$ with a $D^0\bar D^0\to\pi^-\pi^+$ transition have a singularity at the vicinity of the $D^0\bar D^0$ threshold (slightly below 3.73~GeV) if the mass of the $X(3872)$ is above the $D^0\bar D^{*0}$ threshold, and, in such a situation, the branching ratio of the $X(3872)\to\pi^+\pi^-\pi^0$ decay is found to be dominated by the contribution in a small energy region of the $\pi^+\pi^-$ spectrum around this TS energy.
In addition, by looking at the dependence of the $X(3872)\to\pi^+\pi^-\pi^0$ partial width on the energy of the initial state, the enhancement of the decay by the TS is evident with two sharp peaks around the neutral and charged $D\bar D^*$ thresholds.\footnote{While the charged $D^{*+}D^-D^+ + c.c.$ loops are also considered, the contribution from the neutral loop would be the more important one in practice due to the extreme closeness of the $X(3872)$ mass to the neutral threshold and its tiny width.}
The observation of such a characteristic $\pi^+\pi^-$ distribution would be a signal of the TS and the  sequential decay $X(3872)\to  (D^0\bar D^{*0} + c.c.)\to  \pi^0 D^0\bar D^0\to \pi^+\pi^-\pi^0$.

\bigskip

%% file: section6.tex
\section{Triangle singularities in the final state energy spectrum}
\label{sec:6}

In this section, we discuss the structures in the final state energy spectrum with the invariant mass of the initial state  fixed.

\subsection{Charged charmonium-like states}

Structures in the charmonium/bottomonium mass region observed in charged final states are among the most interesting objects to study since they, being electrically charged, must contain at least two quarks and two antiquarks if they are QCD resonances and thus are excellent candidates for exotic hadrons.
Yet, they are either broad or rather close to open-charm thresholds. Moreover, they were all observed in reactions with at least three particles in the final state. The three-body final state interactions, together with the possible coupled channels, introduce complication into the amplitude analysis. Basically for all the so far claimed charged charmonium-like states there are proposals suggesting the importance of certain TSs, which will be reviewed in the following.

\subsubsection{\texorpdfstring{${Z_c(3900)}$}{Zc3900}}
\label{sec:zc_ts}

The BESIII and Belle collaborations almost simultaneously reported the observation of a charged charmonium-like state $Z_c(3900)^{\pm}$ in the $J/\psi\pi^\pm$ invariant mass spectrum from the $Y(4260)\to J/\psi\pi^+\pi^-$ decays in 2013~\cite{Ablikim:2013mio,Liu:2013dau} and confirmed in an analysis of the CLEOc data at $E_{\rm cm}=4.17$~GeV~\cite{Xiao:2013iha}. The existence of the neutral partner $Z_c(3900)^{0}$ was reported later by BESIII in 2015~\cite{Ablikim:2015tbp}, the evidence of which was also reported earlier in Ref.~\cite{Xiao:2013iha}. 
The quantum numbers were determined to be $I^G(J^{PC})=1^+(1^{+-})$~\cite{Collaboration:2017njt} so that it couples to the $D^*\bar D$ in an $S$-wave.
Taking into account that the $Y(4260)$ may contain a large $D_1\bar{D}$ hadronic molecular component~\cite{Wang:2013cya,Cleven:2013mka}, the $Y(4260)\to J/\psi\pi^+\pi^-$ decays can receive contributions from the rescattering process as illustrated in Fig.~\ref{fig:YtoJpsipipi}. 
Although the mass of $Y(4260)$ is slightly below the $D_1(2420)\bar{D}$ threshold, the TS in the  ${J/\psi\pi^\pm}$ invariant mass of the rescattering amplitude is still close to the physical boundary and can influence the $J/\psi\pi^\pm$ invariant mass distribution around the $D^*\bar{D}$ threshold significantly~\cite{Wang:2013cya,Wang:2013hga} (more see below). 
Therefore, in Ref.~\cite{Wang:2013cya}, it is suggested that the $D_1(2420)\bar{D}$ molecular nature of the $Y(4260)$ provides a natural explanation for the appearance of the resonance-like structure $Z_c(3900)$ in $Y(4260)$ decays. 
Despite the importance of the triangle diagram contribution, Ref.~\cite{Wang:2013cya} concluded that a $Z_c(3900)$ resonance was still needed in order to fit to the narrow peak observed in experiments. A similar scenario was also suggested in Ref.~\cite{Liu:2013vfa}. 
Instead of considering the $D_1(2420)\bar{D}D^*$ loop, Szczepaniak suggested that the $Z_c(3900)$ peak could be attributed to the $D_0^*(2300)\bar{D}^*D$ loop,\footnote{The $D_0^*(2300)$ was called $Ds_0^*(2400)$ by the PDG until 2018.} which is in the physical region were the width of the $D_0^*(2300)$ neglected, in Refs.~\cite{Szczepaniak:2015eza,Szczepaniak:2015hya}. When the large decay width of $D_0^*(2300)$ is taken into account, the TS peak becomes broader and yet still has a cusp at the $D^*\bar D$ threshold, analogous to Fig.~\ref{fig:kstarkk}~(a).  

%%%%%%%%%%%%%%%%%%%%%%%%%%%%%%%
\begin{figure}[tb]
  \centering
    \includegraphics[height=6.cm]{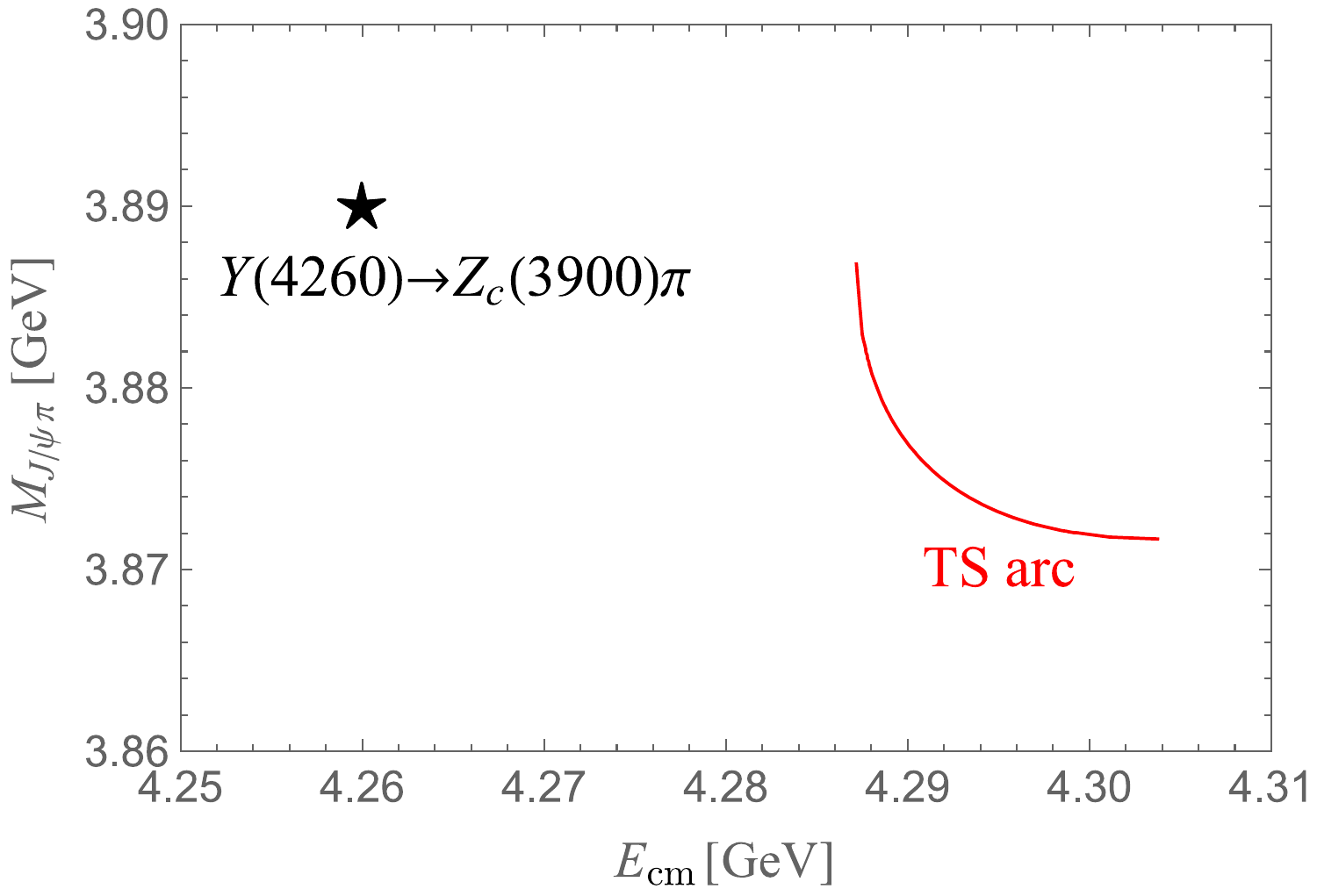} ~~~~
    \includegraphics[height=5.9cm]{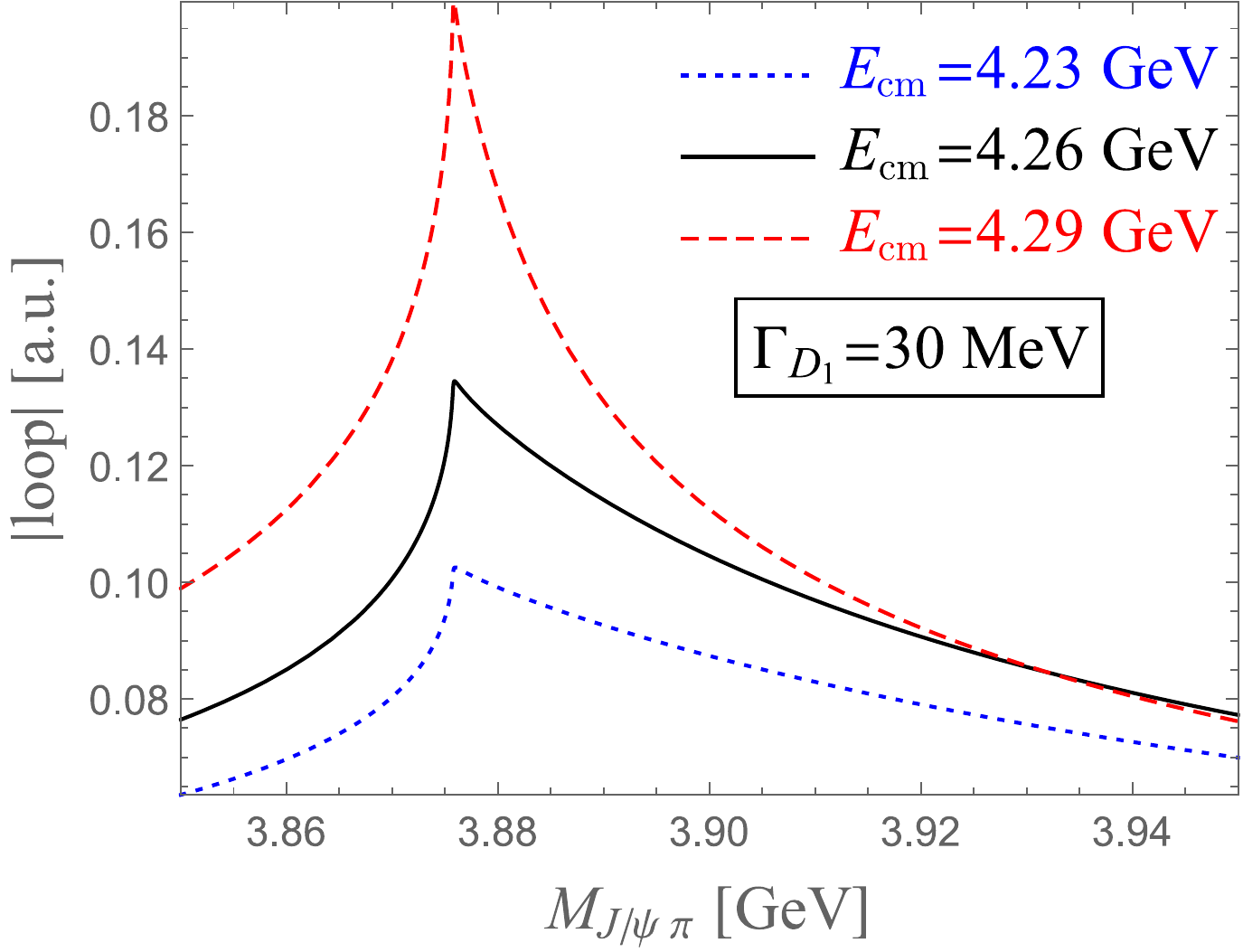}
  \begin{minipage}[t]{16.5 cm}
  \caption{Left: The red curve labeled as ``TS arc'' represents the trajectory in the $E_\text{cm}$--$M_{J/\psi\pi}$ plane along which the TS of the $D_1\bar D D^*$ triangle is on the physical boundary. The coordinates of the two ends of the arc correspond to the bounds of the ranges in Eqs.~\eqref{eq:m23range} and \eqref{eq:mArange}. Right: Dependence of the absolute value of the $D_1\bar D D^*$ triangle loop integral on the incoming energy.
  }
  \label{fig:ytozpi}
  \end{minipage}
\end{figure}

\begin{figure}[tb]
  \centering
    \includegraphics[height=6.2cm]{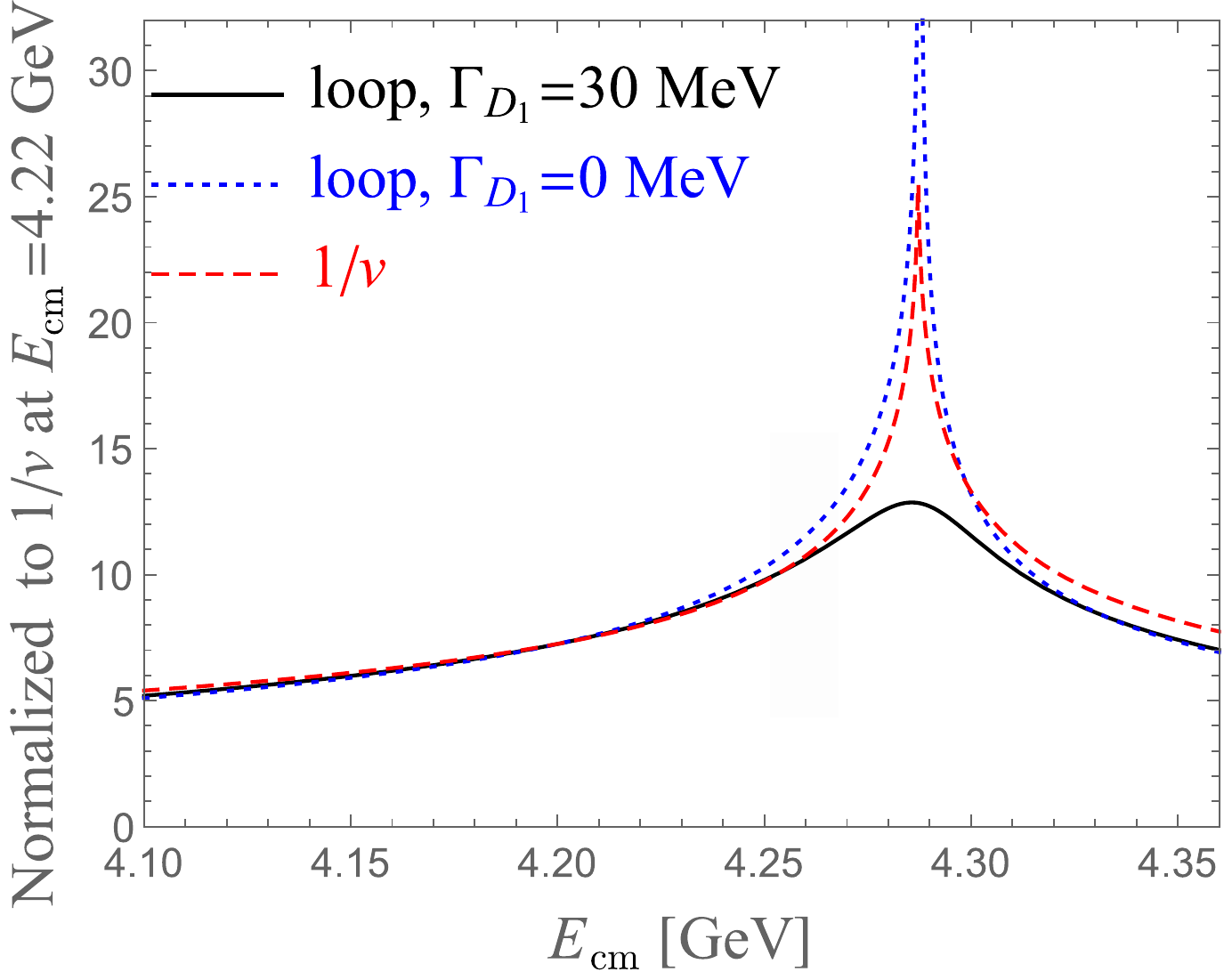}
  \begin{minipage}[t]{16.5 cm}
  \caption{Effect of the $D_1(2420)$ width on the absolute value of the $D_1\bar D D^*$ scalar loop. Here the $J/\psi\pi$ invariant mass is set to 3.886~GeV. The curve of $1/v$ comes from the velocity power counting with $v$ defined in Eq.~\eqref{eq:vpc}, and the $D_1$ width is neglected for this curve. 
  }
  \label{fig:ytozpi2}
  \end{minipage}
\end{figure}
%%%%%%%%%%%%%%%%%%%%%%%%%%%%%%%

Let us consider the $D_1(2420)\bar D D^*$ triangle loop shown in Fig.~\ref{fig:YtoJpsipipi}. By solving Eq.~\eqref{eq:trianglesing} with the width of the $D_1(2420)$ neglected, it is easy to find that the physical region  in the $E_\text{cm}$--$M_{J/\psi\pi}$ plane, where $E_\text{cm}$ refers to the c.m. energy of the $e^+e^-$ producing the $Y(4260)$, is located along the ``TS arc'' depicted in the left panel of Fig.~\ref{fig:ytozpi}. It is clear that the process $Y(4260)\to Z_c(3900)\pi$ is not on this arc, but is only a few tens of MeV away and thus leaves an influence.
Taking a  30~MeV constant width for the $D_1(2420)$, the absolute values of the scalar three-point loop as a function of the $J/\psi\pi$ invariant mass evaluated at three different initial energies $E_{\rm cm}$ are plotted in the right panel of Fig.~\ref{fig:ytozpi}, showing the dependence of the peak on $E_{\rm cm}$. It is evident that the peak becomes more and more modest when one reduces the value of $E_{\rm cm}$.
When $E_\text{cm}=4.29$~GeV, the singularity is away from the physical region only due to the small $D_1(2420)$ width, and the triangle loop produces a sharp peak. Decreasing $E_\text{cm}$, the peak becomes less pronounced since the TS is moving further away from the physical region. Nevertheless, there is always a cusp at the $\bar D D^*$ threshold because they rescatter in an $S$-wave, as discussed in the Section~\ref{sec:3}, and the two-body threshold cusp is a subleading singularity of the triangle diagram. 
The finite width of the $D_1$ does not smoothen this cusp as $D_1$ is in the crossed channel with respect to $D^*\bar D$. The sensitivity of the line shape on the incoming energy is one of the keys to reveal the role of TSs. 
In fact, the dependence on $E_\text{cm}$ in this case can be estimated using a nonrelativistic velocity power counting developed in Refs.~\cite{GUo:2009wr,Guo:2010ak,Guo:2012tg} (for a review see Ref.~\cite{Guo:2017jvc}): away from the TS, the scalar three-point loop integral should scale as $v^5/(v^2)^3 = 1/v$ with $v$ being the typical nonrelativistic velocity for the internal particles, where the factor $v^5$ comes from the nonrelativistic loop integral measure and the $(v^2)^{-3}$ factor is from  three nonrelativistic heavy meson propagators. It was shown in Ref.~\cite{Guo:2012tg} that $v$ should be understood as the average of two velocities defined from the two cuts depicted in Fig.~\ref{fig:triangle}: 
\begin{equation}
v = \frac12\left[\sqrt{|m_{D_1}+m_D - E_\text{cm}|/(2\mu_{12})} + \sqrt{|m_{D^*}+m_D - M_{J/\psi\pi}|/(2\mu_{23})}\right],
\label{eq:vpc}
\end{equation}
where $\mu_{12}$ and $\mu_{23}$ are the $D_1\bar D$ and $D^*\bar D$ reduced masses, respectively. One sees that the nonrelativistic power counting $1/v$ captures the the $E_{\rm cm}$ dependence from the  evaluated loop integral remarkably well except for the small region around the TS.

\begin{figure}[tb]
  \centering
    \includegraphics[width=0.9\textwidth]{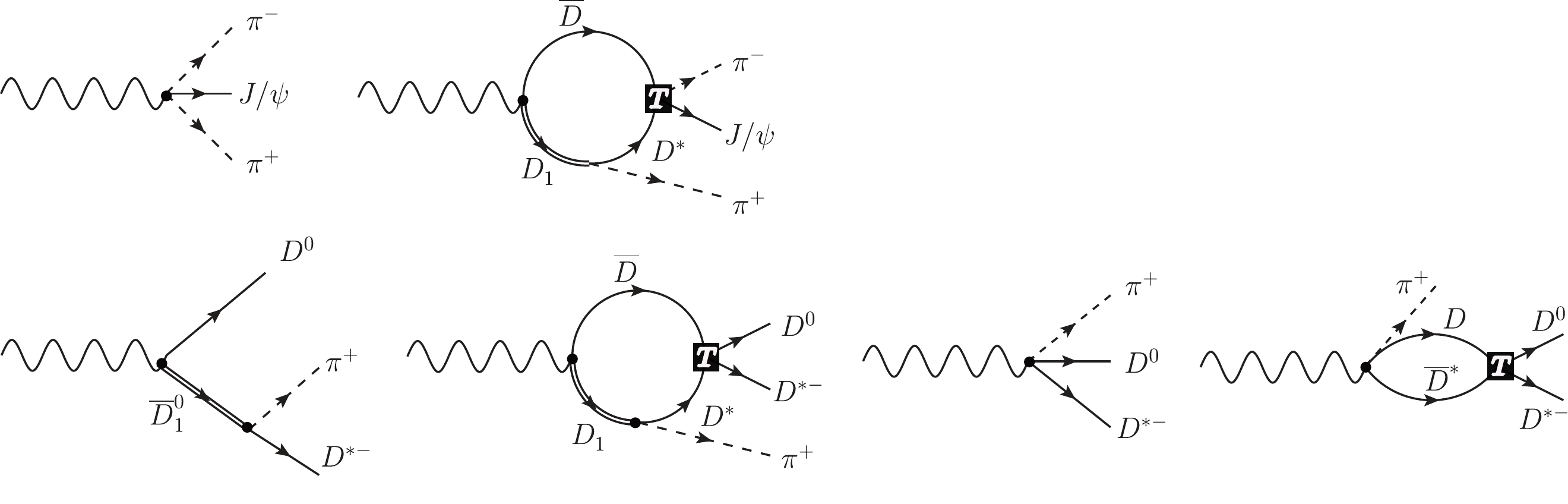}
  \begin{minipage}[t]{16.5 cm}
  \caption{The diagrams considered in Ref.~\cite{Albaladejo:2015lob} in the analysis of the $Y(4260)$ decays into $J/\psi\pi^+\pi^-$ and $D^0 D^{*-}\pi^+$.  The filled squares represent the coupled-channel ($D\bar D^*$, $J/\psi\pi$) scattering $T$-matrix. There are also contributions with the intermediate particles replaced by their charge conjugated ones, which are not shown here.
  }
  \label{fig:zcdiagrams}
  \end{minipage}
\end{figure}
The above discussion implies that the kinematic singularity structure is in fact more complicated than the two-body threshold cusp discussed in Section~\ref{sec:zc_cusp}, and the $D_1\bar D D^*$ triangle diagrams have to be included in a realistic analysis of the experimental data for the $Z_c(3900)$. This is particularly important to establish whether $Z_c(3900)$ has to be included as a resonance pole in the amplitude analysis of all the available data.
Such an analysis of the BESIII data of the $e^+e^-\to J/\psi\pi\pi$~\cite{Ablikim:2013mio} and the $e^+e^-\to D\bar D^*\pi$ at the $e^+e^-$ c.m. energy 4.26~GeV~\cite{Ablikim:2015swa} was done in Ref.~\cite{Albaladejo:2015lob}. In that work, in addition to the triangle diagrams with the $D_1\bar D D^*$ (and their antiparticles) intermediate state, the possibility of having a resonance pole, which can decay into the $\bar D^* D$ and $J/\psi\pi$, is incorporated by constructing a unitarized coupled-channel ($\bar D^* D$ and $J/\psi\pi$) scattering $T$-matrix treating all the heavy mesons nonrelativistically. 
From fitting to the data, the parameters of the $T$-matrix are fixed.
It is found that despite the inclusion of the $D_1\bar D D^*$ loops, the parameters from the best fit still demand the $T$-matrix to have a pole near the $D\bar D^*$ threshold, which can be interpreted as a state corresponding to the $Z_c(3900)$. 
Depending on whether an energy-dependent term is allowed  in the $J/\psi\pi$--$D\bar D^*$ coupled-channel potential, the pole is either a virtual state pole\footnote{The $S$ matrix can have several types of poles depending on where in the complex energy plane they are located. Considering a single-channel problem, the bound state pole is a pole below threshold in the real axis on the first Riemann sheet of the complex energy plane (or on the positive imaginary axis of the complex c.m. momentum plane); the virtual state pole is similar but on the second Riemann sheet (or on the negative imaginary axis of the complex momentum plane); the resonance pole is on the second Riemann sheet but off the real axis (or on the lower half momentum plane off the imaginary axis). 
Contrary to a bound state whose wave function is constrained to a finite region in the coordinate space, the wave function of a virtual state cannot be normalized and spreads over the whole space. Thus, the virtual state cannot be regarded as a normal particle. However, it leads to observable effects, and cannot be distinguished from a bound state above the threshold. A wellknown example is the below-threshold pole of the isovector and spin-0 nucleon-nucleon interaction, leading to a scattering length whose absolute value is as large as $24$~fm. For more discussions, we refer to Ref.~\cite{Guo:2017jvc}.} with respect to the $D\bar D^*$ channel
below its threshold, which could be a few tens of MeV away, or a resonance pole just above the threshold. 

However, debates continued regarding whether the $Z_c(3900)$ pole was really necessary in the analysis of the experimental data. In Ref.~\cite{Pilloni:2016obd}, the JPAC Collaboration performed several fittings to the BESIII data reported in Refs.~\cite{Ablikim:2013mio,Ablikim:2015tbp,Ablikim:2015swa,Ablikim:2015gda}. Both the $J/\psi\pi$ and $D^*\bar D$ channels were taken in to account as well. One difference of this work in contrast to Ref.~\cite{Albaladejo:2015lob} is that a constant $S$-wave coupling was used for the $D_1 D^*\pi$ vertex while a $D$-wave coupling, by treating the $D_1(2420)$ as a $j_\ell^P=\frac32^+$ charmed meson with $j_\ell$ the angular momentum of the light degrees of freedom,  was considered in Ref.~\cite{Albaladejo:2015lob}. 
Several different  scenarios were considered for the $Z_c(3900)$ peak in Ref.~\cite{Pilloni:2016obd}: a pole from QCD dynamics located on various Riemann sheets (II, III or IV), which might be interpreted as hinting at different origins; a pole and triangle diagrams as those in Fig.~\ref{fig:zcdiagrams}; purely kinematic enhancement by suppressing the pole contribution from the previous scenario. It was found that all of the different scenarios could describe the data with a similar quality. This led to the conclusion that the data used in their analysis were not precise enough to distinguish between these hypotheses, which questioned the existence of the $Z_c(3900)$ particle.

The conclusion in Ref.~\cite{Gong:2016jzb} is again different. Based on the results fitting to the BESIII data for $e^+e^-\to J/\psi\pi^+\pi^-$ and $D^{*-}D^0\pi^+$ at different c.m. energies, it is claimed that the $Z_c(3900)$ peak cannot be simulated by just the $D_1\bar{D}D^*$ triangle diagrams, and the $D^*\bar D$ molecular state model of the $Z_c(3900)$ gives a better description. It is pointed out that the updated BESIII data on the $e^+e^-\to J/\psi\pi^+\pi^-$ at $E_\text{cm}=4.23$ and 4.26~GeV~\cite{Collaboration:2017njt} are crucial to distinguish the different scenarios. The event number at 4.23 GeV is higher than that at 4.26~GeV and the peak is more pronounced, while the situation is opposite for the triangle-diagram-induced peaks because $E_\text{cm}=4.23$~GeV is further away from the singular arc, see the left panel of Fig.~\ref{fig:ytozpi}, meaning that the TS should not be less important for the case of $E_\text{cm}=4.23$~GeV than for the case of 4.26~GeV.

One main reason for the difference between the conclusions in Refs.~\cite{Albaladejo:2015lob,Pilloni:2016obd} could be the partial waves of the $D_1D^*\pi$ coupling, as pointed out in Ref.~\cite{Guo:2019cnpc}. Assuming that the $D$-wave coupling can be fixed from the decays of the $D_2(2460)$, the spin partner of the $D_1(2420)$ in the heavy quark limit, the $D$-wave decay width only amounts to about half of the total width of the $D_1(2420)$.
This suggests that both $S$- and $D$-wave couplings need to be taken into account.

Let us emphasize that as long as the $D_1\bar{D}$ pair can be copiously produced in electron-positron collisions, the rescattering process will contribute to the $e^+e^-\to J/\psi\pi^+\pi^-$. If the $e^+e^-$ c.m. energy is lower than about 4.29~GeV (see the left panel of Fig.~\ref{fig:ytozpi}), the kinematic configuration of this $D_1\bar{D}D^*$ loops would only allow the corresponding TS in $M_{J/\psi\pi^\pm}$ to give a peak with its maximum at a cusp exactly at the $D^*\bar{D}$ threshold. 
If more precise data indicates that the peak position of the $Z_c(3900)$ clearly deviates from the $D^*\bar{D}$ threshold, it would imply that the $Z_c(3900)$ has a genuine resonance nature.      

A similar analysis of the data of the $D^*\bar D^*$ and $h_c\pi$ invariant mass distributions for the $Z_c(4020)$, considering  triangle diagrams and allowing for the existence of a pole, is yet to be done.

\subsubsection{\texorpdfstring{${e^+e^-\to\psi(2S)\pi^+\pi^-}$}{e+ e- to psi(2S) pi+ pi-}}
\label{sec:psi2Spipi}

\begin{figure}[tbhp]
	\centering
  \includegraphics[width=0.9\textwidth]{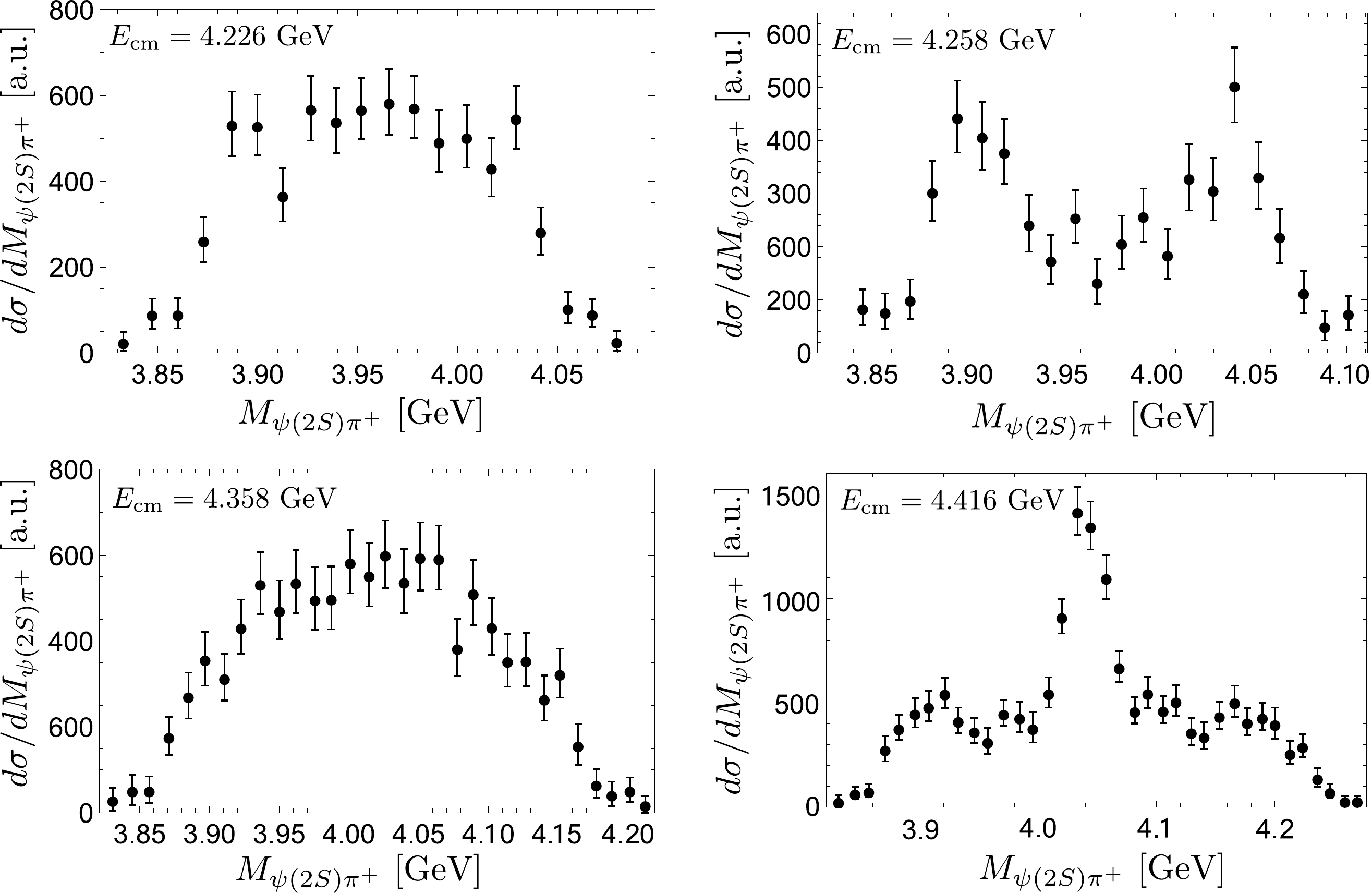}
  \\[5mm]
	\includegraphics[width=0.425\textwidth]{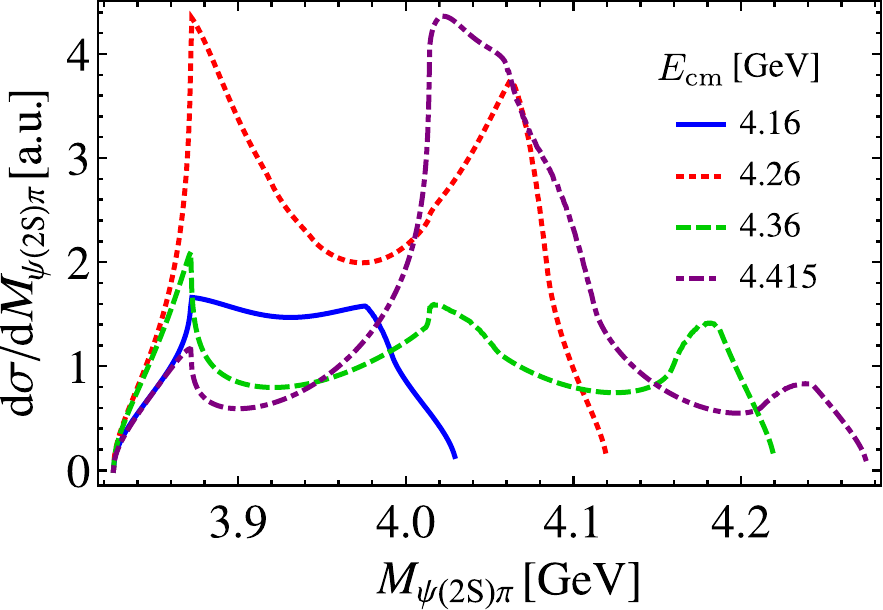}
  \begin{minipage}[t]{16.5 cm}
	\caption{The $\psi(2S)\pi$ invariant mass distributions for the $e^+e^-\to\psi(2S)\pi^+\pi^-$ process at different c.m. energies. The first and second rows: the data from the BESIII experiment~\cite{Ablikim:2017oaf}.  The third row: theoretical predictions according to the open-charm rescattering model; this plot is adapted from Ref.~\cite{Liu:2014spa}.} 
  \label{fig:Psi2Spipi-exp-theory}
  \end{minipage}
\end{figure}

In 2017 the BESIII collaboration reported the measurement of $e^+e^-\to\psi(2S)\pi^+\pi^-$ at 16 c.m. energies ($E_{\rm cm}$) ranging from $4.008$ to $4.6$~GeV~\cite{Ablikim:2017oaf}. 
In the data at $E_{\rm cm}=4.258$, there are two clear peaks around 3.9 and 4.03~GeV in the $\psi(2S)\pi^\pm$ invariant mass spectrum, see Fig.~\ref{fig:Psi2Spipi-exp-theory}, which might be from the $Z_c(3900)$ and $Z_c(4020)$ and their kinematic reflections. The data at $E_{\rm cm}=4.226$~GeV also show signals (but much weaker) for two peaks at these two $M_{\psi(2S)\pi}$ invariant masses. 
A charged charmonium-like structure with a mass around $4030$ MeV is evident in the data at $E_{\rm cm}=4.416$ GeV. However, the BESIII collaboration found that fitting the data with a Breit--Wigner resonance leading to the width parameter to vary in a wide range for different kinematical regions, and claimed that no simple interpretation of the data had been found. 
As can be seen from Fig.~\ref{fig:Psi2Spipi-exp-theory}, the shapes of the $\psi(2S)\pi$ invariant mass distribution are different for all the four c.m. energies. The unusual sensitivity of distribution on the kinematics is an essential feature of these data.

Before the BESIII observation, in 2014 there was a prediction concerning the $e^+e^-\to\psi(2S)\pi^+\pi^-$ process considering the TSs~\cite{Liu:2014spa}. The $e^+e^-\to\psi(2S)\pi^+\pi^-$ process is expected to receive contributions from the rescatterings via the $D_1\bar{D}D^*$, $D_1\bar{D}^*D^*$, $D_2\bar{D}^*D$ and $D_2\bar{D}^*D^*$ triangle diagrams~\cite{Liu:2014spa}. 
The thresholds of $D_1\bar D$, $D_1\bar D^*$ and $D_2\bar D^*$ are different from each other. As a result, in the kinematical region of interest, the physical amplitudes are significantly influenced by the nearby TSs, and the shapes vary dramatically because the distances of different energies to various thresholds are different. 
The $\psi(2S)\pi$ invariant mass distributions at several c.m. energies according to the rescattering model proposed in Ref.~\cite{Liu:2014spa} are shown in the bottom row of Fig.~\ref{fig:Psi2Spipi-exp-theory}.
One can see that the line shapes of the BESIII measurements and those of the theoretical predictions share many similar characters. In this model~\cite{Liu:2014spa}, the relative coupling strengths between different triangle diagrams are constrained by using heavy quark spin symmetry assuming that the initial vector current produces a $D$-wave $c\bar c$, which then couples to the $D_1\bar D$, $D_1\bar D^*$ and $D_2\bar D^*$ in $S$ waves. The single free parameter, an overall coupling, does not change the line shapes of the distributions.  
As discussed, the sensitivity of the distributions on the kinematics is a typical characteristics of the TS mechanism. The observed behavior of the $\psi(2S)\pi^\pm$ spectrum over different  $E_{\rm cm}$  values for the $e^+e^-\to\psi(2S)\pi^+\pi^-$ process may be taken as a strong hint at the important role played by the TSs. 
However, it is worthwhile to notice that the predicted distributions differ from the observed ones in details though the gross feature is captured. In Ref.~\cite{Liu:2014spa}, only one established intermediate charmonium state $\psi(4160)$ was introduced in the rescattering process. Some other higher charmonia, such as the $\psi(4415)$, which could couple to the $D_{1(2)}\bar D^{(*)}$ pairs with different relative strengths, may influence the $\psi(2S)\pi^\pm$ distribution curves significantly, especially in the higher c.m. energy region. 
Furthermore, the $Z_c$ states observed in other processes should also enter the game.

In fact, the $Z_c(3900)$ was taken into account in Ref.~\cite{Molnar:2019uos} in their analysis of the BESIII data in the framework of dispersion relations but without the TSs mentioned above. Considering the $\pi\pi$ final state interaction using the Omn{\`e}s representation~\cite{Omnes:1958hv} with the left-hand cut provided by the $Z_c(3900)$ exchange, a good description of the BESIII data at $E_{\rm cm}=4.226$ and $4.258$ GeV was achieved by introducing complex subtraction constants. For $E_{\rm cm}=4.416$ GeV, a charged charmonium-like state with a mass around $4016$ MeV and a width around $52$ MeV was essential to describe the data in their framework. 
However, they also pointed out that this fitted width was much larger than the width of the $Z_c(4020)$ observed in the $e^+ e^-\to h_c\pi\pi$ and $e^+ e^-\to D^*\bar{D}^*\pi$ reactions~\cite{Ablikim:2013wzq,Ablikim:2014dxl,Ablikim:2015vvn,Ablikim:2013emm}, which was averaged to be $(13\pm5)$~MeV by the PDG~\cite{Tanabashi:2018oca}.

\subsubsection{Other charged charmonium-like states}

In Ref.~\cite{Nakamura:2019btl}, possible explanations of the $Z_c(4430)$ and $Z_c(4200)$ peaks are proposed based on the TS mechanism.
The $Z_c(4430)$ was first observed by the Belle Collaboration in the $\psi(2S)\pi^+$ distribution of $\bar{B}^0\to\psi(2S)\pi^+K^-$~\cite{Choi:2007wga} with favored quantum numbers $J^{PC}=1^{+-}$~\cite{Chilikin:2013tch} and was later on confirmed by the LHCb Collaboration~\cite{Aaij:2014jqa}. The peak of $Z_c(4200)$ with $J^{PC}=1^{+-}$ was reported in the $J/\psi \pi^+$ distribution of $\bar B^0\to J/\psi \pi^+K^-$, $B^0\to J/\psi \pi^-K^+$, and $\Lambda_b^0\to J/\psi p\pi^-$ by Belle and LHCb~\cite{Chilikin:2014bkk,Aaij:2014jqa,Aaij:2016ymb,Aaij:2019ipm}.
These $Z_c(4430)^\pm$ and $Z_c(4200)^\pm$, charged states in the charmonium-mass region, are obvious good candidates of exotic states with a minimal quark content $c\bar c q\bar q$, and some theoretical proposals for the interpretations can be found in the literature (see references in Ref.~\cite{Nakamura:2019btl} and the most recent reviews~\cite{Liu:2019zoy,Brambilla:2019esw}).
As for the $Z_c(4430)$, the Argand plot is also available, and that exhibits a rapid phase motion around the maximum of the magnitude in the counterclockwise direction, which is a typical feature for a Breit--Wigner amplitude and is normally regarded as supporting the resonance interpretation of the peak~\cite{Aaij:2014jqa}.
In Ref.~\cite{Nakamura:2019btl}, for the $Z_c(4430)$ [$Z_c(4200)$], the $\bar{K}^*\psi(4260)\pi^+$ [$\bar{K}^*_2(1430)\psi(3770)\pi^+$] loop, which has a TS at 4.45~GeV (4.2~GeV), is considered.\footnote{There is another rescattering model for the $Z_c(4430)$ considering the triangle diagrams with intermediate mesons $\bar D_{s1}(2700)D \bar D^{(*)}$ for the $\bar B^0\to \psi(2S)\pi^+K^-$~\cite{Pakhlov:2014qva}. However, as can easily checked using Eqs.~\eqref{eq:mArange} and \eqref{eq:m23range}, the corresponding TSs are far away from the physical region, and thus the triangle diagrams are not expected to produce any nontrivial structure around 4.4~GeV. }
While the line shapes are not compared with the experimental data, the Argand plot of $Z_c(4430)$ is fitted well with a non-resonant background included.

The $Z_c(4200)$ was also reported in the analysis of the data for the $\Lambda_b\to J/\psi\pi^-p$ process by LHCb~\cite{Aaij:2016ymb}, which, however does not have any significance for the $Z_c(4430)$. 
It was pointed out in Ref.~\cite{Nakamura:2019btl} that a peak around 4.2~GeV in the $J/\psi\pi$ energy spectrum can be generated from triangle diagrams of the $N^*\psi(3770)\pi^-$ intermediate states. For the $N^*$ being a possible nucleon resonance in range of $1.4-1.8$~GeV, \eg, $N(1440)$, $N(1520)$, and $N(1680)$, the corresponding TSs are at 3.97, 4.004, and 4.116~GeV, respectively, covering the $Z_c(4200)$ region. 
And no $N^*$ candidate can produce a TS peak in the $Z_c(4430)$ region. These features  are consistent with the findings of the LHCb Collaboration.

Two more charged charmonium-like structures $Z_1(4050)$ and $Z_2(4250)$ were explored in Ref.~\cite{Nakamura:2019emd} by considering the TS effect.
These $Z_{1,2}$ peaks were reported in the $\chi_{c1}\pi^+$ distribution of the $\bar B^0\to\chi_{c1}\pi^+K^-$ decay by the Belle Collaboration~\cite{Mizuk:2008me}.\footnote{These peaks were not confirmed at the \babar{} experiment~\cite{Lees:2011ik}.}
The $\bar K^*X(3872)\pi^+$ and $\bar K^*_2(1430)\psi(3770)\pi^+$ loops have singularities around $4.02$ and $4.22$~GeV, respectively.
Particularly, the $Z_1(4050)$ peak around 4.025~GeV, which is in the proximity of the $X(3872)\pi$ threshold, exhibits a peculiar asymmetric shape significantly different from the Breit--Wigner one. That characteristic line shape is well reproduced by the TS effects~\cite{Nakamura:2019emd}.
The Argand plots of these peaks were predicted, which need to be confronted with future experimental measurements to provide further insights into the nature of these peaks.

\subsection{\texorpdfstring{${P_c}$}{Pc}}

Another interesting occurrence of TSs~\cite{Guo:2015umn,Liu:2015fea} is related to the hidden-charm pentaquark candidate $P_c(4450)$, observed  in the $J/\psi p$ invariant mass distribution of the decay $\Lambda_b^0\to J/\psi p K^-$ by the LHCb Collaboration in 2015~\cite{Aaij:2015tga}. It was regarded as a candidate of hidden-charm pentaquark states with masses above 4~GeV that were first predicted in Ref.~\cite{Wu:2010jy}\footnote{See also Ref.~\cite{Hofmann:2005sw} which predicts hidden-charm baryons with much lower masses.} and later on in Refs.~\cite{Wu:2010vk,Wang:2011rga,Yang:2011wz,Wu:2012md,Garcia-Recio:2013gaa,Karliner:2015ina}.
Immediately after the discovery, it was pointed out in Ref.~\cite{Guo:2015umn} that the $P_c(4450)$ mass coincides with the $\chi_{c1}\,p$ threshold and, more interestingly, the TS of the $\Lambda(1890)\chi_{c1}p$ loop diagram followed by the $\chi_{c1}p\to J/\psi p$ rescattering. 
The $\Lambda(1890)$ is a well-established hyperon with quantum numbers $J^P=3/2^+$ and a width of about 100~MeV decaying with a branching fraction of $20$--$35\%$ into $N\bar K$~\cite{Tanabashi:2018oca}. The shape produced by the scalar three-point loop integral well reproduces the peak structure of the $P_c(4450)$ though the overall strength cannot be predicted in the model. 
However, the presence of such a kinematic singularity structure does not exclude the existence of an exotic resonance in addition. It was thus suggested in Ref.~\cite{Guo:2015umn} to search for the $P_c(4450)$ in the process $\Lambda_b\to \chi_{c1}p K^-$. Were the $P_c(4450)$ completely due to kinematical effects, there should be no narrow near-threshold enhancement in the $\chi_{c1}p$ invariant mass distribution because of the Schmid theorem discussed in Section~\ref{sec:Schmid}. Following this suggestion, the LHCb Collaboration measured the branching fraction of $\Lambda_b^0\to \chi_{c1}p K^-$~\cite{Aaij:2017awb}, and the amplitude analysis is on going.
It was pointed out later in Ref.~\cite{Bayar:2016ftu} that for the $\Lambda(1890)\chi_{c1}p$  to produce a narrow peak, the $\chi_{c1}p$ pair needs to be in an $S$-wave, leading to $J^P=1/2^+$ or $3/2^+$ for the quantum numbers of the  $J/\psi p$. Yet, the quantum numbers were not unambiguously determined, and while in the published paper~\cite{Aaij:2015tga} the $3/2^-$ and $5/2^+$ options are preferred, in Ref.~\cite{Jurik:2016bdm} the $3/2^+$ is also one of the preferred quantum numbers.
More triangle diagrams which can produce peaks in the interesting energy region are discussed in Ref.~\cite{Liu:2015fea}, including various combinations of $\Lambda_c^{*}\bar D_{sJ}^{(*)}\bar D^{(*)}$.\footnote{In Ref.~\cite{Mikhasenko:2015vca}, the $D_s^*\Sigma_c D^{*}$ triangle loop is considered for the $\Lambda_b^0\to K^-J/\psi p $. However, in the weak decay vertex $\Lambda_b\to D_s^*\Sigma_c$ which proceeds through $\bar b\to \bar s c\bar c$, the isospin scalar $ud$ quark pair in the $\Lambda_b$ needs to transit into an isospin-vector pair. This breaks isospin symmetry, and thus such a mechanism is expected to have a negligibly small contribution.}

An exciting new discovery from LHCb was reported in 2019 with a one-order-of-magnitude larger data sample~\cite{Aaij:2019vzc}, superseding that in Ref.~\cite{Aaij:2015tga}. It was found that the narrow $P_c(4450)$ in fact consisted of two narrower  narrower structures $P_c(4440)$ and $P_c(4457)$, and a third narrow peak $P_c(4312)$ was discovered.
The $P_c(4312)$ is a few MeV below the $\sigd$ threshold while the higher two are close to (below) the $\sigdstar$ threshold. 
The $J/\psi p$ invariant mass distribution can be well fitted with three narrow resonances, and the resulting masses are close to the predictions in the $\Sigma_c\bar D^{(*)}$ molecular model~\cite{Wu:2012md}.
LHCb also tried to fit to the distribution by considering triangle diagrams. The considered triangles are $D_s^{**}\Lambda_c\bar D^{*0}$  for the $P_c(4312)$ (if the $D_s^{**}$ has a mass of 3288~MeV, then the triangle diagram can produce a peak at 4312~MeV), $\Lambda^*\chi_{c0}p$   for the $P_c(4440)$ (if the $\Lambda^*$ has a mass of 2153~MeV, then this triangle diagram can produce a peak at 4440~MeV) and $D_{s1}^*(2860)\Lambda_c(2595)\bar D^{*0}$ for the $P_c(4457)$. The peaks can be well fitted if all of these exchanged resonances have narrow widths, while with realistic widths the measured $J/\psi p$ invariant mass distribution could not be fitted using these considered triangle diagrams with the $P_c(4457)$ peak being an exception.
They further tried to fit to the data with a model describing the lower two peaks with Breit--Wigner resonances and the $P_c(4457)$ as from the triangle diagram mentioned above, and found that this model was more plausible than the one without introducing any resonance.

It becomes clear that the observation of the three $P_c$ structures has excluded the possibility that the old $P_c(4450)$ was completely due to TSs. 
In particular, their masses are in line with predictions of hadronic molecular models in view of heavy quark spin symmetry interpreting the $P_c(4312)$ as a $\sigd$ molecule and the $P_c(4440)$ and $P_c(4457)$ as $\sigdstar$ hadronic molecules~\cite{Xiao:2013yca,Chen:2016otp,Liu:2018zzu,Liu:2019tjn,Xiao:2019aya,Sakai:2019qph,Du:2019pij}. Furthermore, the data show a hint of another narrow structure at around 4.38~GeV that may correspond to the $\sigstard$ state in this picture as noticed in Ref.~\cite{Xiao:2019aya}, which is strongly backed by the analysis in Ref.~\cite{Du:2019pij}, showing that  the $J/\psi p$ distribution can be described very well using the $\sigh$ molecular model and a narrow peak around 4.38~GeV exists in line with the data. It persists whether or not the data in that region are included in the fit.
However, one needs to understand why the observed $P_c$ states can be produced more easily than the others predicted in the $\sigh$ hadronic molecular model. Given that the TS from the $\Lambda(1890)\chi_{c1}p$ loop is located at the dip between the $P_c(4440)$ and $P_c(4457)$, and the $D_{s1}^*(2860)\Lambda_c(2595)\bar D^{*0}$ triangle is able to describe the peak of $P_c(4457)$, TSs might play an important role in enhancing the production of the observed $P_c$ states through their interference with these resonances.
A complete amplitude analysis and   the forthcoming data to be collected at the LHC Run-3 period will be important to illuminate this issue.
In addition, reactions of different kinematics which are free of the discussed TSs can shed new light on the hidden-charm pentaquark states. Such reactions include the $J/\psi$ photoproduction~\cite{Ali:2019lzf,Wang:2015jsa,Kubarovsky:2015aaa,Karliner:2015voa,Blin:2016dlf,Lin:2017mtz,Cao:2019kst,Wang:2019krd,Winney:2019edt,Rossi:2019szt},
pion induced reactions~\cite{Lu:2015fva,Liu:2016dli} and heavy ion
collisions~\cite{Wang:2016vxa,Schmidt:2016cmd}.
The hidden-charm pentaquarks can also be searched for at $e^+e^-$ machines through reactions such as the $e^+e^-\to J/\psi(\eta_c) p \bar p$, $\Lambda_c \bar D^{(*)}\bar p $ and $\Sigma_c\bar D^{(*)}\bar p$.

\subsection{Light baryons}

An interesting work stimulated by the discovery of $P_c$ in the $J/\psi p$ invariant mass distribution of the $\Lambda_b^0\to K^- J/\psi p$~\cite{Aaij:2015tga} and the proposal of searching for hidden-strange pentaquarks $P_s$ in the analogous reaction $\Lambda_c^+\to \pi^0 \phi p$~\cite{Lebed:2015dca}, was carried out in Ref.~\cite{Xie:2017mbe}.
It was pointed out that the decay could proceed through the $\Sigma^*K^*\Lambda$ and $\Sigma^*K^*\Sigma$ loops with the $K^*\Lambda$ and $K^*\Sigma$ rescattering into the $\phi p$ final state. Such loops have TSs that can produce a peak structure around 2.02~GeV in the $\phi p$ spectrum. Given the tiny phase space for the decay and that the hidden-strange pentaquark mass was expected to be in the same region~\cite{Gao:2000az,Huang:2005gw,Gao:2017hya}, it was concluded that the $\Lambda_c^+\to \pi^0 \phi p$ was not suitable for searching for the hidden-strange pentaquarks.
In fact, there had been measurements of the $\phi p$ invariant mass distribution by the Belle Collaboration~\cite{Pal:2017ypp}. It is interesting to notice that the $\phi p$ invariant mass spectrum produced from the mentioned triangle diagrams agrees with the data without introducing any $\phi p$ resonance. However, the measured structure is still obscure due to large uncertainties, and further experimental examination is helpful to clarify the issue.

Recently in the $\Lambda_c^+\to pK^-\pi^+$ decays, the Belle collaboration reported the observation of a narrow structure in the $K^-p$ invariant mass spectrum. Its mass is 1663 MeV, and the width is 10 MeV~\cite{Shen:2018talk}. 
The signal yields of the $\Lambda_c^+\to pK^-\pi^+$ decays at Belle is about 1.5 million and the bin width of $K^-p$ invariant mass is only 1 MeV, meaning that this is a very precise measurement. 
There are several hyperon resonances whose masses are close to 1663 MeV, such as the $\Lambda(1670)$, $\Lambda(1690)$, $\Sigma(1660)$ and $\Sigma(1670)$, but all of their widths are much larger than 10 MeV. Therefore none of those established hyperons can account for this narrow structure. Interestingly, the peak position of this structure just coincides with the $\Lambda\eta$ threshold ($\sim 1663.5\ \mbox{MeV}$). It is therefore natural to expect that this peak may be related to the $\Lambda\eta$ threshold cusp. However, the two-body unitary cut usually cannot lead to such a narrow peak at the threshold as discussed in Section~\ref{sec:zc_cusp}. Other effects need to exist to narrow down the two-body threshold cusp. The $\Lambda_c^+\to pK^-\pi^+$ decay was studied in Ref.~\cite{Liu:2019dqc}. Two triangle loops composed of the  $a_0^+(980)\Lambda\eta$ and $\Sigma^{*+}(1660)\eta\Lambda$ intermediate states with a fusion of the $\Lambda\eta$ pair into the $\Lambda(1670)$ were considered.
The $\Lambda(1670)$ decays to the $K^-p$ pair in the final state.
While these loops do not develop TSs in the physical region. But the TSs in the complex plane of the second Riemann sheet are close to the $\Lambda\eta$ threshold and can enhance the threshold cusp.
Consequently, a sharp peak is produced at the $\Lambda\eta$ threshold, which is narrower than the $\Lambda(1670)$ width, $25$-$50$~MeV~\cite{Tanabashi:2018oca}. In addition, the Argand plot in the presence of the TS was found to have a circular counterclockwise behavior, but with a cusp due to the $\Lambda\eta$ threshold. As a direct prediction, one may expect to observe a similar phenomenon in $\Lambda_c\to\Sigma \pi\pi$ decays due to the $\eta\Lambda\to \pi\Sigma$ reaction.

\subsection{\texorpdfstring{${B_c\to B_s \pi\pi}$}{Bc to Bs pi pi} }
\label{sec:BctoBspipi}

The TS peak usually appears in the vicinity of the threshold of rescattering particles. Consequently, a genuine dynamic pole that is close to the threshold, such as that of a hadronic molecule, may have a signal mixed with the TS peak. 
This brings ambiguities to our understanding about the nature of some resonance-like peaks observed in experiments such as the $Z_c(3900)$ discussed in Section~\ref{sec:zc_ts}. 
If we can find some ``clean'' processes for which the TS peaks can be distinguished from the genuine resonances, it may help us establish the TS mechanism and clear up fake signals of hadron resonances. 
Since the pole position of a genuine state should not depend on a specific process, while the TS peak is rather sensitive to the kinematic conditions, one would expect that a genuine state should  appear in the processes whether or not kinematic conditions for the TS are fulfilled, as discussed in Section~\ref{sec:distinguish}. 
Or vice versa, if one observes a resonance-like peak in a process at an energy fulfilling the TS kinematic conditions and where one knows that there should no any genuine state, the signal would be likely due to the TS mechanism. 

\begin{figure}[tb]
	\centering
	\includegraphics[width=0.3\textwidth]{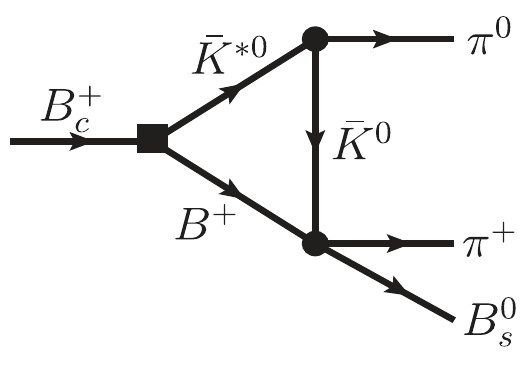} 
  \begin{minipage}[t]{16.5 cm}
	\caption{$B_c^+\to B_s^0  \pi^+\pi^0$ via the triangle rescattering diagram.} 
  \label{fig:BctoBspipi}
  \end{minipage}
\end{figure}

\begin{figure}[tb]
	\centering
	\includegraphics[width=0.6\textwidth]{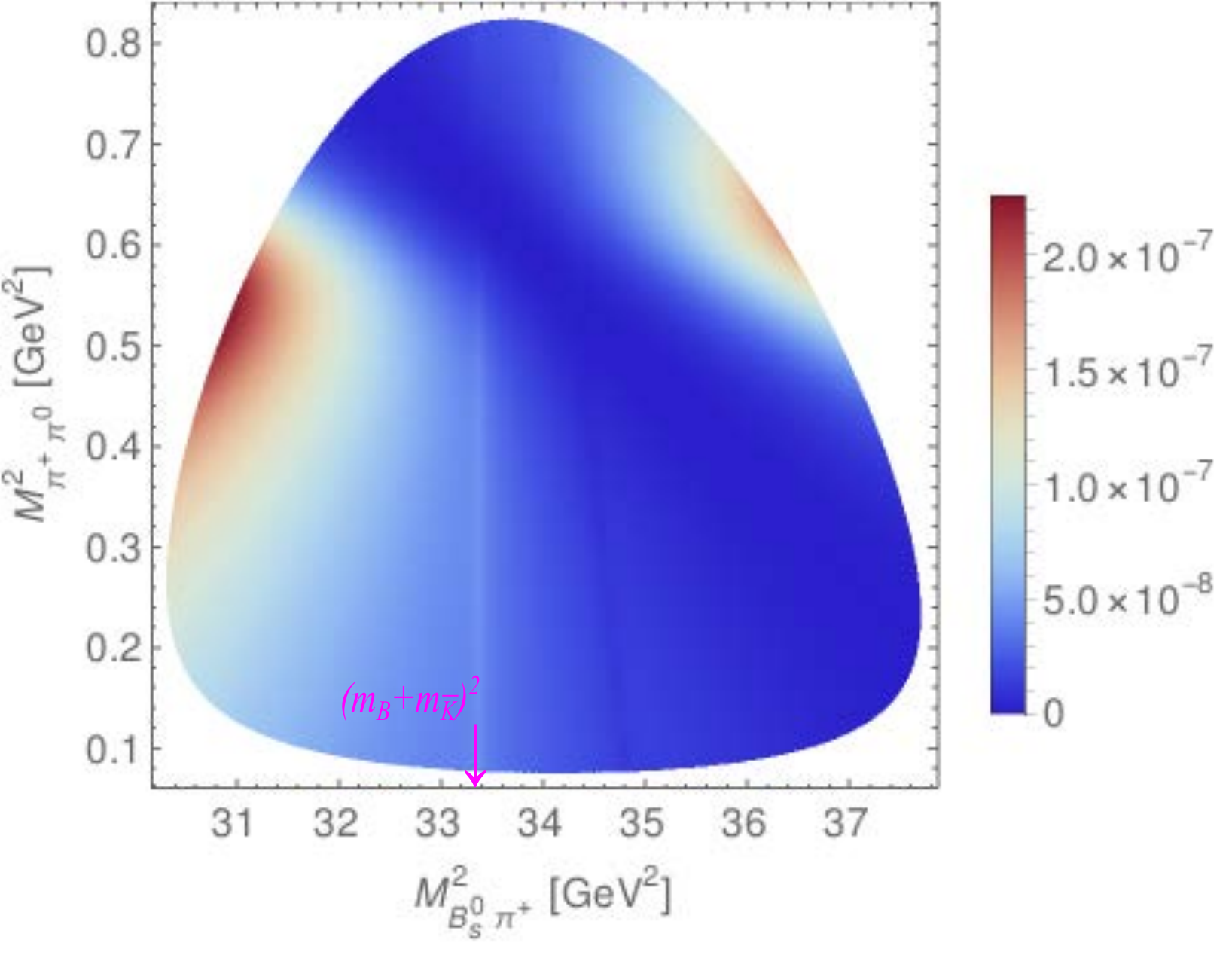} 
  \begin{minipage}[t]{16.5 cm}
	\caption{The Dalitz plot of $B_c^+\to B_s^0  \pi^+\pi^0$ predicted from the TS mechanism depicted in Fig.~\ref{fig:BctoBspipi}. The plot is adapted from Ref.~\cite{Liu:2017vsf}.} 
  \label{fig:DalitzBspipi}
  \end{minipage}
\end{figure}

The decay process $B_c^+\to B_s^0  \pi^+\pi^0$ via the $\bar{K}^*B \bar{K}$ loop as illustrated in Fig.~\ref{fig:BctoBspipi} offers the possibility to observe a TS peak without resonance pollution in the $B_s^0\pi^+$ distribution~\cite{Liu:2017vsf}. On the one hand, for the triangle diagram of Fig.~\ref{fig:BctoBspipi}, the TS in the ${B_s^0\pi^+}$ invariant mass  perfectly falls onto the physical boundary if the width of intermediate state $\bar{K}^{*0}$ is ignored. 
On the other hand, the scattering of the Nambu--Goldstone bosons ($\pi$, $\eta$ and $K$) off the heavy-light mesons ($D^{(*)}_{(s)}$, $B^{(*)}_{(s)}$) has been widely studied in the literature~\cite{Kolomeitsev:2003ac,Guo:2006fu,Guo:2009ct,Liu:2012zya,Albaladejo:2016lbb,Du:2017zvv,Albaladejo:2018mhb,Guo:2018kno,Guo:2019dpg}, and it is found that the $S$-wave $B\bar{K}$ interaction in the isospin-1 channel is rather weak, which does not support the existence of a narrow dynamically generated resonance or bound state near the $B\bar K$ threshold~\cite{Guo:2016nhb,Albaladejo:2016eps}.\footnote{The $X(5568)$ reported by the D0 Collaboration in the $B_s^0\pi^\pm$ spectrum~\cite{D0:2016mwd} is related to this issue. However, there is no theoretical reason from the QCD point of view for such a hadron to exist~\cite{Burns:2016gvy,Guo:2016nhb,Yang:2016sws}. Lattice QCD calculations did not find any signal~\cite{Lang:2016jpk}, and it is also hard to ascribe the observation of $X(5568)$ to the rescattering effects or TS contributions~\cite{Liu:2016xly,Guo:2016nhb}. More importantly, the existence of the $X(5568)$ was not confirmed in several subsequent experiments~\cite{Aaij:2016iev,Sirunyan:2017ofq,Aaltonen:2017voc,Aaboud:2018hgx}. In contrast, the $DK$ and $B\bar{K}$ $S$-wave interactions in the isospin-0 channel are generally supposed to be strong and attractive enough to generate a bound state corresponding to the $D_{s0}^*(2317)$ and its bottom partner, see, \eg, Ref.~\cite{Du:2017zvv}. }
Therefore, if a narrow peak in the $B_s^0\pi^+$ distribution around the $B\bar{K}$ threshold is observed in $B_c^+\to B_s^0  \pi^+\pi^0$ decays, it can be safely identified as a TS peak. The Dalitz plot predicted from the TS mechanism is shown in Fig.~\ref{fig:DalitzBspipi}, where the dominant background from the process $B_c^+\to B_s^0 \rho^+ \to B_s^0\pi^+\pi^0$ is also taken into account. The narrow bright band around the $B\bar{K}$ threshold can be well separated from the $\rho$ resonance.  It is thus desirable to be measured. However, a disadvantage for the proposed processes is that there is a neutral pion in the final state, which poses a severe challenge for its observation at hadron collider experiments such as LHCb.

\bigskip

%% file: section7.tex
\section{Summary}
\label{sec:7}

The threshold cusp and the triangle singularity have been known for more than half a century. They are singularities of the $S$-matrix, in addition to the pole singularities corresponding to hadrons in the QCD context, and can lead to observable effects. 
Poles, \ie, the hadron masses and widths, are a result of the nonperturbative QCD dynamics of the quark-gluon and/or hadron-hadron interactions.
Thus, poles should be located at the same positions in any amplitude that couples to the relevant particles, though their signals vary in reactions due to different couplings and interference with various other contributions and backgrounds. 
In contrast, the locations of Landau singularities depend entirely on kinematical variables including the masses of intermediate particles and the invariant masses of external ones. As a result, a threshold cusp is always fixed at the threshold, while the more complicated Landau singularities such as TSs are very sensitive to the change of kinematic conditions. As discussed, the TS peaks usually appear not far from (and above, if not at) the threshold of the rescattering particles. These features should be a key to distinguishing kinematic singularity effects from genuine resonances.

Although the threshold cusp has been used in making remarkably precise measurements of the $S$-wave $\pi\pi$ scattering lengths, there had not been unambiguous observation of the TS effects. The interest in the threshold cusps and TSs revived in recent years due to the observation of many near-threshold structures. In particular, many of them are prominent candidates of exotic hadrons, the understanding of which is vital towards establishing a clear pattern of the QCD spectrum.

In this article, we have reviewed the threshold cusps and TSs, with a special attention paid to the physical picture and their relations to the new exotic hadron candidates. On the one hand, these singularities lay traps along the way of searching for exotic hadrons, so that some signals might be due to such singularities instead of genuine resonances. On the other hand, they can be used to make precise measurements and enhance the production of hadronic molecules (or other types of hadrons with strong couplings to some intermediate states).
With the data to come from the high-luminosity experiments such as the running LHCb, Belle-II and BESIII, as well as experiments under construction or discussion such as \panda, together with the cooperative efforts from phenomenological investigations and lattice calculations, a better understanding of the QCD spectrum is foreseen.

\bigskip

%% file: appendix.tex
\appendix

% \section{Triangle loop integral in the dispersion formalism}

\section{Triangle singularity in the nonrelativistic formalism}
\label{appendix}

Let us consider the the scalar triangle loop integral.
When all of the three intermediate particles can be treated nonrelativistically, one can express the loop integral Eq.~\eqref{eq:loop_triangle} in terms of elementary functions. Nonrelativistically, $\omega_i (q) = \sqrt{m_i^2+q^2}$ in Eq.~\eqref{eq:scalarIa} can be approximated as $\omega_i(q)\simeq m_i + 2q^2/m_i$.
With this approximation, and performing the contour integration over $q^0$ for Eq.~\eqref{eq:scalarIa}, one gets a convergent integral over three-momentum. Defining
$\mu_{ij}=m_im_j/(m_i+m_j)$, $b_{12} = m_1+m_2-m_A$ and 
$b_{23}=m_2+m_3+E_B-m_A$, in the rest frame of the initial particle, we get
\begin{equation}
  I(k) \simeq \frac{4\mu_{12}\mu_{23} }{N_m} \int\! \frac{d^3\vec 
q}{(2\pi)^3}\left[
  \left(q^2 + c_1 -i\epsilon\right) \left(q^2 + c_2 - 
\frac{2\mu_{23}}{m_3} \vec{k}\cdot\vec{q} - i\epsilon
  \right) \right]^{-1},
  \label{eq:loopinter}
\end{equation}
where $c_1= 2\mu_{12}b_{12}$, and
$c_2=2\mu_{23}b_{23}+q^2\mu_{23}/m_3$ with $q\equiv|\vec q\,|$. The two factors in the integrand correspond to the two unitary cuts shown as the vertical dashed lines in Fig.~\ref{fig:triangle}.
The analytic expression for the above nonrelativistic 
integral can be worked out as~\cite{Guo:2010ak}
%\blue{(check the conditions for this expression)}
\begin{eqnarray}
    I(q) = {\cal N} \frac{1}{\sqrt{a}} \left[
\arctan\left(\frac{c_2-c_1}{2\sqrt{a(c_1-i\epsilon)}}\right) \right. 
%\nonumber\\ && 
    \left.- \arctan\left(\frac{c_2-c_1-2a}{2\sqrt{a(c_2-a-i\epsilon)}}\right)
\right],
    \label{eq:Iexp}
\end{eqnarray}
with ${\cal N}=\mu_{12}\mu_{23}/(2\pi m_1m_2m_3)$, and
$a = \left(\mu_{23}/m_3\right)^2 { q}^2$.

The intermediate particles 1 and 2 are on shell when $q^{2} + c_1 = 0$, and $q^2+ c_2 - 
{2\mu_{23}} \vec{k}\cdot\vec{q}/m_3 = 0$ is the nonrelativistic condition for 
the intermediate particles 2, 3  to be on their 
mass shell. Correspondingly, $\sqrt{|c_1|}$ and $\sqrt{|c_2|}$ define two momentum scales, leading to the averaged velocity used in the nonrelativistic power counting given in Eq.~\eqref{eq:vpc}. 
% When the loop momentum takes values around one of the them, the corresponding intermediate states are close to their mass shells. Although both of $\sqrt{|c|}$ and $\sqrt{|c'|}$ are small compared to $m_i (i=1,2,3)$, their values depend on all of the involved masses, and may be very different from each other. This is the case for some processes involving the $\X$, and will be discussed later.

The expression in Eq.~\eqref{eq:Iexp} can also be written as~\cite{Mehen:2015efa}
\begin{eqnarray}
  I(q) = {\cal N} \frac{1}{\sqrt{a}} \left[
\arcsin\left(\frac{c_2-c_1}{\sqrt{(c_2-c_1)^2+4ac_1-i\epsilon}}\right) - 
\arcsin\left(\frac{c_2-c_1-2a}{\sqrt{(c_2-c_1)^2+4ac_1-i\epsilon}}\right)
\right].
    \label{eq:Iexp2}
\end{eqnarray}
One sees that there is a logarithmic divergence (since the inverse trigonometric functions can be reexpressed in terms of logarithmic functions) at the solution of
\begin{equation}
  (c_2-c_1)^2+4ac_1 = 0 \, .
  \label{eq:nrtrising}
\end{equation}
The equation gives the nonrelativistic version of the TS~\cite{Guo:2014qra}. Because of the nonrelativistic approximation, the singularity is slightly shifted from the exact location. A comparison for a specific example can be found in the appendix of Ref.~\cite{Guo:2014qra}.  

\bigskip